\documentclass[a4,11pt]{article}

\usepackage[utf8]{inputenc}
\usepackage[english]{babel}

\usepackage{siunitx}
\DeclareSIUnit{\nothing}{\relax} 
\DeclareSIUnit\frame{frames}

\usepackage{fancyhdr}

\usepackage{etoc}

\usepackage[
    backend=biber,
    style=authoryear-comp,
    natbib=true,
    sorting=nyt,
    url=false, 
    doi=true,
    maxbibnames=99,
    firstinits=true,
    maxcitenames=2,
    parentracker=true,
    citetracker=true,
    backref,
    backrefstyle=none
]{biblatex}

\AtEveryBibitem{%
    \ifentrytype{inproceedings}{
        \clearlist{publisher}%
        \clearname{editor}%
        \clearfield{url}%
        \clearfield{urlyear}%
        \clearfield{month}%
        \clearfield{eventtitle}%
        \clearfield{location}%
    }{
        \clearfield{url}%
        \clearfield{urlyear}%
        \clearfield{eprintclass}%
    }
}

\DeclareSourcemap{
  \maps[datatype=bibtex]{
     \map{
        \step[fieldsource=doi,final]
        \step[fieldset=eprint,null]
        }
      }
}

\usepackage{longtable}
\usepackage{multirow}

\usepackage{enumitem}
\usepackage{pifont} 
\setitemize{noitemsep,topsep=0pt,parsep=2pt,partopsep=0pt,leftmargin=10pt}
\setitemize[2]{label=\textbullet}

\usepackage{lipsum} 

\usepackage[english]{babel}
\usepackage[autostyle, english = british]{csquotes}
\MakeOuterQuote{"}

\usepackage[most]{tcolorbox}
\newcommand{\myintrobox}[2]{
\begin{tcolorbox}
[colback=lime!5!white,colframe=lime!75!black,title=Coming up #1]
\setitemize[1]{label=\ding{46}} 
#2
\end{tcolorbox}
}

\newcommand{\mysummarybox}[2]{
\begin{tcolorbox}
[colback=orange!5!white,colframe=orange,title=Summary #1]
\setitemize[1]{label=\ding{51}}
#2
\end{tcolorbox}
}

\newcommand{\myRCLbox}[2]{
\begin{tcolorbox}
[colback=black!5!white,colframe=black!50!white,title={Requirements, Capabilities, and Limitations of }#1]
\setitemize[1]{label=\ding{51}}
#2
\end{tcolorbox}
}

\usepackage[a4paper, portrait, left=15mm,right=15mm,top=5mm,bottom=25mm,includehead,nomarginpar,headheight=10mm]{geometry}
\usepackage{titling}

\usepackage{enumitem}
\setenumerate{noitemsep,topsep=0pt,parsep=0pt,partopsep=0pt}

\usepackage{helvet}
\usepackage{relsize}

\usepackage{xspace}
\def\F0{$F_0$\xspace}

\usepackage[dvipsnames]{xcolor}
\usepackage{hyperref}
\hypersetup{
    colorlinks=true,
    linkcolor=Blue,
    urlcolor=Blue,
    citecolor=Blue,
    hypertexnames=true,
    pdftitle={Landscape Report - AI Speech Generation},
    pdfauthor={Simon King},
    pdfkeywords={AI, speech generation, speech synthesis, spoofing, deepfake}
}

\usepackage[acronym,automake,toc]{glossaries}
\setglossarystyle{list}
\setacronymstyle{long-short}

\newglossaryentry{deepfake}
{
    name=deepfake,
    description={Artificially generated content (e.g., speech) intended to deceive the perceiver into believing it is natural}
}

\newglossaryentry{zero_shot}
{
    name=zero-shot,
    description={Generating a speaker identity from a model that did not have any examples of this speaker in the training data, by providing a short (usually a few seconds) sample recording of the target speaker as an input at inference time. Since the model is invariably trained to use input samples of the target speaker, it is not performing a new \textit{task} per se (so "zero-shot" is actually being used inconsistently with the wider field of machine learning)}
}

\newglossaryentry{raw_waveform}
{
    name=raw waveform,
    description={This term is widespread, and intended to contrast with \textit{features} extracted from the waveform. `Raw' is  misleading, because most recorded audio has undergone some processing (e.g., web scraped audio has invariably undergone compression and decompression, so is certainly not "raw"). This report will simply say "waveform"}
}

\newglossaryentry{source_filter_model}
{
    name=source-filter model,
    description={A conceptual model of speech production in which a basic sound source is filtered by the vocal tract}
}

\newglossaryentry{waveform}
{
    name=waveform,
    description={A time series of amplitude against time, representing the signal captured by a microphone}
}

\newglossaryentry{conditioning}
{
    name=conditioning,
    description={Derived from the term "conditional probability", now used more generally to mean providing an additional input to a model to constrain, or otherwise control, its output}
}

\newglossaryentry{meta-data}
{
    name=meta-data,
    description={Information about, or a description of, data. Examples: the location where a speech signal was recorded; the author of a text}
}

\newglossaryentry{voice_cloning}
{
    name=voice cloning,
    description={Creating a synthetic voice with a given speaker identity, using only a limited sample of speech from that speaker}
}

\newglossaryentry{latent}
{
    name=latent,
    description={A hidden variable, such as the activations within a neural network}
}

\newglossaryentry{generative}
{
    name=generative,
    description={A model is described as "generative" if it learns the distribution of some training data, and is then able to generate new data samples from that distribution (i.e., that could plausibly have occurred in the training data, but did not.) }
}

\newglossaryentry{speaker_identity}
{
    name=speaker identity,
    plural=speaker identities,
    description={The identity of a specific individual, as conveyed by their speech}
}

\newglossaryentry{voice_actor}
{
    name=voice actor,
    description={A professional speaker, such as might be used to record data for training an AI speech generation model}
}
\newglossaryentry{studio_conditions}
{
    name=studio conditions,
    description={A quiet, reverberation-free environment with high-quality equipment, used to record speech, generally from a \gls{voice_actor}; contrasts with \gls{found_data}}
}

\newglossaryentry{linguistic_content}
{
    name=linguistic content,
    description={Broadly, the message conveyed by a written sentence or spoken utterance; more narrowly, often used to refer to the words, regardless of \textit{how} they are said}
}

\newglossaryentry{supervised}
{
    name=supervised,
    description={A form of machine learning in which the model learns from labelled data, such as text-speech pairs}
}

\newglossaryentry{unsupervised}
{
    name=unsupervised,
    description={A form of machine learning in which the model learns from unlabelled data, such as speech without a transcript}
}

\newglossaryentry{one_hot}
{
    name=one-hot,
    description={A vector comprising all zeros except for a single entry that is set to one. A common way to represent a label from a closed set}
}

\newglossaryentry{prompt}
{
    name=prompt,
    description={In the context of AI, a user-supplied input that influences a model, especially during \gls{inference}}
}

\newglossaryentry{bottleneck}
{
    name=bottleneck,
    description={A reduced-dimension representation designed to compress information, often with the goal of eliminating irrelevant or unwanted information}
}

\newglossaryentry{leakage}
{
    name=leakage,
    description={Unintentionally exposing information in a model's output; in the context of \gls{VC}, evidence of the \gls{source_speaker} that can be found in the output speech}
}

\newglossaryentry{causal}
{
    name=causal,
    description={In the context of signal processing or AI speech generation, being able to output the current speech sample based only on past and current algorithm or model inputs. Note: sometimes misspelled "casual" in papers!}
}

\newglossaryentry{streaming}
{
    name=streaming,
    description={Being able to incrementally produce output as input is being received}
}

\newglossaryentry{spoofing}
{
    name=spoofing,
    description={Attempting to deceive an \gls{ASV} system using synthetic speech}
}

\newglossaryentry{anti_spoofing}
{
    name=anti-spoofing,
    description={Detecting that speech presented to an \gls{ASV} system is synthetic}
}

\newglossaryentry{spoofed}
{
    name=spoofed,
    description={Synthetic speech intended to deceive an \gls{ASV} system}
}

\newglossaryentry{bona_fide}
{
    name=bona fide,
    description={Genuine. In the context of speech: natural speech and not synthetic. Sometimes written bonafide}
}

\newglossaryentry{hallucination}
{
    name=hallucination,
    description={When a \gls{generative} model produces output that appears to be plausible but is actually incorrect}
}

\newglossaryentry{countermeasure}
{
    name=countermeasure,
    description={A device or algorithm designed to reject \gls{spoofed} speech before it reaches an \gls{ASV} system}
}

\newglossaryentry{end_to_end}
{
    name=end-to-end,
    description={For \gls{TTS}, a model that takes unprocessed text (\textit{not} \gls{normalised_text}) as input and outputs a \gls{waveform}, all with a single model}
}

\newglossaryentry{speech_editing}
{
    name=speech editing,
    description={Automatically modifying the content of a naturally-spoken utterance, typically changing only a small part (e.g., one word) and leaving the rest intact. Variants include speech continuation and infilling}
}

\newglossaryentry{linguistic_specification}
{
    name=linguistic specification,
    description={A specification of how to say a given text. Minimally includes pronunciation, but may also specify \gls{prosody}}
}

\newglossaryentry{prosody}
{
    name=prosody,
    description={A linguistic abstraction which includes the rhythm and intonation of speech: "how" something is said. In speech technology literature, "prosody" is often used to mean pitch, loudness, duration, and voice quality. That usage is technically incorrect: those are the acoustic correlates of prosody, not prosody itself}
}

\newglossaryentry{front_end}
{
    name=front-end,
    description={The set of modules in a \gls{TTS} system that take the input text and produce a \gls{linguistic_specification}}
}

\newglossaryentry{normalised_text}
{
    name=normalised text,
    description={Text comprising only fully-spelled-out words}
}

\newglossaryentry{found_data}
{
    name=found data,
    description={Data that was created for another purpose, but is now being used to train a model, for example an audiobook}
}

\newglossaryentry{acoustic_model}
{
    name=acoustic model,
    description={Given the \gls{linguistic_specification} produced by the \gls{front_end}, the acoustic model predicts the \gls{acoustic_specification} ready for input to the \gls{waveform_generator}}
}

\newglossaryentry{auto_regressive}
{
    name=auto-regressive,
    description={Prediction of the current value of a  (time) series based on its previous values}
}

\newglossaryentry{loss_function}
{
    name=loss function,
    description={A function measuring the difference between a model's output and the desired value, to be minimised during \gls{training}}
}

\newglossaryentry{spectrogram}
{
    name=spectrogram,
    description={A time-frequency representation of a (speech) signal plotting the spectrum against time; "spectrogram" generally means "magnitude spectrogram" with phase having been discarded}
}

\newglossaryentry{neural_network}
{
    name=neural network,
    description={A type of model that can approximate arbitrary functions by using distributed representations, and can be learned from data using simple algorithms to minimise a \gls{loss_function}; inspired by the human brain,  in a very abstract way}
}

\newglossaryentry{training}
{
    name=training,
    description={The process of automatically learning the parameters of a model from data}
}

\newglossaryentry{inference}
{
    name=inference,
    description={The process of generating output from a model, given a specific input}
}

\newglossaryentry{adversarial}
{
    name=adversarial,
    description={In the context of \gls{training} a \gls{generative} model, a type of \gls{loss_function} provided by a real vs. fake classifier}
}

\newglossaryentry{real_time}
{
    name=real time,
    description={The ability of a model (e.g., for \gls{VC}) to generate output without interruption, during the same time that input is received}
}

\newglossaryentry{waveform_generator}
{
    name=waveform generator,
    description={Generates a speech waveform from the \gls{acoustic_specification}. The method used will depend on the form of that specification. For a \gls{mel_spectrogram}, this component is called a \gls{vocoder}. For \glspl{audio_token}, it will be the decoder of the \gls{audio_codec}}
}

\newglossaryentry{acoustic_specification}
{
    name=acoustic specification,
    description={A representation of a speech signal that contains sufficient information to generate a waveform. Over time it has evolved from hand-engineered \gls{parametric_vocoder} features, to the \gls{mel_spectrogram}, and most recently to \glspl{audio_token}}
}

\newglossaryentry{mel_spectrogram}
{
    name=mel spectrogram,
    description={A \gls{spectrogram} in which the frequency axis is on the mel scale (rather than a linear Hertz scale)}
}

\newglossaryentry{audio_codec}
{
    name=audio codec,
    description={An algorithm or model that is able to encode a \gls{waveform} into a compressed representation, and then reconstruct the waveform from that representation; generally, the reconstruction will be lossy (i.e., not exact)}
}

\newglossaryentry{pronunciation_dictionary}
{
    name=pronunciation dictionary,
    description={TO WRITE also mention lexicon}
}

\newglossaryentry{semantic_token}
{
    name=semantic token,
    description={A misleading term often used to refer to a discrete representation obtained from a \gls{SSL} model, intended to convey that information about, say, the \gls{speaker_identity} or \gls{prosody} has been removed; a slightly better term might be "content token"}
}

\newglossaryentry{audio_token}
{
    name=audio token,
    description={A term used to describe an \gls{audio_code} when it is generated from a \gls{SLM}. A speech \gls{waveform} can be encoded as a sequence of audio tokens}
}

\newglossaryentry{text_tokens}
{
    name=text tokens,
    description={Pieces of text, typically larger than a single \gls{character}, and usually learned from data by a text tokeniser}
}

\newglossaryentry{audio_code}
{
    name=audio code,
    description={One of a finite set of discrete codes used by an \gls{audio_codec} to represent speech. See also \gls{audio_token}}
}

\newglossaryentry{vocoder}
{
    name=vocoder,
    description={A "voice encoder" that encodes a speech \gls{waveform} into a representation suitable for transmission, modelling, or manipulation, and then reconstructs a \gls{waveform} from that representation}
}

\newglossaryentry{parametric_vocoder}
{
    name=parametric vocoder,
    description={A form of \gls{vocoder} where the representation is a set of interpretable parameters capturing key properties of the signal, such as the \gls{F0} and the vocal tract frequency response}
}

\newglossaryentry{neural_vocoder}
{
    name=neural vocoder,
    description={A model capable of generating a speech \gls{waveform} from a (mel) \gls{spectrogram}, or other \gls{acoustic_specification}}
}

\newglossaryentry{finetuning}
{
    name=finetuning,
    description={Updating the parameters of an already-trained model (generally a \gls{neural_network}) using additional data, for example to make it better in a specific domain}
}

\newglossaryentry{unit_selection}
{
    name=Unit Selection,
    description={A method for generating the \gls{waveform} for \gls{TTS} that involves selecting small speech \gls{waveform} units, such as diphones, from a large, naturally-spoken speech database, then concatenating them using signal processing}
}

\newglossaryentry{hybrid_unit_selection}
{
    name=Hybrid Unit Selection,
    description={A variant of \gls{unit_selection} in which the selection of units is guided by the predictions of a \gls{SPSS} model}
}

\newglossaryentry{source_speaker}
{
    name=source speaker,
    description={The \gls{speaker_identity} of the speech input to a \gls{VC} system}
}

\newglossaryentry{target_speaker}
{
    name=target speaker,
    description={The desired \gls{speaker_identity} of the speech output by a \gls{VC} or \gls{TTS} system}
}

\newglossaryentry{speaker_embedding}
{
    name=speaker embedding,
    description={A vector representing \gls{speaker_identity}, obtained from a sample of  speech; for example, a \gls{latent} from a \gls{neural_network} that was trained to perform speaker classification}
}

\newglossaryentry{pseudo_speaker}
{
    name=pseudo speaker,
    description={An invented \gls{speaker_identity} intended \textit{not} to belong to any known real person}
}

\newglossaryentry{imposter}
{
    name=imposter,
    description={A user presenting their speech to an \gls{ASV} system, who is \textit{not} the claimed speaker}
}

\newglossaryentry{voice_anonymisation}
{
    name=voice anonymisation,
    description={Removing evidence of the \gls{source_speaker}'s identity from a speech signal, perhaps by replacing it with a \gls{pseudo_speaker}'s identity}
}

\newglossaryentry{naturalness}
{
    name=naturalness,
    description={A term used very widely with regard to synthetic speech, but rarely defined! Usually accepted to mean "human-like"}
}

\newglossaryentry{synthetic_speech}
{
    name=synthetic speech,
    description={Speech generated from a model, not produced by a human}
}

\newglossaryentry{phoneme}
{
    name=phoneme,
    description={An abstract linguistic concept that only exists in the perception of a listener: a phoneme is not a spoken sound, but rather a set of possible sounds that a listener will group into a single category. Changing one phoneme will potentially change a word into a different word (creating a minimal pair). The set of phonemes is therefore language-specific. The term is much abused in speech technology! A succinct definition is "the smallest unit of speech that distinguishes one word from another in a particular language" \citep{MoorePhoneme}. Phonemes can be used to specify the pronunciation of a word in a dictionary. See also \gls{phone}}
}

\newglossaryentry{phone}
{
    name=phone,
    description={A distinct speech sound, such as a specific individual acoustic realisation of a \gls{phoneme}. The set of all possible speech sounds is language-independent. A phonetic transcription of speech comprises phones (not phonemes)}
}

\newglossaryentry{character}
{
    name=character,
    description={A member of a character set, such as ASCII, used in the computer science sense \citep[][Section 4.2.1]{TaylorBook}. See also \gls{grapheme}}
}

\newglossaryentry{grapheme}
{
    name=grapheme,
    description={An abstract linguistic concept: the smallest unit used in writing. The term is used somewhat loosely in speech technology to generally means the same thing as \gls{character}}
}

\newglossaryentry{dataset}
{
    name=dataset,
    description={\todo{TO WRITE} }
}

\newglossaryentry{pitch}
{
    name=pitch,
    description={Pitch is a subjective, perceptual attribute related to the frequency of a sound. In speech technology, the word "pitch" is used synonymously with $F_0$, even though this is technically incorrect}
}

\newglossaryentry{checkpoint}
{
    name=checkpoint,
    description={A neural network saved after \gls{training}, containing all of the model parameters (and optionally other information required to continue training)}
}

\newacronym{SLM}{SLM}{Speech Language Model}

\newacronym{ASR}{ASR}{Automatic Speech Recognition}
\newacronym{TTS}{TTS}{Text-to-Speech}
\newacronym{VC}{VC}{Voice Conversion}
\newacronym{ASV}{ASV}{Automatic Speaker Verification}
\newacronym{SSV}{SSV}{Source Speaker Verification}
\newacronym{HMM}{HMM}{Hidden Markov Model}
\newacronym{DNN}{DNN}{Deep Neural Network}
\newacronym{LLM}{LLM}{Large Language Model}
\newacronym{SPSS}{SPSS}{Statistical Parametric Speech Synthesis}
\newacronym{MOS}{MOS}{Mean Opinion Score}
\newacronym{MUSHRA}{MUSHRA}{MUltiple Stimuli with Hidden Reference and Anchor}
\newacronym{SotA}{SotA}{State-of-the-Art}
\newacronym{SSL}{SSL}{Self-Supervised Learning}
\newacronym{IVR}{IVR}{Interactive Voice Response}
\newacronym{GMM}{GMM}{Gaussian Mixture Model}
\newacronym{PPG}{PPG}{Phonetic Posteriorgram}
\newacronym{RVQ}{RVQ}{Residual Vector Quantisation}
\newacronym{NAC}{NAC}{Neural Audio Codec} 
\newacronym{NLP}{NLP}{Natural Language Processing}
\newacronym{VAD}{VAD}{Voice Activity Detection}
\newacronym{EER}{EER}{Equal Error Rate}

\newacronym{MFCC}{MFCC}{Mel-Frequency Cepstral Coefficients}
\newacronym{CQCC}{CQCC}{Constant Q Cepstral Coefficients}
\newacronym{SVM}{SVM}{Support Vector Machine}
\newacronym{SME}{SME}{Small and Medium-sized Enterprise}
\newacronym{B2B}{B2B}{Business-to-Business}
\newacronym{API}{API}{Application Programming Interface}

\newacronym{F0}{$F_0$}{fundamental frequency}

\makeglossaries

\begin{document}

\pagestyle{fancy}

\fancyhead{}
\fancyfoot{}

\fancyfoot[R]{Page~\thepage}
\fancyfoot[C]{}
\fancyfoot[L]{}

\title{The landscape of AI speech generation and deepfake detection\\
\vspace{10mm}
\mbox{}
\includegraphics[width=10cm]{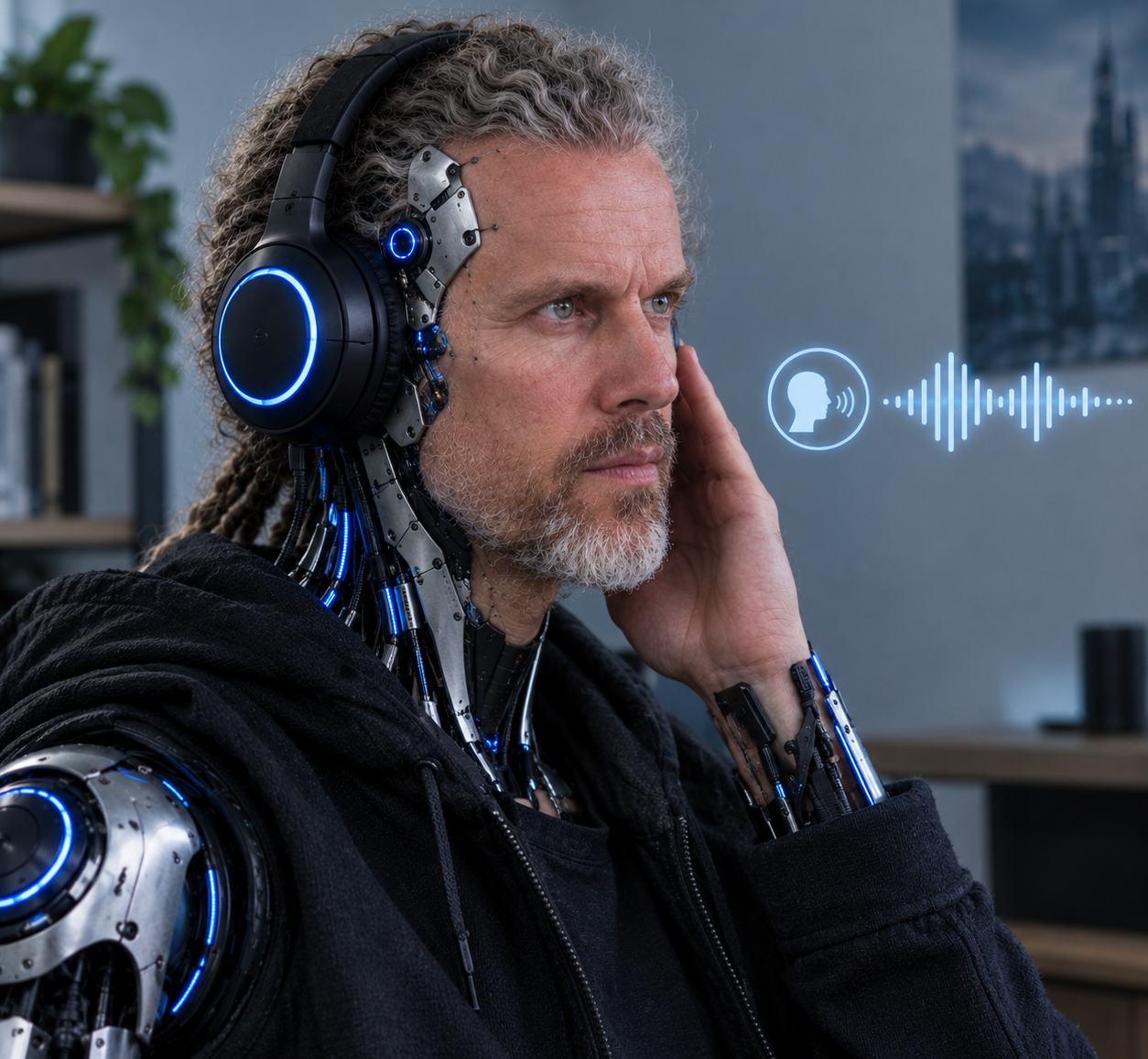}\\
}
\author{Simon King}
\date{May 2025 (minor revisions November 2026)}

\renewcommand{\headrulewidth}{0pt}
\maketitle\thispagestyle{fancy}

\begin{center}
Speech generated by AI can sometimes be hard to distinguish from natural, human speech,\\so automatic methods are required to detect it.

\vspace{10mm}

\emph{This report is my independent opinion and was researched and written entirely by me.\\The only AI-generated content is the image above. AI was not used in any other way.}

\vspace{30mm}
© Simon King

\end{center}

\newpage
\fancyhead[L]{\textbf{AI Speech Generation}}
\renewcommand{\headrulewidth}{1pt}

\begin{tabular}{cp{10cm}}
\smash{\raisebox{-6cm}{\includegraphics[width=6cm]{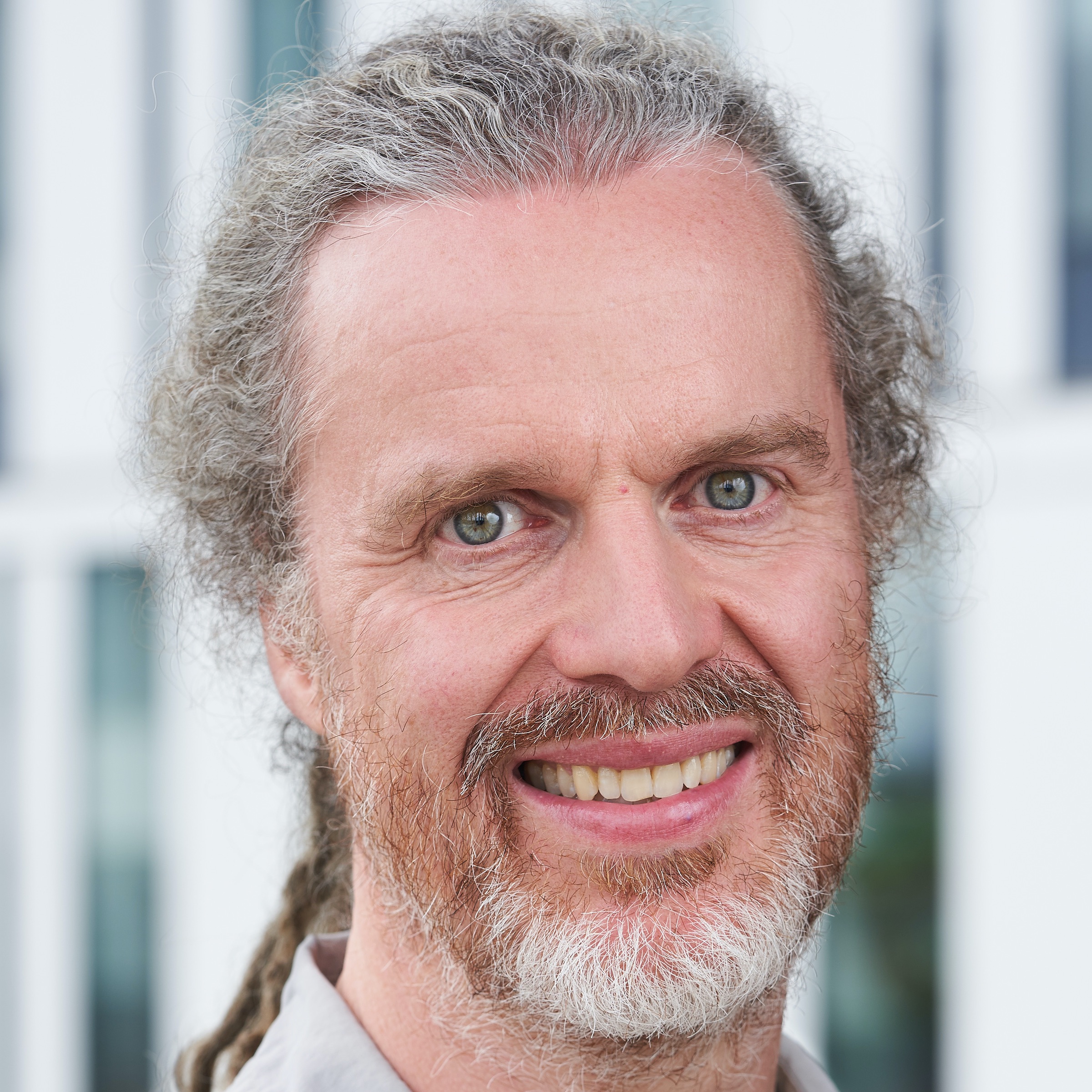}}} & \textbf{Simon King} is Professor of Speech Processing at the \mbox{University} of Edinburgh, where he is director of the \mbox{Centre} for Speech Technology Research and of a long-running Masters programme in Speech and Language Processing. \mbox{Simon} has research interests in speech synthesis, speaker verification spoofing, and signal processing, with over 270 publications. He is a Fellow of the Institute of Electrical and Electronics Engineers (IEEE) and a Fellow of the International Speech Communication Association (ISCA), has led the annual Blizzard Challenge speech synthesis evaluations since 2007, and was Technical Chair of the Interspeech 2023 conference. \newline
He also provides scientific advice to industry, working with companies of every size and at every stage of their lifecycle: major corporations, high-growth specialist companies, start-ups, spin-outs, and pre-commercialisation projects.
\vspace{3mm}
\newline
research website: \href{https://www.cstr.ed.ac.uk}{www.cstr.ed.ac.uk}
\vspace{3mm}
\newline
teaching website: \href{https://speech.zone}{speech.zone}
\vspace{3mm}
\newline
email: \href{mailto:Simon.King@ed.ac.uk}{Simon.King@ed.ac.uk}
\end{tabular}
\vspace{25mm}

\begin{tabular}{lr|p{10cm}}
\multicolumn{3}{c}{Document history}\\\hline
Date & Version & Status \\\hline
2025-03 && literature review complete\\
2025-04 && writing\\
2025-05 &1.0& first complete version\\
2026-01 &1.1& minor editorial changes\\
2026-03 &1.2& removal of "Commercial Pathways" section\\
2026-11 &1.3& minor edits, to prepare for arXiv publication
\end{tabular}

\newpage

\setcounter{tocdepth}{2}
\tableofcontents

\etocsettocdepth{2}
\newlength\tocrulewidth
\setlength{\tocrulewidth}{1.5pt}
\etocsettocstyle{\rule{\linewidth}{\tocrulewidth}\vskip0.5\baselineskip}{\rule{\linewidth}{\tocrulewidth}}

\newpage
\fancyhead[R]{\leftmark}

\section{Executive Summary}

This report informs the reader about the current state-of-the-art in two intertwined topics. The primary topic is \textbf{speech generation}, which is an umbrella term for the automatic creation of synthetic speech with specific, desired properties -- typically the verbal content and speaker identity, but sometimes also the prosody, style, or emotion. Speech generation can be achieved by Text-to-Speech synthesis, Voice Conversion, or automated speech editing. The latest methods for speech generation are currently described as Artificial Intelligence (AI). 

But "AI speech generation" is just a new name for an old problem that has been worked on for many decades and is already used for many commercial applications. So the report provides a history, to set the scene. This will enable the reader to understand why it is a hard problem, and what progress has been made to date. We find that the naturalness of synthetic speech has been steadily improving for many years. 

The report then describes the current state-of-the-art in AI speech generation. This is done in terms of the components used to build a speech generation system, then with regard to the requirements, capabilities, and limitations of current systems, and finishes with a look forward to what might happen in the near future. Current systems can produce very natural-sounding speech, but are less predictable and less stable than older methods. Their behaviour and limitations are not yet fully understood, but are thought to depend more on the data than on the type of model used.

Speech synthesis met the requirements for widespread commercial use a long time ago. The most recent improvements are in one sense nothing more than just another step in a long journey. However, naturalness has now increased to a level that may deceive listeners into believing they are hearing natural speech. This is not incremental, but a step change that opens up the possibility of using synthetic speech to mislead both human listeners and automated systems.

The secondary topic is the automatic detection of speech samples that have been generated by AI, as opposed to naturally spoken by a human. In the context of a bad actor attempting to pass off generated speech as natural, the speech might be referred to as a deepfake, and the detection task as \textbf{anti-spoofing} (when the target is an automated system) or \textbf{deepfake detection} (when the target is a human listener). Detecting speech deepfakes is a much newer field of research than speech generation. We will find that one fundamental problem is detecting a deepfake that was created with a novel form of AI speech generation: one that has never been encountered before. Another fundamental difficulty is that all deepfake detectors will inadvertently label some genuine speech samples as fake.

The report wraps up with
some concluding remarks which include a summary of the benefits and threats of AI speech generation.

\newpage

\section{Introduction and background}
\label{section:introduction}
\glsresetall
\begingroup
\parindent=0em
\localtableofcontents
\endgroup

The report will be inclusive and cover all current methods of speech generation that could be used to create a deepfake, without limiting to certain types of model, or being concerned with a definition of AI. The broad structure of this report is chronological: after some definitions (Section \ref{section:definitions}), brief histories and timelines (Sections \ref{section:history:TTS} to \ref{section:history:spoofing}) are used to set the scene for an overview of the current \gls{SotA} in speech generation (Section \ref{section:SotA:generation}) then \gls{deepfake} detection (Section \ref{section:SotA:detection}).

\textit{The following topics are out-of-scope: non-speech audio generation; music and singing; \gls{deepfake} detection in modalities other than speech; classification tasks other than the detection of synthetic speech.}

\subsection{How to read this report}
This report is intended to be accessible to multiple audiences. For readers unfamiliar with speech technology, there are conceptual explanations of how each technique works; these are supported by further details and references for domain experts. There are "Coming up" and "Summary" boxes to help you navigate. Inevitably, there are many technical terms and acronyms in this report. Those that look like this  -- \glsreset{TTS}\gls{TTS} -- or this -- \gls{deepfake} -- have an entry in the list of acronyms or the glossary of technical terms starting on page \pageref{page:acronyms} and are hyperlinked, as are in-text citations and section cross-references.

\subsection{Definitions}
\label{section:definitions}

\myintrobox
{in this section}
{
\begin{itemize}
\item Definitions of AI speech generation tasks.
\item Relevant speech classification tasks.
\end{itemize}
}

\subsubsection{Speech generation tasks}

The boundaries between many speech generation tasks have become blurred. Conceptually this is simply because, whilst the input varies, the output is always speech. It is the \textit{generation} of a speech signal that is the hardest part, regardless of input, for reasons including: the output speech signal contains information that is not fully determined by the input; human listeners are sensitive to imperfections in speech; our understanding of speech, from both linguistic and technical perspectives, is incomplete.

\paragraph{Text-to-Speech} (TTS) takes plain text as input, which might contain symbols, numbers, and other non-standard words. This contrasts with \gls{normalised_text} which comprises only fully spelled-out words and punctuation symbols. Classical \gls{TTS} systems generate speech for a small set of \glspl{speaker_identity} that are fixed when the system is built. In commercial systems, these speakers are professional \glspl{voice_actor}, recorded under \gls{studio_conditions}. 

Speaker-adaptive \gls{TTS} systems can generate new \glspl{speaker_identity} without substantial changes to the system, and generally using only a small amount of speech (relative to what is required to build a non-adaptive system) from the \gls{target_speaker}. Speaker-adaptive \gls{TTS} can be achieved through various means, depending on the requirements of the use case: applying \gls{VC} to the synthetic speech generated from an existing system; including some data from the \gls{target_speaker} when constructing the system; using a small sample of speech from the \gls{target_speaker} to influence the output of the system during the process of synthesis. This last method is known as \gls{zero_shot} because the adaptation to the new  \gls{target_speaker} occurs during \gls{inference}, without any changes to the system itself. This is currently the most prevalent approach, at least in research publications.

Creating a synthetic voice from a relatively small amount of speech, using either the second or third of these methods, is commonly known as \gls{voice_cloning}. This term was probably coined by Yamagishi and collaborators in their work providing synthetic voice to people with Motor Neurone Disease \citep{yamagishiVoiceCloning}. This is one of the beneficial applications of speech generation technology (page \pageref{page:speakunique}). A brief history of \gls{TTS} is coming up in Section~\ref{section:history:TTS} and the \gls{SotA} can be found in Section~\ref{section:SotA:generation}.

\paragraph{Voice Conversion} (VC) takes speech from a \gls{source_speaker} as input and generates the same utterance in the voice of a \gls{target_speaker}. As with \gls{TTS}, a variety of approaches can be taken. As we will see, the techniques used for \gls{VC} have steadily converged with those of the much larger (and commercially-useful) field of \gls{TTS}, to the point where some systems can perform both tasks. Variants of \gls{VC} include the conversion of properties other than speaker identity, such as speaking style or accent, although the conversion of \gls{speaker_identity} is by far the most common application, for both speech and singing. A brief history of \gls{VC} is coming up in Section \ref{section:history:VC} and the \gls{SotA} can be found in Section \ref{section:SotA:generation}. 

\paragraph{Speech editing, including infilling and continuation}
In these tasks, the input is a speech signal and the output is the same speech signal with some specific, local modifications. A portion of the speech may be edited (changing the words spoken), a missing portion of the speech (e.g., that was inaudible due to noise) may be filled in, or additional speech may be generated to continue the utterance. 

Because the most common use case is the editing of speech via a text transcription, it is typically done with very similar models to those used for \gls{TTS}. A case study is provided in Section~\ref{section:SotA:speech_editing}.

\paragraph{Voice privacy} is a broad term, encompassing topics that are out-of-scope here, such as encryption. But one way to achieve voice privacy is "\gls{voice_anonymisation} [which] is the task of altering the speaker’s voice to hide their identity to the greatest possible extent, while leaving all other speech attributes (traits, states, and spoken contents) intact."~\citep[Section 1]{VoicePrivacy2020Results}. The generated speech will have the identity of a \gls{pseudo_speaker} and the original \gls{source_speaker} will ideally be undetectable by either human listeners or an \gls{ASV} system. Variations include hiding only specific characteristic(s) such as age, accent, or ethnic origin although, just as with \gls{VC}, \gls{speaker_identity} is by far the most common. 

Most methods for \gls{voice_anonymisation} use the same techniques as \gls{VC}, but with a specially-devised \gls{target_speaker} to achieve the requirement of sounding like a \gls{pseudo_speaker} and not like any known real individual. However, some \gls{voice_anonymisation} systems do simply select a  real individual from a database as the \gls{pseudo_speaker}, thus performing impersonation and not anonymisation. \Gls{voice_anonymisation} is included in the sections on \gls{VC}.

\subsubsection{Speech classification tasks}

\paragraph{Speaker identification and verification} are two related tasks with different use cases. Identification involves attributing a speech sample to one of a set of known speakers (with the option that it cannot be attributed to any of them). \glsreset{ASV}\gls{ASV} involves authenticating a speaker's claimed identity based on a speech sample being either \gls{bona_fide} or from an \gls{imposter}.

\paragraph{Anti-spoofing} \textit{\Gls{spoofing}} is the use of speech generation to attack an \gls{ASV} system. These systems can be highly vulnerable to such attacks, and so must include a \gls{countermeasure}. The \gls{countermeasure} may be known informally as "\gls{anti_spoofing}". Current \glspl{countermeasure} are essentially speech \gls{deepfake} detectors, because the \gls{spoofed} speech may have been generated by a deep neural model. The term \gls{countermeasure} implies that the target of the attack is an automated system such as \gls{ASV}, whereas \gls{deepfake} detector implies the target is a human listener.

\paragraph{Speech deepfake detection and attribution} in its simplest form is deciding whether a speech sample is from a natural, human speaker, or not. This is closely related to \gls{anti_spoofing}, although the use case of a \gls{deepfake} detector is perhaps more likely to be alerting a human listener, than a technical \gls{countermeasure} to protect an \gls{ASV} system. Going beyond detection, we may wish to also explain the decision of the detector, or attribute the suspected \gls{deepfake} to a particular source or generation method.

\mysummarybox
{of speech generation and classification tasks}
{
\begin{itemize}
\item There are a number of related \textbf{speech generation} tasks. 
\item Although they differ in their \textbf{input} and use case, similar methods can be used for them all.
\item Several \textbf{speech classification} tasks are relevant to speech generation. 
\item All involve identifying something about \textbf{how} the speech was made.
\end{itemize}
}


\subsection{How speech generation relates to text, image, and video generation}
\label{section:relationship_to_text_image_video}

\myintrobox
{}
{
\begin{itemize}
\item AI-generated text, images, and video are generally intended for human consumption.
\item The most common \textit{input} in all three cases is \textbf{text}.
\item They all require learning from examples = from \textbf{data}.
\item That data was originally created by humans.
\item Methods for generation can be repurposed to \textit{manipulate} existing text, images, or video.
\end{itemize}
}

Historically, there has been a lot of cross-fertilisation between these fields, with apparent paradigm shifts in one often being the borrowing of a technique from another field. Two significant borrowings have occurred recently in speech generation, from text generation and image generation, respectively. 

To understand recent paradigm shifts, we first need to return to the fact that speech generation (from \textit{any} type of input) is an underspecified task:\label{TTS_is_an_underspecified_task} there are, in general, many valid outputs for any given input. For example, there are many different ways of saying "I like apples." but the text does not specify which one to use. There are two solutions to this, both data-driven. The first solution is to say the text in a single "average" way, with "average" being learned from some data. The problem with this approach is that the "average" way might not correspond to any valid rendition by a human speaker. For example, a human speaker might emphasise exactly one of the three words (e.g., "I like APPLES."), but the average of those three options is not something a human would say. So the second solution is to draw a random sample from the possible ways of saying the text. If the model is able to accurately capture the data distribution, this random sample is likely to be a valid way (i.e., something that a human might say), with the disadvantage that the system is now unpredictable: it is \gls{generative}.

The most important recent paradigm shift in speech generation came from borrowing a \gls{neural_network} architecture, the Transformer, from a related sequence-to-sequence task: text-based language modelling. The Transformer is the key neural building block in many popular AI systems and a fundamental component of many \gls{generative} models. Pre-training a model on a large amount of text obtained cheaply by web scraping, with no particular task in mind, results in a very good model of the data distribution: i.e., of language in general. This Generative Pretrained Transformer (GPT) can be used for many different downstream tasks, either by using modest amounts of expensive task-specific data to improve the model (something known as \gls{finetuning}) or by "prompting" the model. Prompting is nothing more than providing some text as context, and using the model to generate a continuation of that text. After the model generates each word\footnote{These models do not actually use words as units, but the distinction is not important to understand this point.}, that word becomes part of the context and so influences which word is generated next. To frame speech generation as language modelling, the input sequence (e.g., text in the case of \gls{TTS}, speech in the case of \gls{VC}) is treated as the context, then the speech output is generated sequentially in an \gls{auto_regressive} fashion as a continuation from that.

The other significant import of techniques into AI speech generation comes from \gls{generative} models of images, of two related neural modelling techniques: diffusion and flows. These are designed to generate both structure and detail in images (and videos). As with the "I like apples." speech example above, it is obvious that averaging is not a good solution to novel image generation: the average of all of Picasso's paintings would not be a good painting. The principal way that diffusion and flow have been applied to AI speech generation is for generating a \gls{spectrogram} which can be treated as an image. But they have also been applied to directly generating a \gls{waveform}. Either way, they are invariably just a component of a larger \gls{TTS} or \gls{VC} system, not a complete solution to the problem.

Historically, cross-fertilisation between fields only meant the exchange of ideas and concepts. This evolved into the exchange of specific techniques, which in the current era means specific \gls{neural_network} architectures (notably the Transformer above), and the associated algorithms for \gls{training} and \gls{inference}. This then went a step further, to the exchange of actual models. For example, a model originally trained to perform text generation could be taken as the basis for speech generation. The most important re-usable models by far are \gls{SSL} models, which can extract representations from speech (for example) that turn out to be useful for a multitude of downstream tasks, such as \gls{ASR}, \gls{TTS}, or \gls{VC}. Specific examples of this will come up later.

Aside from the productive exchanges between generation tasks, the other important connection is that of \gls{deepfake} detection. There is currently far more work on detecting video or audio-visual \glspl{deepfake}, than audio-only. This report is limited to the audio-only case, which is harder than the audio-visual case, where discrepancies between the image and the sound are a useful feature.

\mysummarybox{{of the relationship between AI speech generation and text, image, and video generation}}{
\begin{itemize}
\item \textbf{Adjacent fields} are a rich source of new ideas and techniques.
\item Recent developments in \textbf{generative models} have provided improvements in speech generation.
\item Speech generation has benefitted in particular from:
\begin{itemize}
\item GPT-type models originally developed for \textbf{text-based language modelling};
\item probabilistic models such as diffusion that are more often used for \textbf{image generation}.
\end{itemize}
\item Detecting multi-modal deepfakes is likely to be easier than in a single modality, such as audio-only.
\end{itemize}
}


\subsection{A brief history of Text-to-Speech (TTS)}
\label{section:history:TTS}
\glsresetall

\myintrobox
{}
{
\begin{itemize}
\item A primer on how Text-to-Speech works, to help you understand the material in Section~\ref{section:SotA:generation} onwards.
\item Why TTS remains a hard problem.
\item The different categories of errors that systems make.
\end{itemize}
}

\subsubsection{A Text-to-Speech primer}
\label{section:TTS_primer}

This is the hardest of the speech generation tasks, because text is \textit{ambiguous}: it does not fully specify the spoken form (e.g., the pronunciation of a word must be inferred from its spelling). Most mature commercial \gls{TTS} systems have converged on a single form of solution to this problem: to incrementally disambiguate by inferring the missing information. This can be conceptualised as a pipeline architecture \citep[Section 3.3]{TaylorBook}
in which a sequence of processors each perform a specific task, thus:

\vspace{2mm}
input text $\longrightarrow$ \gls{front_end} $\longrightarrow$ \gls{acoustic_model} $\longrightarrow$ \gls{waveform_generator} $\longrightarrow$ synthetic speech waveform
\vspace{2mm}

This solution is attractive because it easily allows for a variety of specialised methods (models, rules, algorithms, knowledge, data,...) to be combined. For example, some tasks in the \gls{front_end} are best done with simple and reliable hand-written rules (e.g., how to verbalise currency expressions like "£1.99"), whereas the \gls{acoustic_model} generally needs to be trained on a large paired text-speech dataset, but the \gls{waveform_generator} can be trained on a smaller amount of high-quality speech-only data.

The \gls{front_end} itself invariably also comprises a pipeline of processors which incrementally reduce the ambiguity of text (e.g., whether "ISCA" is an acronym or a word) and move closer to the spoken form (e.g., "£1.99" is verbalised as "one pound ninety nine"). Exactly how much information is inferred by the \gls{front_end} varies from system to system: some might simply output a \gls{linguistic_specification} containing only \gls{normalised_text}, others might specify the phonetic pronunciation, or go further and specify some aspects of \gls{prosody}.

The \gls{linguistic_specification} is passed to the \gls{acoustic_model} which performs the main task of \gls{TTS}: predicting the \gls{acoustic_specification}. In general, this is not the final \gls{waveform} but some other representation from which a waveform can later be generated, for reasons explained in the timeline in Section~\ref{section:TTS_timeline}. Models that \textit{do} directly output a \gls{waveform} without an intermediate \gls{acoustic_specification} are sometimes called \gls{end_to_end}, although this term is misleading, because the input to the system is almost never the original text: some text normalisation is necessary.

The form of the \gls{acoustic_specification} has varied considerably over time. In \gls{unit_selection}, it comprised recorded speech units (i.e., exemplars) selected from a database. In \gls{SPSS}, it was the set of parameters required by a \gls{parametric_vocoder} such as STRAIGHT \citep{STRAIGHT}, which employed a handcrafted, expert-designed \gls{acoustic_specification} extracted from speech waveforms using elaborate, bespoke signal processing algorithms, then used in a source-filter model for waveform generation. Acoustic specifications have come a long way since then. In many recent and current systems it is a \gls{mel_spectrogram}, still expert-designed but now trivial to extract from a waveform using simple methods. In the most recent systems, it is a sequence of \glspl{audio_token}.\label{section:TTS_primer:acoustic_specification}

The final step is to generate a \gls{waveform} and the method used will, of course, depend on the form of the \gls{acoustic_specification}. For \gls{unit_selection}, signal processing was used to concatenate the recorded speech units.  \gls{SPSS} systems used a \gls{parametric_vocoder} which most often took the form of a \gls{source_filter_model}, which required a very high degree of skill to implement using classical signal processing techniques. Current systems that employ a \gls{mel_spectrogram} as the \gls{acoustic_specification} require a \gls{neural_vocoder} whereas those employing \glspl{audio_token} simply use the decoder of their \gls{audio_codec}; this will become clearer in Section \ref{section:SotA:waveform_generation}.

Some \gls{TTS} systems take additional inputs, other than the text to be spoken; examples include speaker identity, a style label, or a speech "prompt". These will be covered in Section~\ref{section:SotA:generation}.

\subsubsection{Text-to-Speech errors}
\label{section:tts_errors}

Errors in the \gls{front_end} are generally localised to a portion of the speech output and are most commonly manifested as incorrect words (e.g., "Wed 1st" being verbalised as "wed first" instead of "wednesday the first"), disambiguation failures (e.g., "Polish" being pronounced the same as "polish"), and grossly-incorrect pronunciations (e.g., for proper names like "Buccleuch" where the pronunciation is hard to guess from the spelling).

Despite there being well-developed and highly-effective solutions for the processes performed in the \gls{front_end}, it is easy to find errors in the output from state-of-the-art \gls{TTS} systems. This is for two reasons. The first is obvious: there are genuinely hard cases in text (e.g., "I read the book." is ambiguous without additional information indicating the past or present tense). The second is less obvious: some recent systems either have no \gls{front_end}, or only a very simple one (Section \ref{section:current_approaches:technical}).

Errors in the \gls{acoustic_model} are generally more subtle and not as localised than those in the \gls{front_end}. One class of errors is due to limitations of the model itself. Tell-tale signs include near-identical speech being generated on separate occasions (something that a human talker cannot do), or conversely, unnatural variations may occur within the generated speech such as an inconsistent regional accent that changes within an utterance. The other class of error arises because the \gls{acoustic_model} is trained on a large amount of paired text-speech data: problems with that data will be reflected in the model outputs. Errors in the text element of the \gls{training} data -- notably when the text is not an accurate transcript of the speech element -- will lead to subtle pronunciation errors, slurred, or unclear speech. The \gls{prosody} of the speech may also be inappropriate or otherwise unnatural. Issues with the speech element of the \gls{training} data, such as poor recording quality or background noise, will also be evident in model output. Generative models will randomly reproduce these recording conditions in unpredictable ways; non-generative models may average them out, reducing the quality of all model output.

Errors in the \gls{waveform_generator} of a \gls{TTS} system generally affect the entire speech signal. Almost all methods of waveform generation can do near-perfect\footnote{i.e., imperceptible to non-expert listeners} copy-synthesis, which means transforming a natural recorded speech \gls{waveform} into the waveform generator's internal representation (\gls{acoustic_specification}), then regenerating the \gls{waveform}. However, when used in conjunction with an \gls{acoustic_model}, the input to the \gls{waveform_generator} will deviate from that, due to limitations of the \gls{acoustic_model}. The consequence is that all synthetic speech waveforms contain tell-tale artefacts, regardless of whether these are audible or not.

\mysummarybox{{so far}}{
\begin{itemize}
\item All TTS systems comprise a pipeline of components, which gradually transform text into speech.
\item Even "end-to-end" systems comprise components that correspond to those in pipeline systems.
\item Each component is imperfect and makes errors.
\item The types of error made by each component are somewhat distinct.
\end{itemize}
}


\newpage

\subsubsection{Text-to-Speech timeline}
\label{section:TTS_timeline}

This timeline starts with the arrival of Unit Selection systems in the late 1990s, because these systems were the first to achieve high enough quality for mainstream commercial applications, notably \gls{IVR}. The first \gls{TTS} voices for digital assistants, including Siri and Alexa, used \gls{unit_selection} or \gls{hybrid_unit_selection}.

\begin{longtable}{p{16cm}}

\hline \\

In a paradigm shift, \gls{unit_selection} replaces diphone synthesis \citep[Section 16]{TaylorBook}. Speech is now represented as a large and naturally-varied database with many examples ("units") per type, rather than a single example per type that requires extensive manipulation. For the first time, the data used to build the system comprises recordings of naturally-spoken complete utterances, thus capturing connected-speech effects including co-articulation, as well as prosody. The concept behind \gls{unit_selection} is that, although an infinite database would allow for the synthesis of perfectly-natural speech for any input text, this is not necessary because speech units are, to a sufficient extent, interchangeable between contexts. So good results can be obtained from a finite database of a few 10s of hours of speech from single speaker.

The amounts of explicit modelling, expert knowledge, and signal processing are all reduced compared to earlier systems. Systems comprise a pipeline \gls{front_end} followed by \gls{unit_selection} guided by expert-designed cost functions, then waveform generation using signal processing to concatenate the waveform units. The cost functions express the degree to which units are indeed interchangeable. Example research system: Festival~2 \citep{festival2}. Example commercial systems using (pre-hybrid) \gls{unit_selection}: Laureate from British Telecom; rVoice from Rhetorical Systems (acquired by ScanSoft, who merged with Nuance  then created key components for the first version of Apple's Siri).

\\ \hline \\

The only significant contender to \gls{unit_selection} was \gls{SPSS}, which uses the same \gls{front_end} as contemporary \gls{unit_selection} systems. But the expert linguistic knowledge of phonetics and prosody that form the basis of \gls{unit_selection}'s cost functions is instead used to inform machine learning: regression trees. These are learned from data. During synthesis, they predict the parameters (means and variances) of \glspl{HMM} which generate a parametric \gls{acoustic_specification} for input to a \gls{parametric_vocoder} that produces the waveform. This approach eventually evolved into today's neural systems. But, at the time, it never sounded as good as \gls{unit_selection} because of reliance on a \gls{parametric_vocoder} which involves far more complex signal processing than waveform concatenation, and because the machine learning (regression trees) is too simple. Most systems of this type were built with the HTS toolkit \citep{tokudaSpeechParameterGeneration1995, HTSv2}. \label{section:TTS_timeline:HTS}

\\ \hline \\

Although pure \gls{SPSS} systems could never beat \gls{unit_selection} in terms of naturalness or quality, the \gls{acoustic_specification} they generated was found to be useful \textit{not} for input to a \gls{parametric_vocoder} but rather as a \textit{guide} to selecting recorded waveform units. By using \gls{SPSS} to replace the hand-crafted cost functions of \gls{unit_selection}, whilst retaining concatenation as the method for waveform generation, the "best of both worlds" is developed: \gls{hybrid_unit_selection}. This can be conceptualised as partial synthesis using \gls{SPSS}, to obtain a parametric \gls{acoustic_specification}, then completion of synthesis by selecting recorded waveform units that match this specification. By now, \glspl{front_end} were mature and sophisticated. \Gls{hybrid_unit_selection} was the apogee of \gls{unit_selection} and many commercial systems of this type are deployed; speech from these systems can still be heard today (e.g., on-device, because these systems are computationally efficient). Example systems: Trajectory Tiling from Microsoft  \citep{TrajectoryTiling}; the  voices used in early versions of digital assistants including Apple's Siri (provided by Nuance) and Amazon's Alexa (via the acquisition of IVONA).

\\ \hline \\

\gls{SPSS} based on regression trees and \glspl{HMM} continued to develop alongside the more commercially-successful \gls{hybrid_unit_selection}. Two crucial advances were made that would eventually feed through into current systems. Firstly, \gls{SPSS} could be trained on data not purposely collected for \gls{TTS} under \gls{studio_conditions}: initially this was data originally created for \gls{ASR} but very quickly it became  \gls{found_data} such as audiobooks. \gls{unit_selection} systems generally cannot be created from such data. Secondly, the \glspl{HMM} used in \gls{SPSS} naturally lent themselves to adaptation (by applying data-driven transforms to their parameters). This opened up the possibility of models that could vary \gls{speaker_identity}, speaking style, and so on, often requiring only a limited quantity of data. These features can be found in many \gls{SotA} systems today. Examples of the flexibility of \gls{HMM}-based \gls{SPSS}: speaker adaptation \citep{tamuraHMM-MLLR}; unsupervised training on speech without a transcription \citep{kingUnsupervisedAdaptationHMMbased2008}; ability to work with very small amounts of data \citep{yamagishiRobustSpeakerAdaptiveHMMBased2009}.

\\ \hline \\

Improvements in machine learning, borrowed from adjacent fields, were brought in to \gls{SPSS}: the regression trees were replaced with neural networks, initially driving the same \gls{parametric_vocoder} as before. This early generation of neural \gls{SPSS} did not yet have the adaptation capabilities of the \gls{HMM} systems it replaced. Such systems had little or no commercial deployment, primarily because the vocoder was still unable to match the naturalness of waveform concatenation. \citet{zenStatisticalParametricSpeech2013} are credited with (re)introducing neural networks for \gls{TTS}.
Neural \gls{SPSS} also fed through into contemporary  \gls{hybrid_unit_selection} systems, offering improvements over \glspl{HMM} \citep[e.g.,][]{merrittDeepNeuralNetworkguided2016}.

\\ \hline \\

Expert designed \glspl{parametric_vocoder} were handicapped by their \gls{acoustic_specification}, which was a parametric representation of speech: even copy-synthesis (\gls{waveform} $\rightarrow$ \gls{acoustic_specification} $\rightarrow$ \gls{waveform}) was imperfect. The parameters are difficult to estimate from speech, \textit{and} contain insufficient information to perfectly reconstruct the speech again. The next paradigm shift marked the true beginning of modern neural systems: simplifying the \gls{acoustic_specification} to the \gls{mel_spectrogram}, then training a \gls{neural_vocoder} to generate a waveform from this. Waveform concatenation was rendered obsolete, at last. Examples of early neural vocoders\label{section:TTS_timeline:first_neural_vocoder}: WaveNet \citep{WaveNet,WaveNet-arXiv}, except that, in those first publications, it was intended as a complete model for TTS and was far too slow to deploy; WaveRNN \citep{WaveRNN}.

\\ \hline \\

Next came a period of somewhat unsystematic and haphazard exploration of neural architectures for the \gls{acoustic_model}, to replace the feed-forward or simple recurrent networks in use up to this point. During this time, advances in the \gls{front_end} and the data were neglected.
The consequence was an explosion in the number and variety of \glspl{acoustic_model} (almost always still generating a \gls{mel_spectrogram}), with increasingly complex architectures and training recipes. Nevertheless, a common architectural pattern emerged, comprising an encoder operating at "linguistic timescale" (the input sequence of \glspl{phoneme} or \glspl{character}) and a decoder operating at "acoustic timescale" (the fixed frame-rate of the output \gls{mel_spectrogram}), plus some mechanism to link the two different timescales (to align them during \gls{training}, then to predict duration during \gls{inference}). Although these models can generally be trained on \gls{found_data}, to obtain commercially-acceptable quality they still require purposely-collected data from a professional \gls{voice_actor} recorded under \gls{studio_conditions}. Example models: Tacotron~2 \citep{Tacotron2} which had wide commercial deployment; FastSpeech~2 \citep{FastSpeech2}.

\\ \hline \\

So-called \gls{end_to_end} approaches attempted to do away with all expert knowledge, taking \gls{character} sequences as input and producing a so-called \gls{raw_waveform} as output. The attraction of this approach is that it requires no additional supervision during training, such as an alignment between input and output, or a pre-determined \gls{acoustic_specification} (generally a \gls{mel_spectrogram}). But few, if any, systems were truly \gls{end_to_end} (e.g., they require normalised text input) although the term continues to be used, and remains poorly-defined. Ultimately, this approach did not dominate, for a combination of two main reasons: the waveform is not a good domain in which to maximise \textit{perceptual} similarity to natural speech; the waveform has a very long sequence length (e.g., \qty{24}{\kilo\nothing} samples per second of speech), which is highly-problematic for current models. Examples of systems that claim to be \gls{end_to_end}: Char2Wav actually uses \gls{parametric_vocoder} features as an internal representation \citep{Char2Wav}; DelightfulTTS~2 actually employs a \gls{front_end} end to normalise the text and predict pronunciation, plus external forced alignment for training the duration predictor \citep{DelightfulTTS2}.

\\ \hline \\

The components of successful \gls{acoustic_model}s were recycled and re-used in subsequent models. Many aspects of end-to-end models were adopted, including: the use of \glspl{character} as input, from \gls{normalised_text}; jointly \gls{finetuning} the \gls{acoustic_model} and \gls{neural_vocoder}, starting from two separately-trained models; more sophisticated training methods such as the use of multiple \glspl{loss_function}, or \gls{adversarial} methods. Systems of this period have complex architectures and training methods so, as a consequence, are often hard to replicate and/or unstable to train.

\\ \hline \\

It had already been understood for many years that \gls{TTS} is an underspecified task (cf. Section~\ref{TTS_is_an_underspecified_task}): one text can be spoken in many ways. Until now, solutions to this involved providing the \gls{acoustic_model} with additional inputs to reduce this ambiguity, including methods to discover acoustic variations in the speech that are not explained by the existing \gls{linguistic_specification} input to the model \citep{wattsSentencelevelControlVectors2015}.
The consequence of any unexplained acoustic variation was that the \gls{acoustic_model} would learn to generate \textit{average} output. Taking an average of several speech examples is often not appropriate. Averaging in the \gls{spectrogram} domain results in over-smooth spectral envelopes which lead to muffled or buzzy-sounding speech; averaging \gls{F0} leads to flattened intonation contours.

A new solution to all of these problems emerged: generative models. The fundamental difference with previous models is that generative models accurately capture the distribution of the data (e.g., the multiple ways in which a single text might be spoken) and are able to generate novel samples from that distribution. The first\footnote{\glspl{HMM} are also generative, but their model of the data distribution is poor. There were attempts to sample from \glspl{HMM} for speech synthesis, but results were not very good.} application of generative models was actually for waveform generation, with wider application following a few years later. 

Examples of early generative \glspl{acoustic_model} for \gls{TTS}: GlowTTS \citep{Glow-TTS}; Flowtron \citep{Flowtron}; VITS \citep{VITS}. These were not based on \glspl{LLM}, which emerge later in this timeline. Example generative models of waveforms are listed in the timeline sections on neural vocoders (one above and another below).

\\ \hline \\

In parallel with \gls{acoustic_model} development, there were rapid advances in \glspl{neural_vocoder}, typically trained using adversarial methods to minimise multiple losses in both frequency and time domains. Widespread adoption of any given model was (and still is) highly-dependent on availability of Open Source code and a pre-trained model, because training is computationally expensive. A small number of \glspl{neural_vocoder} have emerged as popular choices because they not only sound good, but take a now-standard 80-band \gls{mel_spectrogram} as input and generate a waveform at the now-standard \SI{24}{\kilo\hertz} sampling rate, and have Open Source code and pre-trained models. Example popular vocoders: \label{section:TTS_timeline:HiFiGAN_is_popular} HiFi-GAN \citep{HiFi-GAN}; WaveGlow \citep{WaveGlow}; BigVGAN \citep{BigVGAN}; MelGAN \citep{MelGAN}.\label{section:TTS_timeline:neural_vocoder} At this time, speech coding researchers were also adopting neural methods. An important model coming from that field is LPCNet \citep{LPCNet}, a waveform generator which takes a classical \gls{acoustic_specification} as input: a low-dimension cepstrum + pitch. This is aligned with the goals of speech coding for compression (whereas the \gls{mel_spectrogram} is not). Speech coding researchers would go on to devise the \glsreset{NAC}\gls{NAC} which appears below in this timeline.

\\ \hline \\

The next paradigm shift was a simplification through reformulation of the problem. By describing the problem of speech generation as language modelling, not regression, it became possible to adopt \glspl{LLM} from text-based \gls{NLP}, which we call \glspl{SLM} when used to generate speech. The complex bespoke architectures of previous generation models were replaced by generic language models. In order to use a \gls{SLM}, it is necessary to use a discrete \gls{acoustic_specification}. This is easily obtained from a \gls{NAC} (see below). A \gls{SLM} driving a \gls{NAC} is the most important \gls{SotA} approach in current use: see Section~\ref{section:SotA:generation}. Example \glspl{SLM}: VALL-E \citep{VALL-E-journal}.

\\ \hline \\

In conjunction with the language model paradigm shift, speech is now represented as a sequence of discrete \glspl{audio_token} instead of a continuously-valued \gls{mel_spectrogram}. The speech-specific \gls{neural_vocoder} is replaced with a neural \gls{audio_codec} (i.e., a \gls{NAC}) which may be speech-specific (by being trained only on speech), or more general purpose (by being trained on speech and other audio). As with the previous generation of \glspl{neural_vocoder}, the \gls{NAC} is trained on natural waveforms, and typically requires orders of magnitude less data than a \gls{SotA} \gls{acoustic_model}. Most \glspl{NAC} are created for audio coding (i.e., compression for storage or transmission), not AI speech generation. We are currently seeing the emergence of specialised \glspl{NAC} for speech generation applications which -- for example -- reduce the symbol rate (to reduce total sequence length), or remove the inconvenient dependency between codebooks created by the use of \gls{RVQ} in many \glspl{NAC}. Example codecs originally designed for general speech or audio compression, but that have been used for speech generation: EnCodec \citep{EnCodec}; Descript Audio Codec (DAC) \citep{DescriptAudioCodec}. Example codecs designed specifically for speech generation: Low Frame-rate Speech Codec \citep{LowFramerateSpeechCodec}; Spectral Codecs \citep{SpectralCodecs}; TiCodec \citep{TiCodec}.

\\ \hline \\

To support \gls{SLM} data demands, it is necessary to automatically collect and curate large-scale (e.g., \qtyrange{10}{100}{\kilo\hour}) paired speech-text datasets. Most companies create in-house "pipelines" to do this, starting from a web crawl of YouTube, podcast feeds, audiobooks, etc., followed by segmentation with \gls{VAD} and/or speaker diarisation, filtering out low quality data using automated metrics, optional speech enhancement or denoising, and automatic transcription using \gls{ASR}. Example data pipeline: Emilia \citep{Emilia}, described on page~\pageref{page:emilia}.

\\ \hline \\

\end{longtable}

\mysummarybox{of the Text-to-Speech timeline}{
\begin{itemize}
\item The overall \textbf{naturalness} of AI-generated speech has been improving continually for decades.
\item \textbf{Key moments} marking a step-change in naturalness, or meeting the requirements for a new use case:
\begin{itemize}
\item Unit Selection in the 1990s with application to automated telephone-based systems (\gls{IVR}).
\item Hybrid Unit Selection in the 2000s that enabled digital assistants in the early 2010s.
\item Initial neural systems in the 2010s, evolving into generative models in the 2020s.
\end{itemize}
\item The required amount of \textbf{data} and \textbf{computational power} has been steadily increasing.
\item Recently-deployed commercial TTS systems actually take a \textbf{backwards step} in text processing.
\item The latest systems are \gls{generative}, meaning that their output randomly varies, even for a fixed input text: this is mostly a positive advance, but does result in occasional audible errors.
\item The previous two points mean that \gls{SotA} systems have started making \textbf{errors} that previous-generation systems do not.
\end{itemize}
}


\newpage
\subsection{A brief history of Voice Conversion (VC)}
\label{section:history:VC}
\glsresetall

\myintrobox
{}
{
\begin{itemize}
\item A primer on how Voice Conversion works and what makes it hard.
\item The types of errors that systems make.
\item This will help you understand the material from Section~\ref{section:SotA:generation} onwards.
\item Non-technical readers can omit the timeline in Section \ref{section:VC_timeline}.
\end{itemize}
}

\subsubsection{A Voice Conversion primer}
\label{section:VC_primer}

In Section~\ref{section:TTS_primer}, we learned that \gls{TTS} is hard primarily because the input text does not fully specify the output speech. \gls{VC} is hard for a similar reason: which aspects of the input should be converted, and which should be preserved, is not well-defined.

For this primer, we will only consider the main use case of \gls{VC}, which is to convert speech from a \gls{source_speaker} to sound like it was spoken by a \gls{target_speaker}. But conceptually everything in this section also applies to other use cases, such as converting speaking style, or indeed converting speaking into singing.

At first sight, it is natural to think of \gls{VC} as a mapping problem. But then what exactly should be mapped (i.e., converted), and what should not (i.e., be copied)? As a simple illustration, consider the use of \gls{pitch} and duration by a speaker, and their perception by a listener. These convey the \gls{speaker_identity} \textit{and} the \gls{linguistic_content} (and other things too). That is, \gls{speaker_identity} and \gls{linguistic_content} are \textbf{entangled}: they will be \textit{fundamentally} hard to separate, given only a speech signal.

\gls{VC} has been a topic of research since the 1980s, initially as a method for expanding the range of voices possible from a \gls{TTS} system%
, but later as a standalone application using natural human speech as the input. Early systems framed the problem as one of mapping speech sounds from a \gls{source_speaker} to a desired \gls{target_speaker}. The mapping was generally applied not to speech \glspl{waveform} but to a parametric representation of speech, such as those of a \gls{source_filter_model} or a \gls{parametric_vocoder}. The form of the mapping varied considerably and could be based on codebooks, \glspl{GMM}, \glspl{neural_network}, or linear regression \citep[e.g.,][]{baudoinTransformationSpeechSpectrum1996}. The mapping approach relies on parallel data: recordings of both source and target speaker speaking the same \gls{linguistic_content}. This restricts the range of practical use cases, so ways have been devised to allow systems to be built without it. One involves artificially constructing pseudo-paired data, the other involves directly training the \gls{VC} system on non-parallel data.

The second of these is dominant, and is currently described as \textit{disentangling} different factors from the speech signal. Such \gls{VC} systems generally encode the input speech into an internal representation in which there is a sufficient degree of disentanglement, at least of those factors to be modified (most commonly \gls{linguistic_content} needs to be preserved and \gls{speaker_identity} will be modified, but in principle other factors could be modified within a similar framework). The output speech is then decoded back from this internal representation, with the desired \gls{target_speaker} identity being imposed.

The most extreme form of internal representation is a text transcription of the input speech, which is assumed to be entirely free of \gls{speaker_identity} (of course it is not, but \gls{VC} rarely if ever attempts to modify traces of \gls{speaker_identity} in the \gls{linguistic_content}; these traces would be of concern in Voice Privacy though -- see Section~\ref{section:SotA:TTS:voice_anonymisation}). This can be achieved with \gls{ASR}, followed by speaker-adaptive \gls{TTS} to generate the output speech in the voice of the \gls{target_speaker}. However, \gls{ASR} is not perfect: gross errors, such as word changes, will occur. Therefore, most systems with an \gls{ASR}+\gls{TTS} \textit{style} architecture for \gls{VC} do not actually use text as the intermediate representation, but stop some way short of that, such as: a phonetic transcription, a \gls{PPG} (essentially a "soft" phonetic transcription), or an internal \gls{latent} representation from within the \gls{ASR} model. Conceptually, these can still be called \gls{ASR}+\gls{TTS} systems, because they can be built from separate \gls{ASR} and \gls{TTS} modules. These can even be pre-existing systems, slightly adapted to use something other than text as their output or input, respectively.

The \gls{TTS} component generates the target speech in the voice of the \gls{target_speaker} from the intermediate representation plus a representation  of the desired \gls{speaker_identity} of the \gls{target_speaker}. For a closed set of \glspl{target_speaker} this may be a simple code, but for an open set it needs to be a \gls{speaker_embedding}; this can be obtained from a pre-trained model (e.g., borrowed from \gls{ASV}), or trained as part of the \gls{VC} system.

The \gls{ASR}+\gls{TTS} approach works \label{section:VC_primer:ASR+TTS} extremely well, provided the \gls{ASR} is accurate. Building such an \gls{ASR} system involves \gls{supervised} machine learning using substantial amounts of expensive transcribed data, which is unavailable for most languages. An \gls{unsupervised} alternative to \gls{ASR} is to use a \gls{SSL} model. \gls{SSL} models are not trained to transcribe speech, but to perform some proxy task such as predicting a masked speech sound, given the preceding or surrounding ones. In learning to do this self-supervised task, these models learn rich internal \gls{latent} representations of speech; these have been found useful in a wide variety of downstream tasks. \gls{VC} systems that use \gls{SSL} look very similar to \gls{ASR}+\gls{TTS} systems, for example by replacing the \gls{PPG} with an \gls{SSL} representation. Generally, explicit \gls{ASR}+\gls{TTS} approaches perform best (with the significant caveat that an accurate \gls{ASR} system is available), closely followed by the \gls{SSL} alternative: a case study is provided in Section \ref{section:SotA:VC:offline}.

\subsubsection{Voice Conversion errors}

Just as we saw earlier with \gls{TTS}, \gls{VC} also turns out to be a poorly-defined task. For \gls{TTS}, the problem was that the text does not fully specify the spoken form, so it is not clear what "perfect" output would be, and there will generally be multiple "correct" possibilities. The parallel in \gls{VC} is that there is no clear definition of what should be retained / converted / discarded from the input speech. In the end, that decision can only be made with regard to a specific use case. Stating a clear use case is surprisingly rare in the research literature for both \gls{VC} and \gls{TTS}: only vague, "general purpose" uses are implied.

The majority of \gls{VC} systems comprise a pipeline of modules. These might be explicit and separately constructed, as in \gls{ASR}+\gls{TTS} architectures, or they might be implicit / internal modules within a neural architecture. Each such module can make errors. 

Errors made in the encoder (e.g., \gls{ASR} errors that are reflected in the \gls{PPG} assigning probability to the wrong phones) will lead to either gross errors such as changing a word, or more likely to pronunciation or articulation errors that reduce intelligibility. If the \gls{source_speaker} and \gls{target_speaker} have different accents, it is unclear what a \gls{VC} system should do, and the vast majority of published work ignores this issue entirely (which is unfortunate, because one of the most widely used datasets, VCTK, deliberately includes a variety of UK accents \citep{VCTK}).

The aspect of \gls{VC} that is most poorly defined is what to do with \gls{prosody}. \Gls{pitch}, energy, duration, and voice quality are acoustic correlates of \gls{prosody} but they are also influenced by \gls{speaker_identity}. So, to define an error in the \gls{prosody} of the output speech requires a clear use case. For use cases requiring  preservation of \gls{prosody}, the most common solution is very simplistic: to copy across \gls{phone} durations, and to perform a simple \gls{pitch} transform to the range of the \gls{target_speaker}.

All \gls{VC} systems need a representation of the \gls{speaker_identity} of the \gls{target_speaker}, obtained either from an external model or learned by the \gls{VC} system itself. This is usually a single vector of fixed dimensionality. As in all other speech technology, this doesn't work equally well for all speakers: some speakers will be better represented, than others. Additionally, imperfect disentanglement will lead to audible variations in target \gls{speaker_identity}, varying with the \gls{source_speaker} (a phenomenon known as source speaker \gls{leakage}) and/or the \gls{linguistic_content}. So, we should expect all \gls{VC} systems to work better for some source-target speaker pairs than others.

The output speech \gls{waveform} for most \gls{VC} systems is generated by a \gls{neural_vocoder} or neural \gls{audio_codec}, so the errors described in Section~\ref{section:tts_errors} will occur.

\mysummarybox{{so far}}{
\begin{itemize}
\item \glsreset{VC}\gls{VC} systems share many characteristics with \glsreset{TTS}\gls{TTS} systems.
\item The prototypical architecture of a \gls{VC} system comprises:
\begin{itemize}
\item an encoder to extract the \gls{linguistic_content} from the input speech
\item a decoder to generate speech corresponding to that \gls{linguistic_content}, using the \gls{speaker_identity} of a \gls{target_speaker}.
\end{itemize}
\item Some \gls{VC} systems are built from pre-existing components, such as \gls{ASR} for the encoder, \gls{TTS} for the decoder, and a speaker representation from an \glsreset{ASV}\gls{ASV} system.
\end{itemize}
}


\subsubsection{Voice Conversion timeline}
\label{section:VC_timeline}

This timeline starts just before the field transitioned from using parallel data (i.e., where the \gls{source_speaker} and \gls{target_speaker} speak the same \gls{linguistic_content}) to non-parallel data.

\begin{longtable}{p{16cm}}

\hline \\

\gls{VC} was initially considered to be a mapping problem, with the mapping learned from paired data. The focus was on the mapping function, with a wide variety of methods being tried, including \glspl{GMM}, regression functions of various kinds, exemplar-based (so-called "non-parametric") methods, and so on. The vast majority of these operated on a parameterised representation of speech, and therefore required a \gls{parametric_vocoder} to reconstruct the waveform, with STRAIGHT \citep{STRAIGHT} being a predominant choice, as did contemporary \gls{SPSS} approaches to \gls{TTS} (refer to the entry in the TTS timeline on page~\pageref{section:TTS_timeline:HTS}). Example \gls{GMM}-plus-vocoder system from this period: \citet{stylianouVCjournal1998}.

\\ \hline \\

As mapping functions matured, the reliance on parallel data was initially sidestepped by methods that created pseudo-parallel data, which could be used to train existing \gls{VC} systems that relied on parallel data. Conceptually, creating pseudo-parallel data is very simple: find source-target pairs not of whole utterances but of smaller speech units, possibly right down to the frame level. Example method for creating pseudo-paired data: INCA \citep{INCA}.

\\ \hline \\

At this point, it became clear that the \gls{parametric_vocoder} was a limiting factor on output speech quality, exactly as it was in contemporary \gls{SPSS} approaches to \gls{TTS}. The emergence of \glspl{neural_vocoder} offered a way forward: refer to the \gls{TTS} timeline on page~\pageref{section:TTS_timeline:first_neural_vocoder}

\\ \hline \\

With the much higher waveform quality now available from \glspl{neural_vocoder}, and the focus firmly on using non-parallel data, there was a rapid expansion in neural architectures for \gls{VC}. Two common patterns emerged: cycle methods, and encoder-decoder architectures.

\\ \hline \\

The simplest way to train a neural model for \gls{VC} would be from paired source-target data. \label{page:VC_timeline:cycle_methods} The model would be trained, given a source utterance, to minimise the loss when generating the corresponding target utterance. \textbf{Cycle} methods offer a clever way to do this even when no target utterance is available to compute this loss. Cycle methods simply involve making a "round trip" source$\rightarrow$target $\rightarrow$source, then minimising the loss of the final, reconstructed source speech. Additional constraints must be imposed to prevent the model merely copying source to source. CycleGAN-VC \citep{CycleGAN-VC} uses an adversarial loss to steer the model's internal representation to match the features of natural speech, whereas Cycle~VAE \citep{tobingLowlatencyRealtimeNonparallel2021} uses a loss to encourage the model to use the same latent representation for both the source$\rightarrow$target and target$\rightarrow$source (and indeed source$\rightarrow$source) directions, thus encouraging it to be speaker-independent (i.e., disentangled from speaker).

\\ \hline \\

The other common pattern that has emerged is an \textbf{encoder-decoder} architecture. The representation linking the two should capture only the \gls{linguistic_content} plus those (ill-defined!) aspects of prosody from the source speech that are desired to be mapped to the target speech (exactly what is desirable depends on the use case), but that are \textit{not} part of the \gls{source_speaker}'s identity. Initially, this was most commonly achieved by placing an information "bottleneck" on that representation (i.e., limiting its representational power, such as by quantising it, to "squeeze out" speaker information), but later systems instead supervise this representation. This supervision might involve borrowing a pre-existing model to use as the encoder (e.g., an \gls{ASR} or \gls{SSL} model), or by training the encoder to \textit{emulate} such a model (e.g., in student-teacher fashion). Example systems: \gls{PPG} from an \gls{ASR} system \citep{PPG-VC}; ASR+TTS \citep{liuAnytoManyVoiceConversion2021}, which is the case study in Section~\ref{section:SotA:VC:offline}; StreamVC \citep{StreamVC} has the same architecture as the SoundStream \gls{NAC} \citep{SoundStream} --and is therefore also fast and causal-- but with its latent representation encouraged to emulate the representations from HuBERT \citep{HuBERT} (which is not causal). A case study is provided in Section~\ref{section:SotA:VC:streaming}.

\\ \hline

\end{longtable}

\mysummarybox
{of the Voice Conversion timeline}
{
\begin{itemize}
\item \gls{VC} was initially framed as a data-driven mapping problem.
\item Today it involves explicitly disentangling the factors of interest so they can be manipulated.
\item Techniques for \gls{VC} have been steadily converging with:
 \begin{itemize}
 \item \gls{ASR} and \gls{SSL}, for extracting the \gls{linguistic_content};
 \item \gls{TTS}, for generating the target speech.
 \end{itemize}
\item \gls{VC} systems are often described as "X-to-Y" where X and Y can be: "one", "many", "any". This indicates the number of source (X) and target (Y) speakers it supports, where "many" is a closed set and "any" is an open set.\label{section:VC_timeline:x_to_y}
\end{itemize}
}


\subsection{A brief history of the interplay between speech generation and speaker verification}
\label{section:history:spoofing}
\glsresetall

\gls{TTS} did not originally consider speaker identity explicitly, since concatenative approaches such as Unit Selection almost always used a single-speaker database, so could not sound like any other speaker. With the advent of speech synthesis \textit{models} trained on data from many speakers, it became of interest to confirm that the generated speech had the desired identity. This originated in the Blizzard Challenge (see Section~\ref{section:challenges}) where a subjective evaluation of speaker similarity was introduced to check this. 

Early in the development of \gls{HMM}-based speech synthesis (Section \ref{section:TTS_primer}) it was realised that the generated speech could be used to attack \gls{ASV} very successfully. This was because the underlying model used to generate speech (a \gls{GMM} operating on \gls{parametric_vocoder} features) was very similar to the model being used to verify speaker identity (also based on \glspl{GMM}). The same held true for speech generated by \gls{VC}. This initiated a fruitful line of work that continues to this day: building \glspl{countermeasure} to protect \gls{ASV} against such \gls{spoofing} attacks and, to a lesser extent, designing speech generation (e.g., \gls{VC} or \gls{TTS}) systems specifically to evade \glspl{countermeasure}. One of the most productive researchers, Yamagishi, actually works on both speech generation, and \glspl{countermeasure}, although most researchers specialise in only one or the other.

This line of work has evolved and branched out into related classification tasks including \gls{deepfake} detection and attribution, and voice privacy, as defined in Section~\ref{section:definitions}. The timeline of challenges and competitions coming up in Section~\ref{section:challenges} will illustrate how this has unfolded through the sequence of ASVspoof and VoicePrivacy challenges, and we will cover the \gls{SotA} in Section~\ref{section:SotA:generation}.

\mysummarybox
{of the interplay between speech generation and speaker verification}
{
\glsreset{TTS}\glsreset{VC}\glsreset{ASV}
\begin{itemize}
\item Early \gls{ASV} systems were highly vulnerable to \gls{TTS} or \gls{VC} attacks, even though the quality of the synthetic speech was poor by modern standards.
\item So there was rapid development of \glspl{countermeasure} to protect \gls{ASV} against such \gls{spoofing} attacks.
\item Very recently, the field has expanded and diversified, leading to a cluster of inter-related tasks including:
\begin{itemize}
\item \gls{deepfake} detection, effectively the same as an \gls{ASV} \gls{countermeasure};
\item \gls{deepfake} attribution to determine which \gls{TTS} or \gls{VC} system was used to create the deepfake;
\item \gls{voice_anonymisation} methods to hide \gls{speaker_identity} (or other speaker attributes);
\item source speaker tracing, to defeat \gls{VC} used for \gls{voice_anonymisation};
\item specialised AI speech generation designed to defeat \glsfirst{countermeasure}s or \gls{deepfake} detectors.
\end{itemize}
\item Advances in one task often stimulate accelerated development in another: an "arms race" or a "virtuous circle", depending on your point of view.
\end{itemize}
}

\subsection{Evaluation, challenges and competitions}
\label{section:history:challenges}
\glsresetall

As we have seen already, all speech generation systems make errors, but not all errors are equal. For example, a \gls{TTS} system's \gls{front_end} might incorrectly predict the pronunciation of a word, or its \gls{vocoder} might produce an audible artefact. These are very different types of error. So it may come as a surprise that the overwhelming majority of evaluations reported in the literature do not make any such distinction. Instead, they measure only holistically, using some global dimension such as \gls{naturalness} and reporting only global statistics such as the average and standard deviation.

The motivations for evaluating AI-generated speech depend on what the goals are. By far the most common reason to evaluate is to demonstrate the superiority of one system over another, which is widely seen as a pre-requisite for scientific publication: it is indeed much harder to get work published without such a result.\footnote{The bias against publishing negative results, however interesting they may be, is not unique to this field, of course.} This need to demonstrate superiority is the key driving factor for the continued use of simplistic, holistic evaluations.

\myintrobox
{in the first part of this section}
{
\begin{itemize}
\item The Gold Standard for evaluation is a \textbf{subjective} listening test involving human listeners.
\item Automated \textbf{objective} evaluation is an attractive alternative.
\end{itemize}
}

\subsubsection{Evaluation of AI-generated speech}
\label{section:history:challenges:evaluation}

\paragraph{Subjective evaluation}

Since AI-generated speech is generally intended for human consumption, the most obvious way to evaluate it is by listening. This immediately requires decisions to be made: which listeners? what should they listen to? what are they being asked to judge? These are actually fundamentally-difficult questions, and remain the subject of active research \citep{wagnerSpeechSynthesisEvaluation2019}. Here, we give only a brief survey of standard approaches used in the field. The \gls{SotA} in evaluation is covered in Section~\ref{section:SotA:evaluation}.

Expert listeners,\footnote{But not \textit{forensic} analysis of AI speech, which is out-of-scope.} most often the developers themselves, may be able to narrow down the cause of errors (e.g., a mispronunciation is not usually caused by the \gls{vocoder}) which is very useful during system development. But the developers are biased listeners, so reporting the results of expert listening is rare in the literature. Instead, naive listeners are employed, and only general, overall judgements are asked of them. Where there is a specific use case or product in mind, these listeners represent a sample of the end-user population, so they are normally, but not exclusively, native speakers of the language concerned.

The materials played to listeners in a listening test are, almost without exception, generated from short sentences and played as isolated utterances. This is for good reasons of listener attention span and avoiding fatigue, but does place limitations on what can be discovered. For example, it is much harder to detect within-utterance inconsistencies in speaker identity or accent with short samples. The test sentences are very often chosen randomly. They are synthesised by the system under evaluation, and at least one other system for comparison. If available, natural recordings of the same sentences are also included, as another "system" (i.e., without the listeners being aware they are recordings), to provide an upper bound.

Listeners are most often asked to judge just two things about each speech sample, which is almost always presented in isolation: \gls{naturalness}, and similarity to the target speaker. The former is usually not further described to the listeners (they must use their own interpretation of what this means), and there is no reference. The latter is also not further described, but a reference recording of the target speaker is provided. Commonly, listeners provide judgements for isolated utterances on a 1--5 Likert scale, which are aggregated across all listeners and utterances and reported as \gls{MOS}. It is very important to understand that listener judgements, even when made on isolated samples played without any context, exhibit a considerable degree of relativity because listeners will make implicit comparisons between samples: the scores given to samples from one system are strongly influenced by the other samples within the test. In the literature, any \gls{MOS} value reported out of context, or presented as an absolute score, or compared with a \gls{MOS} value from a different listening test (e.g., from another paper), is misleading and should be ignored. An alternative paradigm to \gls{MOS} is to present listeners with a pair (or more) of samples, and ask them to rank them. This can give more clear-cut results, but does not scale well for comparisons of many systems, and still does not allow comparisons between different listening tests.

Historically, listeners were also used to evaluate intelligibility, generally by transcribing the words they heard. This is still done in some evaluations, but it has become more common to use an objective method for this, described in the next section.

The form of evaluation described above generally fails to \textit{explain} why one system was found to be superior to another by the listeners. For example, some evaluations report a higher value for \gls{naturalness} on the synthetic speech than for recorded natural speech. This is most likely due to careless experimental design: for example the recorded natural speech samples have been inexpertly extracted from a longer recording so their prosody is inappropriate for presentation as an isolated utterance, or they are low-quality recordings (e.g., background noise).

In summary, subjective evaluation methods have not developed at the same pace as the methods for AI speech generation. The \gls{MOS} paradigm is in fact no longer fit for purpose. But there has been a recent resurgence of interest in better evaluation, which will be discussed in Section~\ref{section:SotA:evaluation}.

\mysummarybox
{of subjective evaluation}
{
\begin{itemize}
\item Evaluation involves implicit or explicit comparisons among two or more systems.
\item Recorded natural speech is often included as a "system" to provide a benchmark.
\item A typical evaluation measures \gls{naturalness}, speaker similarity, and intelligibility.
\item \Gls{naturalness} has no clear definition and is left to listeners' own intuition; they are likely to conflate it with, for example, audio quality.
\item Listeners are often crowdsourced from services such as Prolific or Amazon Mechanical Turk.
\item Test materials are often chosen randomly and without sufficient care.
\item Subjective evaluation is far from being a "solved problem".
\end{itemize}
}

\paragraph{Objective evaluation}

Subjective evaluation is laborious and expensive. There are also limits on what naive human listeners can be asked to judge. It is also probable that they are not actually very good at some tasks, such as judging speaker similarity for unfamiliar voices. A variety of objective methods have been developed, or borrowed from other fields, to use as an alternative or complement to subjective evaluation of synthetic speech. 

It is important to make a distinction between methods designed to estimate synthetic speech naturalness, and those for measuring speech signal \textit{quality}. The latter come from the field of speech coding, transmission, and enhancement, and include PESQ (ITU-T P.862 \href{https://www.itu.int/rec/T-REC-P.862}{www.itu.int/rec/T-REC-P.862}) and POLQA (ITU-T P.863 \href{https://www.polqa.info}{www.polqa.info}); these do not generally correlate well with listeners' ratings of naturalness, and were never intended for that purpose.

\begin{itemize}
\item Objective measures of \gls{MOS} naturalness: these are models trained on data from past subjective evaluations, which comprise pairs of speech utterances and listener ratings. Given a new speech utterance, the model predicts a score on a 1--5 scale. These scores correlate moderately well with what would be obtained in a subjective test, and are widely reported in publications. But generalisation to new forms of speech generation remains a challenge (see Section~\ref{section:SotA:evaluation}). Examples\footnote{Google's AutoMOS \citep{AutoMOS} was an earlier example but no model was released, so it was of no practical use to the field.}:  MOSNet \citep{MOSNet}; UTMOS \citep{UTMOS}, an entry to the VoiceMOS Challenge (Page~\pageref{page:VoiceMOS}).
\item \gls{ASR} can be used to transcribe synthetic speech and thus estimate intelligibility.  Modern \gls{ASR} is remarkably robust to recording conditions, speaker accent, mispronunciations, etc., so is likely to provide rather optimistic estimates of how intelligible the speech would be to listeners. Nevertheless, \gls{ASR} is very widely used for \gls{TTS} and \gls{VC} evaluation. Example: Whisper \citep{WhisperASR}.
\item Methods developed for \gls{ASV} have been repurposed for measuring speaker similarity. A typical approach involves obtaining a speaker embedding from natural reference speech, and comparing this to an embedding from synthetic speech. The cosine similarity, which ranges from 0 to 1, is reported. Examples: ECAPA-TDNN \citep{ECAPA-TDNN}, Resemblyzer\footnote{\href{https://github.com/resemble-ai/Resemblyzer}{github.com/resemble-ai/Resemblyzer}}; see also speaker encoding in Section~\ref{section:SotA:speaker_encoder}.
\item \citet{VCC2018Benchmark} repurposed an \gls{anti_spoofing} \gls{countermeasure} for use as an objective measure for \gls{VC}.
\end{itemize}

Up to this point, we have only considered the evaluation of speech intended for human consumption. There are two other cases: speech that has been generated to attack \gls{ASV}, and natural speech that has been manipulated to protect the speaker's privacy when submitting their speech to, for example, an untrusted \gls{ASR} system. 

\mysummarybox
{of evaluating AI speech generation}
{
\begin{itemize}
\item Most AI-generated speech is intended for human consumption, except in special cases:
\begin{itemize}
\item \gls{voice_anonymisation} to achieve privacy in human-to-machine applications is coming up in Section~\ref{section:SotA:TTS:voice_anonymisation};
\item \gls{spoofing} attacks on \gls{ASV} are coming up in Section~\ref{section:sota:deepfake_detection}.
\end{itemize}
\item Since \gls{ASV} models have long outperformed humans at speaker verification/identification, they \textit{probably} do so for synthetic speech too.
\item objective measures rely on models, learned from data:
\begin{itemize}
\item these models only attempt to predict listener ratings;
\item they do not simulate human listening, so are not sensitive to the same things;
\item for example, \gls{ASV} models have been found to rely entirely on low-level acoustic features.
\end{itemize}
\end{itemize}
}

\subsubsection{Challenges and competitions}
\label{section:history:challenges:challenges}
\label{section:challenges} 

\myintrobox
{next}
{
\glsreset{SotA}
\begin{itemize}
\item Evaluation entails making comparisons with other systems published in the literature.
\item Replicating such systems can be near-impossible due to incomplete system descriptions, closed-source code, proprietary data, computational cost, technical difficulty, etc.
\item It is generally incorrect to compare results between listening tests: to compare systems, they must be evaluated within the same test.
\item As a consequence, making fair comparisons is hard or impossible for individual research teams.
\item One solution is a centrally-organised challenge or competition to compare systems.
\end{itemize}
}

Challenges and competitions serve an important role in this field. They enable the direct comparison of competing approaches by 1) establishing a level playing field, typically comprising common data and a standardised evaluation methodology, and 2) directly comparing systems. They can raise visibility of a newly-emerging area, attract researchers to work on it, and thus accelerate progress. But there are risks of stifling innovation: participating teams may converge on a small number of proven approaches rather than risk trying a novel method in a high-stakes public contest; the performance measures employed by a challenge may be adopted across the field and continue to be used beyond their useful lifespan.

Some challenges report their overall results in semi-anonymous fashion, without linking specific participating teams to the reported performance measures. The participating teams are then free to choose whether to reveal which result was theirs. This was first done by the Blizzard Challenge, to encourage industry participation. It is often possible to link most results to participating teams by reading each team's paper. Not all challenges require participating teams to submit a paper describing their system; this substantially devalues those challenges, since it becomes far harder to learn scientific lessons, such as which approaches were most successful. The three primary challenge series are:
\vspace{2mm}

\begin{itemize}
\item Blizzard Challenge for evaluating \glsreset{TTS}\gls{TTS}.
\item Voice Conversion Challenge (VCC) for evaluating \glsreset{VC}\gls{VC}.
\item ASVspoof (\glsreset{ASV}\gls{ASV} \gls{spoofing}) for testing the effectiveness of \gls{anti_spoofing} \glspl{countermeasure}.
\end{itemize}
\vspace{2mm}

\noindent These have been joined in more recent years by new series that represent emerging tasks:
\vspace{2mm}

\begin{itemize}
\item VoicePrivacy challenge for \gls{voice_anonymisation}.
\item VoiceMOS challenge for automatic estimation of synthetic speech \gls{naturalness}.
\item ADD (Audio Deep synthesis Detection) for generating, detecting, localising, and attributing speech \glspl{deepfake}.
\end{itemize}
\vspace{5mm}

\begin{longtable}{lp{2cm}|p{2cm}|p{11.5cm}}
Year & Challenge & Task & Key findings \\\hline

2005 & Blizzard & \gls{TTS} & First large-scale listening test; release of US English ARCTIC databases, 4 speakers $\times$ $\sim$\SI{1}{\hour}; established \gls{MOS} as the standard for evaluating naturalness; mostly naive listeners; 6 entries (i.e., participating research teams); \citet{Blizzard2005Summary,Blizzard2005Evaluation}. \\\hline

2006 & Blizzard & \gls{TTS} &  Larger database, US English, single speaker; 14 entries; highest intelligibility from \gls{HMM}-based \gls{SPSS} but highest naturalness from \gls{unit_selection}; \citet{Blizzard2006Summary}.\\\hline

2007 & Blizzard & \gls{TTS} & Even larger database, still US English, single speaker, \SI{8}{\hour} with a sub-task on $\sim$\SI{1}{\hour}; 16 entries; first use of speaker similarity testing, to ensure entries did actually use the common dataset; best system was \gls{unit_selection} from the company IVONA that would later be acquired by Amazon to create the first Alexa voice; \citet{Blizzard2007Summary,Blizzard2007Analysis}. \\\hline

2008 & Blizzard & \gls{TTS} &  
Even larger database for English, but now UK accent, single speaker, \SI{15}{\hour}; added Mandarin, single speaker, \SI{6.5}{\hour}; 19 entries; included new listener types (older; visually-impaired); over 1000 listeners in total; first time benchmark systems were included, to facilitate year-on-year comparisons; highest intelligibility from \gls{HMM}-based \gls{SPSS} driving a \gls{parametric_vocoder}, but highest naturalness from \gls{hybrid_unit_selection} using waveform concatenation -- a pattern that would persist for nearly 10 subsequent years, until the arrival of \glspl{neural_vocoder} put \gls{unit_selection} out of the running; \citet{Blizzard2008Summary}. \\\hline

2009 & Blizzard & \gls{TTS} & UK English and Mandarin again, similar size databases to 2008; 19 entries; sub-task in which synthetic samples were processed with a telephone line simulator before subjective evaluation, revealing that synthetic speech intelligibility is degraded more than natural speech; \citet{Blizzard2009Summary}. \\\hline

2010 & Blizzard & \gls{TTS} &  UK English again, this time commercial-quality recordings of a professional speaker, plus Mandarin again; first tentative use of paid commercial crowdsourced listeners (Amazon Mechanical Turk); sub-task in which synthetic samples had noise added before subjective evaluation in which one system achieved \textit{higher} intelligibility than natural speech, by simulating the Lombard effect; another sub-task using higher-sampling rate speech;
the most natural systems are again (hybrid) \gls{unit_selection}; \citet{Blizzard2010Summary}.\\\hline

2011 & Blizzard & \gls{TTS} & US English only, single professional speaker,\SI{17}{\hour}; 9 entries, suggesting fading interest in repeating a similar task every year; very clear pattern: all the higher-naturalness systems used (hybrid) \gls{unit_selection} and all the lower-naturalness systems were \gls{HMM}-based \gls{SPSS}, with roughly the reverse pattern for intelligibility; \citet{Blizzard2011Summary}.\\\hline

2012 & Blizzard & \gls{TTS} &  US English only; again, only 9 entries, confirming that the field was in need of fresh challenges; substantial number of listeners recruited using Amazon Mechanical Turk and found to be less reliable than some other listener types; \citet{Blizzard2012Summary}.\\\hline

2013 & Blizzard & \gls{TTS} & A "reboot" of the challenge with substantial changes in the data: \SI{300}{\hour} of US English, single speaker, reading \textit{audiobooks} - the use of large quantities of "found" data is the start of an important trend in the field; 15 entries; added 4 Indian languages $\times$ $\sim$\SI{1}{\hour}; first attempt to go beyond evaluating naturalness and measure multiple dimensions; majority of systems are now much better than the benchmark systems, illustrating considerable progress over the last few years; \citet{Blizzard2013Summary,Blizzard2013SummaryIndian}.\\\hline

2014 & Blizzard & \gls{TTS} & Indian languages (only), 6 languages $\times$ $\sim$\SI{2}{\hour}. One task involved code-switching (to English). Only 7 entries, suggesting limited interest in such small datasets -- the field was already moving on to larger, found data; \citet{Blizzard2014Summary}. A summary of the first 10 years of the Blizzard Challenge can be found in \citep{BlizzardKingLoquens2014}.\\\hline

2015 & Blizzard & \gls{TTS} &  Very similar to 2014, with 3 languages $\times$ $\sim$\SI{4}{\hour} plus 3 languages $\times$ $\sim$\SI{2}{\hour}, again monolingual and code-switching tasks, and 8 entries; \citet{Blizzard2015Summary}.\\\hline

2015 & ASVspoof1 & \gls{anti_spoofing} & Human vs. \gls{VC}/\gls{TTS}; 16 teams. Relatively simple attackers (i.e., \gls{VC} \& \gls{TTS} systems). Near-perfect detection of seen attackers; very poor detection of an \textit{unseen} \gls{TTS} system, because detectors over-fitted to low-level signal processing artefacts of seen systems. Teams mostly focussed on features, not classifiers; \citet{ASVspoof2015Summary, ASVspoof2015Journal}. Post-challenge work introduces Constant Q Cepstral Coefficients \citep{todiscoConstantQCepstral2016}, widely adopted in many subsequent systems.\\\hline

2016 & Blizzard & \gls{TTS} & A return to English (this time, British) with a domain-specific database of professionally-read children's stories from a well-known publisher, attracting 13 entries. Added a third benchmark: the neural-network system Merlin \citep{Merlin}. Picked up the multidimensional listening test design from 2013. Most entries were \gls{unit_selection} with many of those, including the top 3 most natural, being \gls{DNN}-driven \gls{hybrid_unit_selection}; \citet{Blizzard2016Summary}.\\\hline

2016 & VCC & \gls{VC} & The first ever challenge on \gls{VC}, modelled on the Blizzard Challenge (common data and a single subjective evaluation for all systems, here a lab-based listening test with paid listeners). Monolingual parallel data from 4 source and 4 target speakers of US English each saying 162 utterances, from the DAPS dataset \citep{DAPS}. Attracted 17 entries, which generally took a mapping approach and used a \gls{parametric_vocoder} such as STRAIGHT \citep{STRAIGHT}; \citet{VCC2016Summary, VCC2016Analysis}.\\\hline

2017 & Blizzard & \gls{TTS} & Essentially a re-run of the 2016 challenge, with slightly more data and 13 entries. Whilst the most natural system was once again \gls{hybrid_unit_selection}, for the first time two systems employing a vocoder were also very highly ranked. Both used \gls{DNN} \glspl{acoustic_model}, with one driving STRAIGHT \citep{STRAIGHT} and the other driving a \gls{neural_vocoder} based on recently-announced WaveNet \citep{WaveNet}; \citet{Blizzard2017Summary}.\\\hline

2017 & ASVspoof2 & \gls{anti_spoofing} & Audio replay detection. Very popular, with 49 teams. Best system \citep{ASVspoof2017SystemS02} is an ensemble of multiple features and classifiers. Subsequently,  flaws in the challenge data were discovered -- digital silence in some recordings \citep{ASVspoof2017Analysis} and telephony artefacts \citep{ASVspoof2017ArtefactsJournal} -- which render the task trivial in some cases: an important lesson for future challenges; \citet{ASVspoof2017Summary}.  \\\hline

2018 & Blizzard & \gls{TTS} & Another run of the same task as 2016 and 2017, with 10 entries. The majority of entries were now \gls{DNN}-based, and the few concatenative entries were all  \gls{hybrid_unit_selection}. The great majority of systems have comparable intelligibility, suggesting convergence on "solving" this aspect of \gls{TTS}; \citet{Blizzard2018Summary}. \\\hline

2018 & VCC & \gls{VC} & Using the same data as 2016, but at a higher sampling rate and allowing the use of additional external data, for both parallel and non-parallel tasks, attracting 23 entries. The field had advanced considerably since 2016, with the emergence of the \gls{PPG} approach (essentially an \gls{ASR} model as the encoder, a speaker-dependent \gls{acoustic_model} to perform the conversion, followed by a \gls{neural_vocoder} to generate the waveform). \glspl{PPG} now became the \gls{SotA} for non-parallel \gls{VC}. For the first time in a \gls{TTS} or \gls{VC} challenge, \gls{ASR} was used to measure intelligibility, not listeners; this would later become common practice; \citet{VCC2018Summary}. \\\hline

2019 & Blizzard & \gls{TTS} &  The Challenge finally tackles Chinese! Unsurprisingly, a record number of entries were received: 24. Most entries use a neural \gls{acoustic_model}, generally an encoder-decoder architecture (including many based on Tacotron \citep{Tacotron2}), driving a \gls{neural_vocoder} such as WaveNet \citep{WaveNet} or WaveRNN \citep{WaveRNN}. Intriguingly, "speech expert" listeners rated one system to be as natural as recorded speech, although this was not the case across other listener types; \citet{Blizzard2019Summary}. \\\hline

2019 & ASVspoof3 & \gls{anti_spoofing} & Human vs. \gls{TTS}, \gls{VC}, replay. Increasing popularity to 63 teams; as in 2017, the best system \citep[Section III.B]{ASVspoof2019Journal} is an ensemble, as are most other entries; \gls{anti_spoofing} is now well-established (e.g., regular mainstream sessions at conferences); \citet{ASVspoof2019Summary}.\label{page:ASVspoof2019}  \\\hline

2020 & Blizzard & \gls{TTS} & Again, Chinese data was used, but this time with tasks on Mandarin and the Shanghainese dialect, with 16+8 entries respectively. For the first time, all entries were neural, with most employing sequence-to-sequence models such as Tacotron \citep{Tacotron} or Tacotron~2 \citep{Tacotron2}, and all using a \gls{neural_vocoder}. Four systems were as intelligible as recorded speech, confirming again that this aspect really is closer to being "solved" than naturalness; \citet{Blizzard2020Summary}. \\\hline

2020 & VCC & \gls{VC} &  Adds cross-lingual and semi-parallel tasks, now using data from the EMIME project. The focus remains on using very small amounts of data from the source and target speakers, plus external data. The large number of 30 entries indicates growing popularity of \gls{VC}, which is being driven by a convergence of \gls{VC} and \gls{TTS} techniques, such as using the \gls{TTS} model Tacotron as the decoder (synthesiser) within "\gls{ASR}+\gls{TTS}" or \gls{PPG}-based architectures for \gls{VC}, and the use of precisely the same \glspl{neural_vocoder} for both \gls{VC} and \gls{TTS} tasks. This was the last speech VCC: subsequent SVCC challenges in 2023 and 2025 would be for singing; \citet{VCC2020Summary}. \\\hline

2020 & VoicePrivacy & \gls{voice_anonymisation} & The first in a new series, the task being hiding speaker identity whilst retaining utility for downstream tasks (here: \gls{ASR} and human listening), attracting entries from 25 teams. Data from VoxCeleb \citep{VoxCeleb,VoxCeleb2}, LibriTTS \citep{LibriTTS}, and VCTK \citep{VCTK}. Systems took either an x-vector-plus-speech synthesis, or a signal processing approach, the former offering better privacy. 
Results indicate that there is a false sense of privacy against standard \gls{ASV} because an informed attacker (an anonymisation-aware \gls{ASV} system) is often able to identify the source speaker via information leakage. It was also found that objective metrics were \textit{not} correlated with the subjective opinion of listeners; \citet{VoicePrivacy2020Results}.\label{section:history:challenges:voice_privacy} \\\hline

2021 & Blizzard & \gls{TTS} &  European Spanish data from a professional speaker, and including a sub-task on synthesising sentences including English words. 14 entries from a more diverse range of countries than many previous challenges. Around this time, the amount of data being used to train models had started to rise steeply, which was out of reach for many academic teams, so a limit of \SI{100}{\hour} of speech data was imposed. The amount of speech from the target speaker was under \SI{10}{\hour}, mirroring the typical commercial use case (still the case today): models (pre)trained on large amounts of multi-speaker data (typically pre-existing "found" data) as the basis for generating a small number of known speaker identities, e.g., by \gls{finetuning}. Most entries used one of the two dominant model classes of the time: Tacotron~2 \citep{Tacotron2} or FastSpeech~2 \citep{FastSpeech2}; this was in conjunction with a Generative Adversarial Network (GAN)-based \gls{neural_vocoder}, with HiFi-GAN \citep{HiFi-GAN} being the most popular choice, for reasons explained in the \gls{TTS} timeline on page~\pageref{section:TTS_timeline:HiFiGAN_is_popular}; \citet{Blizzard2021Summary}. \\\hline

2021 & ASVspoof4 & \gls{anti_spoofing}, \gls{deepfake} detection & Signals underwent telephone transmission or compression (mp3, etc.). Continued popularity with 54 teams. But all \gls{deepfake} detectors failed to generalise well to unseen methods: analysis reveals that almost all are relying on  artefacts in non-speech regions \citep[Figure 3(c)]{ASVspoof2021Journal}. Lower error rates, setting a new \gls{SotA}, are achieved post-challenge \citep{takAutomaticSpeakerVerification2022} by using a pre-trained \gls{SSL} model (not allowed in ASVspoof4, but subsequently became commonplace), with \gls{finetuning} on \gls{spoofed} speech, then a sophisticated neural classifier trained with data augmentation; \citet{ASVspoof2021Summary, ASVspoof2021Journal}. \\\hline

2022 & VoicePrivacy & \gls{voice_anonymisation} & The main innovations since 2020 are in devising a better \gls{pseudo_speaker}. ECAPA-TDNN \citep{ECAPA-TDNN} supersedes x-vectors as a speaker representation. An \gls{ASR}+\gls{TTS} pipeline performs best. Once again, subjective results are at odds with objective metrics. The challenge design has become excessively complex, resulting in a small number of only academic participants, a lesson for all challenges; \citet{VoicePrivacy2022Plan,VoicePrivacy2022Results}. \\\hline

2022 & ADD & \gls{deepfake} detection \& generation & Follows from ASVspoof 2021, but with different organisers and using Mandarin speech. \label{page:ADD_challenge} The key innovation is to approach real-world conditions with "low-quality" fakes (in fact high-quality \gls{deepfake} speech, mixed with background noise and music), and partially-fake speech from the HalfTruth dataset \citep{HalfTruth}. There is also a \gls{deepfake} generation task. Immediately popular, with 42 entries. The summary paper lacks summary information on the entries; only a few detection entries and  one generation entry appear to have a paper. "Only 3 of the 42 participating teams produced systems that outperformed the best baseline" \citep[Section 5.2]{ADD2022Summary} and at least one of those was from the challenge organisers \citep{xueAudioDeepfakeDetection2022} -- see Section~\ref{section:current_approaches_deepfake:technical:xueAudioDeepfakeDetection2022}. Detection of "low-quality" fakes was surprisingly hard (best \glspl{EER} above 20\%) but detection of partially-fake speech was much easier (best \glspl{EER} below 5\%); \citet{ADD2022Summary}. \\\hline

2022 & VoiceMOS & \gls{MOS} \mbox{prediction} & The first ever challenge on predicting human ratings of \gls{TTS} and \gls{VC} speech, i.e., constructing an objective measure (Section~\ref{section:history:challenges:evaluation}), with 22 teams participating; training data came from the subjective listening test results for English samples collected over 10~years in the Blizzard Challenges, Voice Conversion Challenges, etc.; although task is \gls{MOS} prediction, main results are for the task of ranking, not absolute score prediction; all top-performing systems used \gls{SSL} models plus \gls{finetuning}; however, on Out-Of-Domain test samples (in Chinese), this is less clear-cut; \citet{VoiceMOS2022Summary}.\label{page:VoiceMOS} \\\hline

2022 & SASV & spoofing-aware ASV & Integrated systems that perform both \gls{anti_spoofing} and \gls{ASV}, with 23 teams; the top systems were all ensembles of independent \gls{ASV} and \gls{anti_spoofing} \gls{countermeasure} sub-systems; \citet{SASV2022Summary}. \\\hline

2023 & Blizzard & \gls{TTS} & European French, with $\sim$\SI{50}{\hour} from the main speaker and $\sim$\SI{2}{\hour} from another speaker, attracting 18 entries. Many innovations were made in the evaluation methods, both generally such as multi-stage testing to discriminate amongst the very best systems, and language-specific. This is part of an emerging consensus in the field that standard protocols, such as \gls{MOS}, are no longer fit for purpose (noting that the Blizzard Challenge was itself responsible for making these the standard, nearly two decades ago) -- we will follow up on this in Section~\ref{section:SotA:evaluation}; \citet{Blizzard2023Summary, Blizzard2023Journal}. \\\hline

2023 & SVCC & singing \gls{VC} & Followed on from the VCC series, but now for singing synthesis, with 24 teams; as with \gls{VC}, most systems used \gls{ASR}+\gls{TTS} architectures, with either \gls{PPG} features from an \gls{ASR} model, or \gls{SSL} features, for the \gls{linguistic_content} representation; \citet{SVCC2023Summary}.\\\hline

2023 & ADD & speech \gls{deepfake} detection \& generation & Added new tasks: locate regions within partial-fakes from HalfTruth dataset \citep{HalfTruth}; deepfake algorithm recognition (to attribute fakes to specific \gls{deepfake} generators). Both tasks contribute some level of explainability. Remained very popular, with around 50 entries, of which only around half had their paper accepted for publication. The best algorithm classifier is described in Section~\ref{section:current_approaches_deepfake:requirements:xieGeneralizedSourceTracing2024a} and used similar components to the best \gls{deepfake} detector \citep{wuUSTCNERCSLIPSystemTrack2023}; \citet{ADD2023Summary}. \\\hline

2023 & VoiceMOS & \gls{MOS} \mbox{prediction} & Increased task difficulty with a focus on rating Out-Of-Domain samples; also added singing and enhanced noisy speech tasks; the top-ranked team used different methods for each task with \gls{SSL} again being best for rating synthetic speech \citep{VoiceMOS2023_T06_LE-SSL-MOS}; another highly-ranked system used the existing \gls{MOS} predictor UTMOS \citep{UTMOS} plus a very simple model with only 768 parameters; 10 teams; \citet{VoiceMOS2023Summary}. \\\hline

2024 & ASVspoof5 & \gls{spoofing}-robust \gls{ASV}; \gls{deepfake} detection & Added more attack types; open condition task allowed use of external data and models, dramatically reducing error rates, especially for best system T45; most systems better than top systems from 2021; \citet{ASVspoof2024Summary}. \\\hline

2024 & VoicePrivacy & \gls{voice_anonymisation} & Challenge design simplified, therefore more participants; stronger baselines; stronger attackers; use of large pre-trained models and much larger data; again, \gls{ASR}+\gls{TTS} pipeline provides near-perfect anonymisation (but not emotion preservation); \citet{VoicePrivacy2024Plan,VoicePrivacy2024Results}. \\\hline

2024 & VoiceMOS & \gls{MOS} \mbox{prediction}  & Shifted focus to only the most natural synthetic speech in order to "zoom in" and discriminate amongst the very best \gls{TTS} and \gls{VC} systems; 8 teams -- steadily declining since 2022; \citet{VoiceMOS2024Summary} \\\hline

2024 & SVDD & singing \gls{deepfake} detection & Two tracks, one with controlled, clean singing samples which 77 teams entered, the other with "in the wild" samples from user-generated content websites, which no teams entered; the best systems were all ensembles with multiple features and/or classifiers, with the top system fusing 3 distinct sub-systems trained on multiple singing datasets \citep{SVDD2024_top_system}; \citet{SVDD2024Summary}.\label{page:SVDD}\\\hline

2024 & CoVoC & \gls{TTS} & A "Conversational Voice Clone" challenge sponsored by a commercial data provider (MagicData), with 11 teams; used designed test material to capture conversational and connected-speech effects; 10 "professional" listeners, no crowdsourcing; somewhat suspiciously, the organiser's own official system, based on TorToise \citep{TorToise}
driving a \gls{NAC}, was ranked top; \citet{CoVoC2024Summary}. \\\hline

2024 & VoicePrivacy Attacker & \gls{ASV}  & Attacker systems against voice anonymisation: using \gls{ASV} to determine the \gls{source_speaker} for anonymised speech. \textit{In progress at the time of writing (summary paper not yet available)}. \\\hline 

2024 & SSTC & \gls{SSV} & The  Source Speaker Tracing Challenge is similar to VoicePrivacy Attacker, but using a database of \gls{VC} speech rather than purposely-anonymised speech, with 6 teams; the \textit{tracing} task is narrowly defined as \textit{verifying} whether two \gls{VC} samples originated from the same source speaker; the top-ranked system employed speaker-contrastive learning \citep{SSTC2024_top_system} and is the case study for Section~\ref{section:current_approaches_deepfake:requirements:source_speaker_tracing}; \citet{SSTC2024Summary}. \\\hline

2024 & ICAGC & \gls{voice_cloning} & Inspirational and Convincing Audio Generation Challenge, with entry rules copy-pasted from the Blizzard Challenge; evaluated \gls{TTS} systems for "persuasiveness and acceptability" in two tasks using Chinese data, Inspirational Emotion Modulation (5 teams), and Convincing Background Audio and Speech Fusion Generation (only 2 teams); simple \gls{MOS} evaluations only; this challenge is part of a trend for Chinese replications of existing challenges (in this case, of Blizzard); \citet{ICAGC2024} \\\hline

2024 & LIMMITS & \gls{zero_shot} \gls{TTS} & Few- and zero-shot voice cloning for both \gls{VC} and \gls{TTS} in multiple Indian languages, with 9 teams, who were required to base their systems on a pre-trained YourTTS model \citep{YourTTS}; simple \gls{MOS} evaluations only; \citet{LIMMITS24} \\\hline

2025 & SVCC & singing \gls{VC} & \textit{In progress at the time of writing} \\\hline

2025 & Blizzard & \gls{TTS} & The first Blizzard Challenge on a genuinely low-resource language, Bildts, and a focus on scaled-down systems. \textit{In progress at the time of writing.}  \\\hline

\end{longtable}

\mysummarybox
{of challenges and competitions}
{
\begin{itemize}
\item Challenges and competitions have a high value for newly-emerging fields or use cases.
\item They sometimes become too complex, resulting in low participation.
\item Long-running challenges can become highly-influential, for better or worse (e.g., use of \gls{MOS} in the Blizzard Challenge).
\end{itemize}
}

\newpage

\section{The current state-of-the-art in AI speech generation}
\label{section:SotA:generation}
\glsresetall
\begingroup
\parindent=0em
\localtableofcontents
\endgroup


\myintrobox
{in this section}
{
\begin{itemize}
\item With the high rate of publication in this field, it is easy to become overwhelmed.
\item A single, comprehensive literature review would be long and indigestible.
\item Instead, this section offers two complementary views of the \gls{SotA}:
\begin{itemize}
\item Section~\ref{section:current_approaches:technical}: technology \textbf{components} ("building blocks"); 
\item Section~\ref{section:current_approaches:requirements}: system \textbf{requirements} (e.g., data, compute), \textbf{capabilities}, and \textbf{limitations};
\end{itemize}
\item A selection of survey articles, which complement this report, is listed in Section~\ref{section:generation_reviews_and_surveys}.
\end{itemize}
}


\subsection{Current approaches to speech generation: the building blocks}
\label{section:current_approaches:technical}

\myintrobox
{in the "building blocks" section}
{
\begin{itemize}
\item Our first view of the \gls{SotA} is by looking at the components of current speech generation systems.
\item Although kept separate in the early part of this report, \gls{TTS} and \gls{VC} are now covered together, because they are converging on common approaches and shared components.
\item Each component might be an explicit and well-defined part of a system.
\item Or it may only exist conceptually, in an integrated architecture.
\end{itemize}
}

\subsubsection{The Text-to-Speech front-end}
\label{section:SotA:text_processing_and_the_front_end}

\paragraph{Text normalisation}
Despite a long history of work in text processing for \gls{TTS}, and well-established solutions for resolving much of the ambiguity of text, current systems reported in research publications often use very basic text processing, typically limited to rudimentary text normalisation using inadequate methods such as regular expressions to detect and expand abbreviations. They rely on machine learning to solve everything else, using the same paired text-speech data used to train the \gls{acoustic_model}. Regular expressions are not the \gls{SotA} method for text normalisation! They do not work very well, so many current systems now make what \citet{sproatRNNApproachesText2017} call "silly errors": things that classical approaches would never get wrong. This situation is likely to end soon. Because standard methods for evaluation (e.g., \gls{MOS} naturalness, or using \gls{ASR} to estimate intelligibility) have saturated, evaluation is receiving a long-overdue overhaul. We are witnessing a rise in more linguistically-sophisticated, multidimensional evaluations that will catch front-end errors: see Section~\ref{section:SotA:evaluation}.

\paragraph{Resolving ambiguity in text}
A classical \gls{front_end} does a lot more than text normalisation (Section~\ref{section:TTS_primer}). It attempts to resolve all \textit{ambiguity} before predicting pronunciation. This is done primarily through syntactic cues (e.g., "read" is ambiguous, until the tense is specified, at which point the correct pronunciation can be retrieved from a dictionary), with some recourse to shallow semantics such as word sense disambiguation (e.g., whether "bass" is a fish or a musical term). Attempting to learn all of this from paired text-speech data will probably never work well enough: even when that data is very large, it will not include sufficient examples of all the cases that a classical \gls{front_end} can easily handle. The problem can be partially solved in a purely data-driven way, using implicit representations of structure and meaning derived from text-only models, starting with BERT, and extending to the latest \glspl{LLM} of text, which are very strong models of structure and meaning. Using a text-only \gls{LLM} as the starting point for a speech-generating \gls{SLM} (i.e., the pre-training -- post-training paradigm) may bring with it those implicit representations.

\paragraph{Pronunciation}

Some systems, notably those from groups with mature commercial \gls{TTS} systems (e.g., Microsoft), do still phonemise the input. But many \textit{research} systems reported in the recent literature take \gls{normalised_text} as their input and omit explicit prediction of pronunciation. They rely on learning that, and everything else, from data. In such a \gls{TTS} system, which does not phonemise the text, it is common to find a component labelled "text encoder" which performs the conversion of \gls{normalised_text} to another representation suitable for passing to the \gls{acoustic_model} (Section~\ref{section:SotA:acoustic_model}). 

Learning pronunciation from data does not mean relying \textit{solely} on paired text-speech data: even a million hours of speech only contains around a billion sentences, which is tiny compared to the corpora used for text models. There are two ways in which current systems leverage large text-only corpora to escape this limitation: using a pre-trained text tokeniser in the front-end, as described next; or using a pre-trained text-only \gls{LLM} as the starting point for a \gls{SLM}, as in LLasa or Spark-TTS (Section~\ref{section:SotA:TTS:controllable}). 

\Gls{normalised_text} is a sequence of \glspl{character}, which is a very simple representation of language; using it as the input to a \gls{TTS} model requires the model to find common \gls{character} sequences (e.g., "th" or "oach"), \textit{then} learn how they map to sounds. These common \gls{character} sequences can be discovered from text alone. This is a standard step in text-based models and is called tokenisation; this is generally task-agnostic (i.e., not specialised for any particular task such as machine translation). So, a \gls{TTS} model can simply borrow a pre-trained tokeniser. Tokenised text is, in some sense, closer to pronunciation than the original \gls{character} sequence, so it should be easier to learn how to predict pronunciation from it. 

Few \gls{character}- or tokenised-text-input models have any identifiable internal representation of pronunciation.
This means that there is no obvious way to control their pronunciation or correct errors: the latter is critical for most commercial use cases. Therefore, it may be desirable for this class of models to employ some representation via which pronunciation \textit{can} be controlled or corrected. An obvious choice is a \gls{SSL} speech representation such as HuBERT \citep{HuBERT} which maps speech into a sequence of codes (or alternatively latent vectors) at a fixed rate of, for example, \SI{50}{\hertz}. \gls{SSL} representations are task-agnostic so, as mentioned in Section~\ref{section:relationship_to_text_image_video}, are useful for many different tasks.

One of the main arguments against phonemising the input is to avoid making unrecoverable errors, particularly for words not in the pronunciation dictionary. The main argument in favour of phonemising the input is precisely to take advantage of a pronunciation dictionary because this guarantees correct word pronunciation, notably for words where the pronunciation cannot easily be inferred from the spelling. A reasonable compromise solution is called \textit{representation mixing}, where the input text is partially-phonemised: for example, words found in the dictionary are phonemised and the remainder are left as their spelling. The input to the model  is an interleaved sequence of \glspl{character} and \glspl{phoneme}.

\mysummarybox
{of current approaches to the TTS front-end}
{
\begin{itemize}
\item Traditional approaches analysed the syntax (structure) and semantics (meaning) of the input text.
\item This analysis was generally only used internally, to disambiguate the text, then discarded.
\item Explicit syntactic and semantic analyses are now uncommon:
\begin{itemize}
\item some systems revert to older, inadequate methods for text normalisation, leading to "silly errors";
\item other systems replace them with task-agnostic representations learned from text alone.
\end{itemize}
\item Traditional approaches represent pronunciation explicitly using \glspl{phoneme}, which remains the preferred approach in commercially-deployed systems.
\item Task-agnostic \gls{SSL} representations learned from speech alone offer an alternative to \glspl{phoneme}, but are more commonly employed in \glsreset{VC}\gls{VC} than \glsreset{TTS}\gls{TTS}.
\end{itemize}
}

\subsubsection{The Voice Conversion encoder}
\label{section:SotA:vc_encoder}

\gls{VC} and \gls{TTS} systems have converged on similar neural architectures, at least for what we might call the  "backbone" \gls{acoustic_model}  (Section~\ref{section:SotA:acoustic_model}). Input to this core component is from either a text encoder (for \gls{TTS}) or a speech encoder (for \gls{VC}). This is an elegant and flexible system architecture because the interface between the encoder and the acoustic model in both cases is simply a sequence of tokens. Whether those are \gls{text_tokens}, \glspl{semantic_token}, \glspl{audio_token}, or indeed a  sequence of multiple token types (e.g., as for "zero-shot" TTS in Section~\ref{section:SotA:TTS:zero-shot}), does not change the architecture.

As we saw in Section~\ref{section:VC_primer}, the currently-dominant approach to \gls{VC} is to disentangle the input speech into representations that can either be passed through directly to the \gls{acoustic_model}, discarded and replaced, or transformed (e.g., in the case of \gls{prosody}).

The vast majority of \gls{VC} systems are designed to change \gls{speaker_identity} by preserving \gls{linguistic_content} and discarding any evidence of the \gls{source_speaker}'s identity. The encoder in such a \gls{VC} system is therefore responsible for extracting a representation of the \gls{linguistic_content} which contains minimal evidence of \gls{speaker_identity}. The handling of \gls{prosody} varies from system to system and depends on the use case: refer to the list in Section~\ref{section:SotA:start_of_use_cases}. The current \gls{SotA} for the encoder is a form of automatic speech recognition, and there are two broad approaches to creating this.

One approach is \gls{supervised} learning: training an explicit \gls{ASR} model to transcribe speech into text or \glspl{phone}. Such a model can then be employed as an encoder in various ways. Its final output can be used, either as "hard" symbols (in the extreme case, a word-level transcription, but more usually a phonetic transcription) or a "soft" probability distribution over \glspl{phone}, known as a \gls{PPG} (a widely-used representation). 

Generally, using symbols imposes a narrower information \gls{bottleneck} which is less likely to have \gls{leakage} of information about the \gls{source_speaker}, but has a much higher risk of error in conveying the \gls{linguistic_content}: changing just one symbol might change the meaning of the utterance. Conversely, using a distribution over supervised classes, which is most commonly a \gls{PPG}, mitigates the risk of making gross errors in the \gls{linguistic_content}, whilst very slightly increasing the risk of speaker leakage. Going a step further, instead of obtaining a \gls{PPG} from the output of an \gls{ASR} model, an internal \gls{latent} representation may be extracted from within the model.  A significant downside of the \gls{ASR} approach is that supervised learning requires labelled data, in amounts that are only available for well-resourced languages.

The other approach is \gls{unsupervised} learning, either from a pre-existing \gls{SSL} model, or learned as part of the full \gls{VC} model. This "implicit \gls{ASR}" is attractive since the model can be trained on speech without transcriptions, making a multilingual or language-independent model easier to achieve. Again, this representation might comprise "hard" symbols (e.g., codes from HuBERT k-means) or be "soft" or "distributed", such as a \gls{latent} vector taken from a middle layer of an \gls{SSL} model.

\pagebreak

A few \gls{VC} systems employ more than just a \gls{linguistic_content} encoder. They also explicitly extract other aspects of the source speech, such as prosody or style, so that these can be transferred to the target speech. One use case for this is \gls{VC} for \gls{voice_anonymisation}, covered in Section~\ref{section:SotA:TTS:voice_anonymisation} and the task in the 
VoicePrivacy challenge listed on page~\pageref{section:history:challenges:voice_privacy}.

\mysummarybox
{of current approaches to the VC encoder}
{
\begin{itemize}
\item Most encoders amount to some form of \gls{ASR}, whether explicit or implicit, to extract the \gls{linguistic_content} of the source speech.
\item In most use cases, treatment of other aspects (e.g., prosody, style) is less sophisticated than for the \gls{linguistic_content}.
\end{itemize}
}

\subsubsection{The speaker encoder}
\label{section:SotA:speaker_encoder}

\paragraph{Simple speaker representations}

For the restricted case where there are speaker labels available for the entire training data, including data for the \gls{target_speaker}(s), the simplest speaker representation is a \gls{one_hot} vector. This is the approach taken by Microsoft in their top-rated \gls{TTS} entries to the Blizzard Challenges of 2021 and 2023 (see Section~\ref{section:challenges}) \citep{DelightfulTTS, MuLanTTS} and in many other contemporary models. But in the general case, a method is required for representing unseen speakers (i.e., those not available when building the system). This may be an external model or an integrated part of the system.

\paragraph{External speaker representations}
\label{section:SotA:speaker_encoder:external_speaker_representations}

The most obvious choice of model here is one trained to perform \gls{ASV}. These are readily available, and are typically trained on a corpus with a \textit{very} large number of speakers, such as VoxCeleb1~or~2 \citep{VoxCeleb,VoxCeleb2}, and so are able to produce useful representations for unseen speakers. Since these models are trained specifically and only to discriminate between speakers from short audio samples, they tend to rely on low-level acoustic cues related to the vocal tract physiology of the speaker. They will not generally capture longer-term features such as accent or the speaker's idiosyncratic prosody. In general, \glspl{speaker_embedding} capture a rather limited picture of speakers \citep{rautenbergFeatureImportanceInterpretability2023}.

Relatively few \gls{TTS} models actually use external models to provide a speaker representation. One example is CosyVoice \citep{CosyVoice} which uses an x-vector from the CAM++ \gls{ASV} model. As \glspl{SLM} and other generative models emerged, the field of \gls{TTS} moved rapidly from simple \gls{one_hot} representations, bypassed external \gls{speaker_embedding} models, and started using a speech sample to "prompt" the model (see below), marking the rise in popularity of "zero-shot TTS": see Section~\ref{section:SotA:TTS:zero-shot}.

The situation in \gls{VC} is different, but also varies by use case, especially whether the target speaker is from a closed (i.e., known during training) or open set, as first introduced on page~\pageref{section:VC_timeline:x_to_y}. For a closed set of target speakers, it is normal for the model to have its own speaker encoder. For an open set, an external speaker representation is most common. Historically this would have been an x-vector \citep{x-vectors} but is now more likely to be from the successor model, ECAPA-TDNN \citep{ECAPA-TDNN}. Speaker embeddings from such models are also used in objective measures of speaker similarity: see Section~\ref{section:SotA:evaluation:speaker}.

\paragraph{Speech prompts as speaker representations}

In \gls{TTS}, there is a strong trend towards discarding any form of \textit{explicit} speaker encoding and instead providing a powerful \gls{acoustic_model} such as a \gls{SLM} with a speech \gls{prompt}, which is generally a few seconds of speech from the target speaker, typically as either a \gls{mel_spectrogram} or encoded as a sequence of \glspl{audio_token}. Examples of this trend can be found in the evolution of CosyVoice \citep{CosyVoice} into CosyVoice~2 \citep{CosyVoice2}, or in the VALL-E model family \citep{VALL-E-journal, VALL-E2}. Using an audio sample in this way allows the model to, in principle, adapt to \textit{any} previously-unseen \gls{target_speaker} during inference. Because no model parameters change, this is called "zero-shot TTS". This has become a very popular task for many recent models: see Section~\ref{section:SotA:TTS:zero-shot}.

\mysummarybox
{of current approaches to the speaker encoder}
{
\begin{itemize}
\item Current speaker encoders capture physical and physiological characteristics, not sociological ones (e.g., accent), primarily because of the very short speech samples used.
\item A single vector from an \gls{ASV} model is not sufficient to reconstruct a speaker's identity.
\item But it is sufficient to provide \gls{conditioning} to a model that already has the ability to generate that (or a similar-sounding) speaker.
\end{itemize}
}

\subsubsection{Acoustic modelling}
\label{section:SotA:acoustic_model}

Our definition of \gls{acoustic_model} here encompasses both \gls{TTS} and \gls{VC}. The \gls{SotA} in \gls{VC} is an \gls{ASR}+\gls{TTS} approach, or an \gls{SSL}-based equivalent of that (Section~\ref{section:VC_primer:ASR+TTS}), so the \gls{acoustic_model} will be essentially the same as that in a \gls{TTS} system, with the type of input being the only difference. 

The \gls{acoustic_model} receives input from either the \gls{TTS} front-end, or the \gls{VC} encoder. Given this input, and some optional further inputs or conditioning (e.g., something that provides the \gls{target_speaker} identity), it generates an \gls{acoustic_specification} that the subsequent \gls{waveform_generator} can use to generate the final output speech \gls{waveform}.

The \gls{acoustic_model} is the most data-hungry component of the system, because it requires training on substantial quantities of labelled data (minimally, text-audio pairs, for \gls{TTS}). Section~\ref{section:provenance_of_data} discusses the provenance of this data. (Other components can be trained on unpaired data: e.g., speech-only for the \gls{waveform_generator}.)

A variety of neural architectures are in current use for the  \gls{acoustic_model}. For \gls{TTS}, the model must predict the duration of each input token (sometimes called "upsampling" because the output sequence length is generally longer than the input); for \gls{VC}, the duration of the input speech is available and many (but not all) systems simply copy that: see Sections~\ref{section:SotA:VC:offline} and \ref{section:SotA:VC:streaming} for use cases.

It remains an open question where exactly speaker \gls{conditioning} should be injected into a system. Injecting it into the \gls{acoustic_model} may enable modelling of speaker-dependent prosody and some aspects of accent. Injecting it  later, into the \gls{waveform_generator}, limits those possibilities -- this is a common design in \gls{VC}: a speaker-conditioned \gls{neural_vocoder}.
 
Injecting it earlier, into the \gls{TTS} front-end, should theoretically enable the model to learn speaker-dependent effects in text processing (e.g., how a given speaker pronounces the year "2011"), and pronunciation (e.g., accent). Few, if any, systems attempt this explicitly. But the use of a speech \gls{prompt} as the speaker representation -- as described in Section~\ref{section:SotA:speaker_encoder} -- particularly with a \glsreset{SLM}\gls{SLM} (see below), does implicitly open up this possibility, especially when using \gls{character} input.

\paragraph{Encoder-decoder architectures}

"Encoder" and "decoder" here refer to two sub-parts of the \gls{acoustic_model} architecture, and should not be confused with other usages of these terms elsewhere. The classical way to deal with the upsampling task is to separate the \gls{acoustic_model} into two parts: the encoder runs at the "clock rate" of the input tokens (e.g., \glspl{character}, for a \gls{TTS} system without phonemisation); the decoder runs at the same rate as the \gls{acoustic_specification} (e.g., the frame-rate of a \gls{mel_spectrogram} or an \gls{audio_codec}). Between the two components, there is a mechanism to perform upsampling. This is either explicit duration prediction, as in models like FastSpeech~2 \citep{FastSpeech2} or implicit duration modelling using attention, as in Tacotron~2 \citep{Tacotron2}. During training, upsampling entails aligning the input and output sequences. During inference, upsampling involves predicting the duration of each input token in terms of output frames or tokens.

Models that have an explicit duration predictor need to train this component. One way to do this is with external supervision, using data with the duration of each input token labelled (e.g., a forced-alignment of a phonetic transcription). The need for supervision can be avoided by making the duration predictor a differentiable function, so it can be learned along with the encoder and decoder of the \gls{acoustic_model}. That is, the model must learn how to align input and output sequences, which is a hard task during the early stages of training, when the encoder and decoder are not yet well-trained themselves.  Models that use attention for implicit duration modelling also use that mechanism to learn the input-output alignment during training, although this may optionally be guided by external alignments. Alignment errors are a common failure mode of models that learn to align without external supervision: during \gls{training}, these manifest as slow or unstable convergence; during \gls{inference}, they manifest as skipped or repeated words.

\paragraph{Speech Language Models}

Many of the latest \glspl{acoustic_model} borrow a neural architecture first used in text \glsreset{LLM}\glspl{LLM}. When applied to speech generation, we will call these \glsreset{SLM}\glspl{SLM}. Other terms can be found in the literature, including "Large Speech Model" or "Large Speech Language Model". This class of models takes a conceptually different approach to encoder-decoder architectures, framing the problem as one of \textit{language modelling}. Given the input sequence as context (i.e., the "history" or "left context" in language modelling terms), the model performs a single step of inference to predict \textit{one} output token: e.g., the first \gls{audio_token}. This is then placed in the history, and inference runs again to predict the next output token: the model is auto-regressive. There is no explicit prediction of duration, so during inference the model must also predict when to stop generating output: this is a common source of error with this class of models (e.g., stopping speech generation before the end of the text, or appending extraneous speech-like sounds), as it was with previous-generation models that also only implicitly modelled duration, such as Tacotron~2 \citep{Tacotron2}, where it was called "stop token failure".

A key attraction of the \gls{SLM} approach is that the core, or "backbone", model is fairly generic: it is borrowed from text modelling, and can benefit very quickly from advances made in that field, whether that be better architectures or methods for training or inference. The philosophy of the approach is to take an "off-the-shelf", large, high-capacity, general-purpose sequence model (at the time of writing, a stack of Transformers) and to focus efforts on defining the task and curating suitable data. The most prominent example of the \gls{SLM} approach to \gls{TTS} is VALL-E \citep{VALL-E-journal}, used as a case study for "zero-shot" \gls{TTS} in Section~\ref{section:SotA:TTS:zero-shot}. This approach is also taken by \citet{zhouCodecLanguageModelBased2024} who use the same \textit{architecture} as the Llama text-only model. Llasa \citep{Llasa} goes a step further and not only adopts an \gls{LLM} \textit{architecture} for its \gls{SLM}, but it starts from a Llama  \textit{checkpoint}, as does Spark-TTS \citep{Spark-TTS} which is the case study in Section~\ref{section:SotA:TTS:controllable}. 

\paragraph{Flow and diffusion}

Using \glspl{SLM} for speech generation requires representing speech as a sequence of discrete symbols. Flow and diffusion models, in contrast, directly generate a continuous representation such as a \gls{mel_spectrogram} or waveform samples. These models generate novel samples by transforming a simple probability distribution (e.g., a Gaussian) through a sequence of differential equations. Models of this type are generally not auto-regressive, which makes it easier to parallelise computation. On the other hand, inference involves many steps of computation to gradually transform noise into the output sample. F5-TTS \citep{F5-TTS} is a good example for \gls{TTS}. The core of the model is the Diffusion Transformer (DiT), with a final flow component to generate a \gls{mel_spectrogram}. A pre-trained \gls{neural_vocoder} is used to generate a waveform from this.

\glspl{SLM} and flow/diffusion have complementary capabilities, and it is common to see both in a single model, especially for \gls{TTS}. The auto-regressive nature of an \gls{LLM} is well-suited to generation of (spoken) language sequences. Flow and diffusion are well-suited to generating the fine details of natural speech signals, so it is actually more common to see them employed in \glspl{neural_vocoder} than \glspl{acoustic_model}: see below in Section~\ref{section:SotA:waveform_generation}.

\mysummarybox
{of current approaches to the acoustic model}
{
\begin{itemize}
\item Encoder-decoder architectures are used for both \gls{TTS} and \gls{VC}. 
\item They may permit control over duration and other features.
\item \glspl{SLM} are good for generating sequences of relatively coarse-valued outputs, such as \glspl{audio_token}. They are typically employed as the core, or "backbone" of a system, in conjunction with additional components to prepare the input, and to convert their output to a waveform.
\item They are often controlled via a \gls{prompt}.
\item Flow and diffusion are good for generating continuously-valued outputs containing fine details. They are most often found in a specific \textit{component} of a larger model, such as the decoder of an encoder-decoder architecture.
\end{itemize}
}

\subsubsection{Waveform generation}
\label{section:SotA:waveform_generation}

Despite a number of attempts to construct \gls{end_to_end} systems for \gls{TTS}, the overwhelming majority of well-performing current systems are modular: they use a separate \gls{neural_vocoder} to get from an \gls{acoustic_specification} to the \gls{waveform}. This is also often the case for \gls{VC}.

The ideal vocoder should be able to convert speech waveforms into an \gls{acoustic_specification} and then achieve near-perfect reconstruction of the \gls{waveform} from that. This would be in itself almost trivial, except that constraints are placed on the  \gls{acoustic_specification}: primarily that it is something an \gls{acoustic_model} can generate, given text (in the case of \gls{TTS}) or speech (in the case of \gls{VC}) input. This constraint immediately creates tensions that in turn mean that the design choices for any practical waveform generator are a compromise. Furthermore, the design choices will depend on the capability of the \gls{acoustic_model}. There may also be secondary constraints on the \gls{acoustic_specification}, such as being a useful representation of a speech \gls{prompt}. So, it is currently the case that no single \gls{neural_vocoder} is the "perfect" choice for all AI speech generation use cases.

A practical illustration of this is how VALL-E~2  uses "the Encodec [sic] model to tokenise speech signals [i.e., the speech \gls{prompt}] and the Vocos decoder to generate target high-quality speech signals" \citep[Section~2.2]{VALL-E2}. Different vocoders might be used because (although not stated in the paper) the requirements for representing the speech \gls{prompt} are different to those for the \gls{waveform_generator}. EnCodec is described in \citet{EnCodec} and Vocos is described in \citet{Vocos}. The remainder of this section is structured by the principal design choices that must be made for any \gls{SotA} \gls{neural_vocoder}.

\paragraph{Handcrafted vs. learned acoustic specification}

The history described on page~\pageref{section:TTS_primer:acoustic_specification} has led us to a greater diversity of \glspl{acoustic_specification} than perhaps ever before, because they can now be learned from data.

The classical \gls{parametric_vocoder} and its source-filter
\gls{acoustic_specification} have been superseded by the only handcrafted \gls{acoustic_specification} still in widespread use: the \gls{mel_spectrogram}. This needs only a few hyperparameters to be selected: the frame size and shift (from within a fairly narrow range) and the number of mel filter bands (which increases for higher sampling rate waveforms). It is the only truly universal \gls{acoustic_specification} and offers the system builder a wide range of pre-trained \glspl{neural_vocoder} that all accept the same input. Some popular vocoders accepting \gls{mel_spectrogram} input were listed in the TTS timeline on page~\pageref{section:TTS_timeline:neural_vocoder}.

The alternative to the \gls{mel_spectrogram} is a learned representation. Usually, an encoder-decoder architecture is trained, using natural speech \glspl{waveform} as both input and output (i.e., an auto-encoder) and with the \gls{acoustic_specification} being the representation linking the two. Constraints are placed on this, to make it suitable for modelling, notably a frame rate comparable to handcrafted \glspl{acoustic_specification}, typically \qtyrange[range-units=single]{25}{100}{\frame\per\second}. Such models are usually trained with multiple \glspl{loss_function}, including \gls{adversarial} losses computed by discriminators that are concurrently trained to distinguish natural waveforms from synthetic (encoded-decoded) ones. After training, the model can be split into encoder and decoder components that can be deployed separately. Because this class of models came from the field of speech coding, where the primary goal is compression (the smallest possible \gls{acoustic_specification} for a desired reconstruction quality), the \gls{acoustic_specification} is most often also discrete. Such a model is known as a \glsreset{NAC}\gls{NAC}.

\paragraph{Continuous vs. discrete acoustic specification}

A \gls{NAC} with a discrete \gls{acoustic_specification} is the most common configuration, and is well-suited for the \gls{SLM} paradigm of acoustic modelling: see Section~\ref{section:SotA:acoustic_model}. An important design choice in any such \gls{NAC} is the codebook size used for discretisation. Rather than a single large codebook, most \glspl{NAC} adopt an approach well-known from classical speech coding: residual coding. For a \gls{NAC} this is \glsreset{RVQ}\gls{RVQ} with typically \numrange{8}{12} layers of coding from coarse to fine \citep[e.g.,][]{EnCodec,DescriptAudioCodec}. The residual nature of these layers creates a dependency between the codes for a single frame, adding complexity when generating these codes from an AI speech generation model since they must be generated in a specific order. Spectral Codecs \citep{SpectralCodecs} uses multiple codebooks in a different way -- one codebook per frequency band -- which avoids this dependency and allows, for example, an \gls{SLM} to generate all the codes for a single frame at once.

Discretisation is a powerful tool for compression but is, by definition, lossy. Discrete codes are also incompatible with \glspl{acoustic_model} that generate continuous values. It is straightforward to use the continuously-valued latents from a \gls{NAC} instead of the codes. For example, the "TorToise trick" \citep[Section~2.2.2]{TorToise} involves initially training an \gls{acoustic_model} to generate codes, then a final \gls{finetuning} stage that bypasses the discretisation and couples the \gls{acoustic_model} directly to the waveform generator: see under "Universal vs. matched to the acoustic model" below.

\paragraph{Speech-specific vs. general audio}

In classical speech coding, maximum compression is obtained when using a speech-specific method, accepting that it will give lower quality for other signals, such as music. For a \gls{NAC}, besides some architectural design decisions that are mildly-specific to speech, such as the \gls{waveform} sampling rate, frame rate, number of \gls{RVQ} layers, or codebook sizes, the training data is the main factor. For example, the low bitrate monophonic version of EnCodec is trained on "speech, noisy speech, music and general audio" but the higher quality stereo version is trained "on only \SI{48}{\kilo\hertz} music" \citep[Section~4.1]{EnCodec}, reflecting the intended use cases of the two versions.

\paragraph{Compression vs. generation}

Because of their origin in speech coding, most \glspl{NAC} are designed to achieve high rates of compression (relative to an uncompressed \gls{raw_waveform}). They can be operated at a variety of bitrates, from very low values which suffice for voice communication but are inadequate for \gls{TTS} or \gls{VC}, to higher rates that are good enough for high-fidelity music. There is nothing in their construction, nor specifically in the design of their \glspl{acoustic_specification}, that is tailored for speech generation from an \gls{acoustic_model}. Nevertheless, general-purpose \glspl{NAC} \citep{DescriptAudioCodec,EnCodec,SoundStream} perform very well for speech signals and have been widely adopted for speech generation.

A key challenge in acoustic modelling, and acutely so for autoregressive language models, is the long sequence length of the \gls{acoustic_specification}: for example, a \SI{5}{\second} utterance may result in a sequence of \numrange{250}{400} frames or \glspl{audio_token}. This is a very long context for even a large and powerful \gls{SLM}. This situation is exacerbated for \gls{RVQ} codes: a naive treatment of the \numrange{8}{12} layers is to flatten them along the time axis, effectively multiplying the length of the code sequence by the number of layers.

Therefore, a few \glspl{NAC} attempt to reduce the sequence length. The most obvious way is to reduce the number of frames per second. HALL-E \citep{HALL-E} reduces EnCodec's original frame rate of \SI{48}{\frame\per\second} to just \SI{8}{\frame\per\second} by learning to reorganise its codes. The large speech foundation model Moshi \citep{Moshi} operates at \SI{12.5}{\frame\per\second} and the Low Frame-rate Speech Codec \citep{LowFramerateSpeechCodec} operates at \SI{21.5}{\frame\per\second}.  Another way to reduce the effective sequence length, or in a more general sense the modelling burden placed on the \gls{SLM}, is to reduce the number of \gls{RVQ} layers. Language Codec \citep{Language-Codec} does this by encouraging better codebook utilisation, thus reducing the number of layers to 4.


\paragraph{Universal vs. matched to the acoustic model}

The ideal waveform generator might be a "black box", trained only on natural speech recordings and universally able to reconstruct the \gls{waveform} when driven by \textit{any} \gls{acoustic_model}. However, no \gls{acoustic_model} is perfect and each will leave its own unique pattern of errors in the generated \gls{acoustic_specification}. To compensate for this, it is possible to perform \gls{finetuning} of the waveform generator, using \gls{acoustic_model} output as its input and natural speech \glspl{waveform} as the target. Alternatively, the \gls{acoustic_model} and waveform generator can be jointly finetuned. In either case, this is computationally expensive because of the long sequence length of a \gls{waveform}, so is almost always done after training the two components separately.

For \glspl{acoustic_model} that generate a \gls{mel_spectrogram}, \gls{finetuning} is common. It is less common when generating \gls{NAC} codes, although the "TorToise trick" described earlier is one example of this. In practice, the benefits of \gls{finetuning} the waveform generator vary considerably from one \gls{acoustic_model} to another. It is best viewed as something to be done only as a last resort, not least because it makes the waveform generator model specific not just to a class of \glspl{acoustic_model} but to a specific \gls{checkpoint}, which is inconvenient.

\mysummarybox
{of current approaches to waveform generation}
{
\begin{itemize}
\item \Glspl{neural_vocoder} can be characterised by their design choices.
\item The principal classes of waveform generator currently in use are:
 \begin{itemize}
 \item \gls{mel_spectrogram}-input \glspl{neural_vocoder};
 \item \glsreset{NAC}\glspl{NAC} designed for general audio compression;
 \item \gls{NAC} variants with design choices to make them better suited to speech generation.
 \end{itemize}
\item Every waveform generator leaves tell-tale artefacts in the \gls{waveform}.
\item These artefacts are what current deepfake detectors (over-)rely on, coming up in Section~\ref{section:sota:deepfake_detection}.
\end{itemize}
}

\subsubsection{Evaluation}
\label{section:SotA:evaluation}

\paragraph{Subjective evaluation}

After a long period of little change in the methods used, subjective evaluation (Section~\ref{section:history:challenges:evaluation}) is finally receiving renewed interest for \gls{TTS}. (Subjective evaluation for \gls{VC} lags behind, so there is little to say about that here.)

There are two fundamental limitations of the \gls{MOS} paradigm: the stimuli are presented in isolation without any context; the score of a system under test is affected by the other \gls{TTS} systems included in that test (Range-Equalizing Bias \citep{cooperInvestigatingRangeEqualizingBias2023}). Contextualised testing has long been called for \citep{wagnerSpeechSynthesisEvaluation2019}, but is still uncommon: the vast majority of subjective evaluations reported in the literature involve playing isolated utterances to listeners, with no context provided and no clear use case in mind. Although \gls{MOS} scores should be treated (and reported) with caution, the field as a whole continues to use this test and we can find many papers even today that quote \gls{MOS} scores as if they were absolute values. There is recent solid empirical evidence that  the \gls{MOS} paradigm really does have severe limitations for \gls{TTS} evaluation \citep{LeMaguerLimitsMOS2024}, something that was already recognised in the field of audio quality assessment where \gls{MOS} originated \citep{zielinski2008on}.

The current \gls{SotA} in subjective evaluation is a period of change. \gls{MOS} continues to be used by some researchers, whilst others are moving forward with a number of alternative evaluation paradigms. 

In terms of replacing \gls{MOS}, the obvious candidate is \gls{MUSHRA} (borrowed from the field of lossy \gls{audio_codec} testing). This is an explicit relative judgement / ranking task, first applied to synthetic speech evaluation by \citet{henterGaussianProcessDynamical2012}. \citet{Blizzard2023Journal} is a serious attempt to move the field past the \gls{MOS} paradigm by promoting the \gls{MUSHRA}-like paradigm, and further innovating with multi-stage testing to make ever-finer distinctions between the best systems.

Ultimately, there is no single replacement for \gls{MOS}, but rather it should be recognised that a suite of evaluation methods will be needed, reflecting the multidimensional nature of AI speech generation, of human perception of spoken language, and the varied requirements of different use cases (Section~\ref{section:current_approaches:requirements}).

Another recent development is the rediscovery of linguistically well-motivated  test materials, including texts that contain specific phenomena to more directly test one well-defined aspect of a \gls{TTS} system, such as homograph disambiguation by the front-end \citep[e.g.,][]{Blizzard2023Journal}, now being picked up by other researchers who are finding that \gls{MOS} is no longer able to provide useful information, e.g., Sesame\footnote{\href{https://www.sesame.com/research/crossing_the_uncanny_valley_of_voice}{www.sesame.com/research/crossing\_the\_uncanny\_valley\_of\_voice}}\label{page:sesame}. \citet[Section 4.1]{ELLA-V} introduce "100 hard sentences" which "included challenging phonetic patterns, alliteration, and unusual (abnormal) combinations of words". Unfortunately, inspection of these materials\footnote{\href{https://ereboas.github.io/ELLAV}{ereboas.github.io/ELLAV}} reveals that they are artificial and not ecologically-valid (i.e., unlikely to occur in any real use case). In contrast, \citet[Section~3.3]{BASE_TTS} hand-created a set of test sentences containing "compound nouns, emotions, foreign words, paralinguistics, punctuations, questions, and syntactic complexities" to test the "emergent abilities" of systems that "demonstrate natural prosody on textually complex sentences". These sentences are much more realistic than those of \citet{ELLA-V}, and their division into linguistic categories indicates which aspects of a \gls{TTS} system they would test. These attempts to pay more careful attention to test materials (even if currently limited to the text to be generated) are early signs of change that will spread across the field.

\paragraph{Objective evaluation}

For \gls{TTS} or \gls{VC} \textit{evaluation}, the purpose of an objective measure is to replace expensive subjective testing using human subjects and listening tests. It is distinct from other applications of objective methods, such as \gls{deepfake} \textit{detection}. Some objective evaluation measures are already useful (but still far from perfect), and were mentioned in Section~\ref{section:history:challenges:evaluation}. 

For an objective evaluation measure to mimic human judgements, we require data from subjective evaluations. As in many areas of machine learning, generalisation is the hardest challenge \citep{cooperGeneralizationAbilityMOS2022}. Because most current approaches involve training a model on specific synthetic speech samples that have human ratings, they are invariably less reliable for types of synthetic speech which were unseen in training (the same is true for \gls{deepfake} detectors, of course). This means that there is currently no prospect of fully replacing subjective evaluations: as the \gls{SotA} of AI speech generation advances, we will  continue to need at least \textit{some} subjective evaluation data from human listeners for objective measures to keep pace.

Currently, only objective measures of \textit{naturalness} attempt to mimic human judgements. All other recent objective measures (for speaker similarity or intelligibility) simply borrow a model from another domain, as already explained in Section~\ref{section:history:challenges:evaluation}, and so will continue to advance as and when those other domains provide better models. For example, Wespeaker \citep{Wespeaker} may supersede current speaker embeddings, and be used to measure speaker similarity. \label{section:SotA:evaluation:speaker}

Naturalness is the primary dimension of interest, and objective measures use models trained on pre-existing datasets of human \gls{MOS} ratings, including those collected by Blizzard Challenges and Voice Conversion Challenges (Section~\ref{section:challenges}). This was the case for MOSNet \citep{MOSNet} (trained only on human-rated \gls{VC} speech, not \gls{TTS}) and the popular UTMOS \citep{UTMOS}, mentioned in Section~\ref{section:history:challenges:evaluation}. More recent developments include: LDNet \citep{LDNet} which takes account of listener-dependent score variation; RAMP+ \citep{RAMP+} which employs a retrieval step to find similar (in \gls{SSL} feature space) speech samples that already have human ratings; SpeechLMScore \citep{SpeechLMScore} which uses \gls{SLM} perplexity as a proxy for naturalness -- uniquely, this does not require human-rated data for training, and so does not attempt to estimate \gls{MOS}, but can provide a useful system ranking.

\mysummarybox
{of current approaches to evaluation}
{
\begin{itemize}
\item Objective measures of \textbf{naturalness} make  predictions that are usable, with caution.
\item But subjective testing (i.e., listening tests) remains the Gold Standard for evaluating naturalness.
\item The limitations of the \gls{MOS} paradigm are clear.
\item If a single replacement is desired, use \gls{MUSHRA}.
\item However, a suite of evaluation methods would be better.
\item Objective measures are the most common choice for \textbf{speaker similarity} and \textbf{intelligibility}.
\item But there is some doubt whether they predict human judgements. 
\end{itemize}
}


\subsection{Current approaches to speech generation: requirements, capabilities, and limitations}
\label{section:current_approaches:requirements}
\glsresetall

\myintrobox
{{in this section}}
{
\begin{itemize}
\item Illustrative use cases of AI speech generation.
\item The focus is on their \textbf{requirements}, \textbf{capabilities}, and \textbf{limitations}.
\item Each use case is exemplified by a small number of carefully-selected systems from the literature.
\end{itemize}
}

\label{section:SotA:start_of_use_cases}

\subsubsection{Use case: Text-to-Speech controlled via natural language}
\label{section:SotA:TTS:natural_language_control}

Parler\footnote{\href{https://github.com/huggingface/parler-tts}{github.com/huggingface/parler-tts}} \citep{LythNaturalLanguageGuidance} is chosen for this first case study because it illustrates a number of important trends. First, the \gls{acoustic_model} is a \gls{SLM}, casting the \gls{TTS} task as \textbf{autoregressive language modelling}, generating discrete \gls{NAC} codes. Second, Parler is trained on a very large corpus of speech. This illustrates the most important current trend: curation of large amounts of found data using an automated \textbf{data pipeline} (cf. Emilia on page~\pageref{page:emilia}). Third, Parler's \gls{SLM} is a generative model, and so inference involves sampling: it is non-deterministic and can generate \textbf{multiple renditions} for a fixed input. 

Instead of using a speech \gls{prompt} to specify the desired speaker identity and style, Parler accepts a natural language description such as "\textit{A female speaker delivers a slightly expressive and animated speech with a moderate speed and pitch. The recording is of very high quality, with the speaker's voice sounding clear and very close up.}" although it can easily be adapted to accept direct categorical labels instead of indirect natural language prompts, if the use case requires that.

So, the training data must also be annotated with these natural language descriptions (or labels). Parler's data pipeline uses a variety of methods to do that, from simple signal-based measures for pitch range, to pre-trained models for gender, accent, or recording quality. InstructTTS \citep{InstructTTS} used a much smaller proprietary database with expensive human annotations. PromptTTS++ \citep{PromptTTS++} uses a FastSpeech~2-like encoder-decoder architecture, although with a generative diffusion-based decoder, but was only trained on a small audiobook corpus.

Inference-time sampling is analogous to drawing samples from the training data distribution, so there will always be some variation (even if not desired). This is both a strength and a weakness, likely to apply to all generative models. For example, even after \gls{finetuning} a multi-speaker model with single-speaker data, Parler still generates unwanted variation in speaker identity. Inference hyperparameters need to be chosen to control the diversity of samples. Setting these incorrectly can result in poor performance, including \gls{hallucination}. The training data are never perfect, and inference sampling will reflect this. For example, the model will occasionally deviate from the specified input text (presumably because this happens in the training data). The standard (but rarely reported!) solution is to pass all inference samples through an \gls{ASR} filter.

\myRCLbox
{Text-to-Speech controlled via natural language}
{
\begin{itemize}
\item Requirements: a very large data dataset (e.g., \SI{50}{\kilo\hour}), curated using an automated pipeline. Relatively expensive to train from scratch (e.g., 8 GPUs for 1 week).
\item Capabilities: sampling during inference creates multiple high-quality renditions with natural variation in, e.g., prosody; model is able to create some combinations (e.g., speaker + recording quality) that did not co-occur in the training data; \gls{finetuning} on a few hours of speech is very fast and therefore cheap. 
\item Limitations: because the model is inherently generative, there is no guarantee of obtaining a rendition with exactly the desired properties. 
\end{itemize}
}

\subsubsection{Use case: Zero-shot Text-to-Speech}
\label{section:SotA:TTS:zero-shot}

As we already know, the task of \gls{TTS} is under-specified. This can make model training unstable because of seemingly-random variation in the target speech that is not explained by the input text. For a generative model, inference will therefore produce random variations, which may or may not be desirable. 

This can be mitigated by providing more than just text as input. One option is also to provide the target speech (via a \gls{bottleneck}, to prevent the model merely copying it to the output). During \gls{inference} the target speech is not available, so some other speech sample must be provided. It was found that the model tends to copy some properties of this sample (or "prompt"), such as speaker identity or prosody. These resulted in\footnote{The author thinks: by happy accident, not by design.} "\gls{zero_shot} \gls{TTS}" and "prosody transfer", respectively.

At the time of writing, \gls{zero_shot} \gls{TTS} is the most popular use case found in the literature, generally with a single, very short (e.g., \SI{3}{\\second}) speech prompt. Note that any commercial use case would require the permission of the target speaker, in which case it would generally be possible to obtain a much larger speech sample (enough to perform \gls{finetuning}).

Our case study is VALL-E \citep{VALL-E-journal}. Like Parler, this is a \gls{SLM} generating discrete \gls{NAC} codes, trained on \SI{50}{\kilo\hour} of speech. It takes a short \SI{3}{\second} speech prompt to specify the desired speaker identity. The model also "preserve[s] the speaker's emotion and acoustic environment from the prompt" \citep[page~1]{VALL-E-journal} although whether the latter is desirable is debatable.

The demonstration samples\footnote{\href{https://aka.ms/valle}{aka.ms/valle}} reveal that the claimed in-context learning abilities (i.e., that the model can learn new \textit{tasks} at inference time) are doubtful: the model cannot reproduce Scottish accents, for example. It seems likely that the model is limited to sampling from its training data distribution after all.

The subsequent model VALL-E~2 represents a less positive trend in the field. It claims "achieving human parity for the first time" \citep[page~1]{VALL-E2} but this is based on \gls{MOS} reported by other authors across many different listening tests, which is not valid (as explained in Section~\ref{section:history:challenges:evaluation}). 

\myRCLbox
{Zero-shot Text-to-Speech}
{
\begin{itemize}
\item Requirements: a very large dataset (e.g., \SI{50}{\kilo\hour}), obtained from audiobooks. Computationally expensive to train from scratch.
\item Capabilities: able to preserve some (but not all) properties of the speech \gls{prompt}.
\item Limitations: a very short speech \gls{prompt} is insufficient for the claimed in-context learning, so the model is likely to only be sampling from the training data distribution; no control over what is, and what is not, preserved from the prompt.
\end{itemize}
}

\subsubsection{Use case: Speech editing and multi-tasking}
\label{section:SotA:speech_editing}
\label{section:SotA:TTS:controllable}

VoiceCraft \citep{VoiceCraft} is a clear example of the flexibility of the \gls{SLM} approach to speech generation. The insight is very simple: the \gls{SLM} can be trained to do any language modelling task, simply by arranging the training data in a way that expresses this task. This generally involves arranging the input and output sequences accordingly, noting that these can be interleaved sequences of text and audio tokens (or anything else that can be represented as tokens, such as task instructions). Here, the tasks are speech infilling and continuation. The former can be used to edit speech by masking out the portion to be edited and instructing the model to generate that portion from text. The latter can be used for \gls{zero_shot} \gls{TTS}, by continuing the speech \gls{prompt}.

Spark-TTS \citep{Spark-TTS} also takes advantage of the flexibility of the \gls{SLM} approach. Starting from a pre-trained text-only \gls{LLM}, the model is post-trained to learn several new tasks. The task to be performed is specified to the model using an input token, and includes control over speaker attributes and prosody. This demonstrates the real power of the \gls{SLM} approach: the input is a sequence of tokens, not just representing inputs or controls, but now also instructions specifying the \textit{task} that the model should perform. One can imagine straightforward extensions of a model like Spark-TTS to include \gls{VC}, \gls{TTS}, and speech editing in a single \gls{checkpoint}. Such a model could create speech \glspl{deepfake} from an existing \gls{bona_fide} speech sample, using \gls{TTS} to change a few words, \gls{VC} to modify the prosody, and speech editing to splice this into the remaining, unchanged speech.

\myRCLbox
{{speech editing and multi-tasking}}
{
\begin{itemize}
\item Requirements: a generic \gls{SLM} backbone model; carefully curated data that represents the task to be performed.
\item Capabilities: any task that can be cast as language modelling, for which data can be found.
\item Limitations: like all generative models, the system is non-deterministic and will occasionally generate bad output.
\end{itemize}
}

\subsubsection{Use case: Offline Voice Conversion}
\label{section:SotA:VC:offline}

As we saw in the timeline in Section~\ref{section:VC_timeline}, most \gls{VC} systems have converged on a single conceptual architecture in which speech from the \gls{source_speaker} is first encoded into a \gls{latent} representation that captures the \gls{linguistic_content}, then decoded back into speech with the identity of the \gls{target_speaker}. The encoder may use a supervised model borrowed from \gls{ASR}, or an unsupervised counterpart. Likewise, the decoder may be a complete \gls{TTS} model, or something simpler. 

The two systems in \citet{liuAnytoManyVoiceConversion2021} are good examples of the \gls{ASR}+\gls{TTS} approach. The first system is referred to as Seq2seqPR-DurIAN in the paper, in which the \gls{ASR} system is a speaker-independent \gls{phoneme} recogniser (PR), and the \gls{TTS} system is DurIAN \citep{DurIAN}. The \gls{target_speaker} identity is input to the \gls{TTS} model as either a one-hot vector for "any-to-many" \gls{VC} or as a speaker embedding for "any-to-any" \gls{VC}. The second system is referred to as BNE-Seq2seqMoL in the paper (but as BNE-PPG-VC in other papers). This time, rather than use a \gls{phone} transcription from the \gls{ASR} system, \gls{bottleneck} features are extracted, which are used in the same way as \gls{PPG} features: they represent the current \gls{phone}, without making a hard decision. The relatively simple \gls{TTS} system takes these as input, along with the \gls{target_speaker} identity.

\myRCLbox
{Offline Voice Conversion}
{
\begin{itemize}
\item Requirements: pre-existing, trained \gls{ASR} and \gls{TTS} models, or the data, compute power, and skill required to train them.
\item Capabilities: \gls{ASR}+\gls{TTS} architectures can in principle modify any aspect of the speech, including duration.
\item Limitations: the most data-hungry in training, and most computationally-expensive for both training and inference, approach to \gls{VC}; only suitable for well-resourced languages; not intended for low-latency applications (but refer to "Real-time Voice Conversion" below).
\end{itemize}
}

\subsubsection{Use case: Real-time Voice Conversion}
\label{section:SotA:VC:streaming}

One use case of \gls{VC} is for human-human communication. This requires the system to operate in \gls{streaming} mode, which in turn requires that the methods and models used are \gls{causal} and can operate in \gls{real_time} (or faster). 

A model that typifies the previous generation is that in
\citet{tobingLowlatencyRealtimeNonparallel2021}, which is a classical encoder-decoder system of a type which would normally require parallel training data, but can be trained on non-parallel data using the cycle method (page~\pageref{page:VC_timeline:cycle_methods}). By keeping the model simple, a very low latency of \SI{24}{\milli\second} is achieved on CPU.

But currently, the obvious and most common approach to real-time \gls{VC} is to make modifications to a high-quality offline system in order to meet these requirements. Models are simplified to reduce the computational cost during \gls{inference}, and architectures modified so that they are \gls{causal}: that is, they need no knowledge of future inputs (or, more realistically, a very limited look-ahead) to generate the current output. The DualVC sequence of models \citep{DualVC, DualVC-2, DualVC-3} exemplifies this approach; the latest DualVC~3 model starts with the \gls{SSL} model wav2vec2.0 \citep{wav2vec2.0} as a \textit{non}-causal content encoder, and uses this as the teacher to train a simpler, causal encoder. The output of this represents the \gls{linguistic_content} (in this paper, and many others, this is misleadingly called "semantic information"). A language model predicts the missing future  \gls{linguistic_content} so that the decoder has access to both past and future context, from which it predicts a \gls{mel_spectrogram}. The authors claim \SI{50}{\milli\second} latency running on CPU, which is low enough for human-human conversation. The StreamVoice sequence of models \citep{StreamVoice,StreamVoice+} takes a similar teacher-guidance approach, but uses a \gls{SLM} as the backbone model, modified to be causal.

StreamVC \citep{StreamVC} comes from a different direction and has an elegant design pattern. It starts with the SoundStream \gls{NAC} \citep{SoundStream}, which inherently supports real-time \gls{streaming} (even on edge devices such as smartphones) because it is \gls{causal} and computationally-efficient. The representation between encoder and decoder is repurposed: instead of aiming for audio compression using \gls{RVQ}, the encoder is trained to be the "recognition" component, under the supervision of a pre-existing, external, speaker-independent \gls{SSL} model (essentially a student-teacher arrangement). StreamVC represents the \gls{SotA} in real-time \gls{VC}, with a claimed latency of \qty{71}{\milli\second}.

\myRCLbox
{Real-time Voice Conversion}
{
\begin{itemize}
\item Requirements: causal models, optionally with a small amount of lookahead.
\item Capabilities: low-latency; streaming.
\item Limitations: limited lookahead, so prosody can only be copied from source to target with little or no change to duration and only simple transformation of \gls{F0}; the streaming requirement severely limits the complexity of models that can be used, but student-teacher arrangements can import some benefits from existing large non-causal models (especially for the encoder).
\end{itemize}
}

\subsubsection{Use case: Voice anonymisation for privacy}
\label{section:SotA:TTS:voice_anonymisation}

As noted in the entry for the VoicePrivacy Challenge on page~\pageref{section:history:challenges:voice_privacy}, information leakage can result in information about the \gls{source_speaker} being present in the \gls{VC}-generated speech. Privacy~PORCUPINE \citep{Privacy-PORCUPINE} is our case study here, since it offers a deep insight into one mechanism for this leakage. A standard way to remove speaker information is to quantise the \gls{linguistic_content} using Vector Quantisation: a codebook. The speech from each \gls{source_speaker} will utilise the codebook entries with a particular distribution. The consequence is that the pattern of codes (e.g., how frequently a particular code is used, or whether certain codes are never used) depends on the \gls{source_speaker}. This is a \textit{meta-feature}: see Section~\ref{section:deepfake_detection:meta_features}. Privacy~PORCUPINE attempts to construct a codebook for which this is not the case, reducing leakage and improving privacy.

\myRCLbox
{Voice anonymisation for privacy}
{
\begin{itemize}
\item Requirements: methods to quantify privacy, e.g., the  ZEBRA framework \citep{PrivacyZEBRA}.
\item Capabilities: the latest voice anonymisation systems can achieve good levels of privacy.
\item Limitations: better privacy currently always comes at the cost of reduced downstream task utility.
\end{itemize}
}


\subsection{Review and survey articles on speech generation}
\label{section:generation_reviews_and_surveys}

There are many review articles on speech generation, but some are better than others. This list is intended to help the reader select those worth reading (and avoid the others!), so is ordered roughly on that basis.

\begin{itemize}

\item \citet{sisman2020overviewVC} "An Overview of Voice Conversion and Its Challenges: From Statistical Modeling to Deep Learning" provides a comprehensive, authoritative, and peer-reviewed overview of \gls{VC} up to late 2020. This adds depth to the brief history in the current report. \citet{bargumReimaginingSpeechScoping2023} continues on and brings things closer to the present, whilst \citet{stylianou2009VCsurvey} could be read for historical completeness.

\citet{tanSurveyNeuralSpeech2021} "A Survey on Neural Speech Synthesis" and related conference tutorials \citep{tan2021tutorial, tan2021tutorial_ijcai2021, tan2021microsoft} are comprehensive up to the point in time just before \glspl{SLM}.

\item \citet{kimNeuralSpeechAudio2024} provides a deep and authoritative tutorial on \glspl{NAC} from an audio coding perspective.

\item \citet{williamsNewChallengesContent2022} "New Challenges for Content Privacy in Speech and Audio" is not a review as such, but a useful position paper.

\item \citet{cuiRecentAdvancesSpeech2024} "Recent Advances in Speech Language Models: A Survey" attempts to classify and catalogue \glspl{SLM}, with a main focus on speech classification tasks but some coverage of speech generation. 

\item \citet{aroraLandscapeSpokenLanguage2025} "On The Landscape of Spoken Language Models: A Comprehensive Survey" defines Spoken Language Models as foundation models that can later be specialised for a specific task such as \gls{ASR}, speech translation, etc.

\item \citet{maReviewHumanEmotion2024} "A Review of Human Emotion Synthesis Based on Generative Technology" uses a formal literature review methodology so is likely to have captured a fair picture of the \gls{SotA}.

\item \citet{torcoliObjectiveMeasuresPerceptual2021} "Objective Measures of Perceptual Audio Quality Reviewed: An Evaluation of Their Application Domain Dependence" is a helpful comparison of measures that are often included in data pipelines (Section~\ref{section:provenance_of_data}) and sometimes used as evaluation measures (Sections~\ref{section:history:challenges:evaluation}~and~\ref{section:SotA:evaluation}), although is also promoting the authors' own 2f-model.

\item SpeechTrident is an attempt to catalogue a large number of "representation learning, neural codec, and language models that contribute to speech/audio large language models" \citep{SpeechTrident}. It is useful only as a source of links to new and recent models, but is not a critical review.

\item\citet{xieControllableSpeechSynthesis2024} is essentially just a list of papers; the associated git repository adds tables to group systems by common components. Useful as a source of new and recent papers.

\end{itemize}


\subsection{Remaining problems, trends, and predictions for speech generation}
\label{section:generation:research_trends_and_predictions}
\glsresetall

\myintrobox
{next}
{
\begin{itemize}
\item A look forward to the near future of AI speech generation.
\item Structured into the main themes of: data, text processing, controllability, acoustic modelling, and waveform generation.
\end{itemize}
}


\subsubsection{Data for speech generation}
\label{section:provenance_of_data}

A catalogue of common datasets is out-of-scope, but the list of data used by Spark-TTS \citep{Spark-TTS} would be a good place to start, since this model was trained on almost everything publicly available in English and Mandarin! That is itself a serious problem, not unique to this model, for two linked reasons: 1) no-one has properly analysed exactly what is in those datasets (e.g., which speakers are present); 2) it leaves little or no data available for testing. For example, VCTK \citep{VCTK} is widely used for evaluating \gls{zero_shot} \gls{TTS}, but was included in the Spark-TTS training data, so cannot be used for evaluating that model. Likewise, it is common to showcase models like Spark-TTS generating "celebrity" voices\footnote{Ethics are out-of-scope in this report, but clearly this is a dubious practice.} without checking whether that person's speech was in the training data (which is highly likely).

\paragraph{Data acquisition and curation pipelines}

The scale of data required by some models, at least for training "from scratch" (i.e., starting from randomly-initialised model parameters, not from an existing \gls{checkpoint}), requires a fully-automatic approach to data acquisition and curation. Acquisition of large-scale data (say, \SI{10}{\kilo\hour} and above) invariably means scraping data from online sources and passing it through a pipeline of processes to transform it into a usable dataset for model training.

Simply acquiring data is not enough: it needs curation. Minimally, this will involve labelling which parts contain speech, and transcribing the words. Emilia \citep{Emilia} \label{page:emilia} is a typical pipeline, using web scraping for the initial data acquisition from YouTube videos, audiobooks, and so on, followed by curation using \gls{VAD}, speaker diarisation, and \gls{ASR} to segment and transcribe the speech, then a final filtering step for quality control. It is important to be aware that the data available online is ever-changing (not only growing, but also sometimes removed); any dataset created by such a pipeline will vary each time the procedure is run. A dataset comprising \SI{100}{\kilo\hour} of speech spanning 6 languages was created and released by Emilia's creators to mitigate this problem.

The final filtering step in Emilia and comparable proprietary pipelines used in industry, is vital. The typical approach is to automatically estimate various measures, and then apply a threshold to discard poor data (e.g., overlapping speakers, excessive background music or noise) and/or annotate each segment with labels for use in modelling (e.g., Parler \citep{LythNaturalLanguageGuidance}). Classical speech (and general audio) quality measures often require a clean reference (they are called "intrusive"), which of course is not available here. The solution is to use neural models that have been trained to \textit{estimate} these measures in a reference-free ("non-intrusive") fashion. Popular choices include: DNSMOS \citep{DNSMOS,DNSMOS_P.835}; TorchAudio-Squim (PESQ, STOI, SI-SDR) \citep{TorchAudio-Squim}; Brouhaha (SNR, C50 reverberation) \citep{Brouhaha}; UTMOS \citep{UTMOS}. New models are coming out all the time and we can expect to see these being applied to data curation: e.g., SESQA \citep{SESQA}, or NISQA \citep{NISQA} that estimates four speech quality dimensions: Noisiness, Coloration, Discontinuity, and Loudness.

These models have uses beyond data curation. Many of them are employed as objective measures for evaluating synthetic speech (Section~\ref{section:history:challenges:evaluation}) and some are even used immediately after inference from a generative model to filter out poor samples. If the measure is implemented as a differentiable model, it can be used as a \gls{loss_function} for model training, for example NOMAD \citep{NOMAD}.

Although the amount of data used to train speech generation models will never reach the scale of text models, we have already reached a point where comparable problems are emerging. The most immediate of these is that it is very hard to be sure exactly what is, and it not, in the training dataset of some large models. A good illustration is Spark-TTS \citep{Spark-TTS} mentioned above: it is unclear how to evaluate this model, given the lack of knowledge about what the model's training data contains.

Automatic data curation is imperfect. \gls{ASR} transcriptions are generally sufficiently accurate, and can now be near-verbatim and include fillers or non-speech events such as laughter; they can be checked against a manual transcription if necessary, as can automatic language identification. But other labels are harder to verify. Accurate automatic labelling of language variety/dialect/accent remains a challenge. Identifying whether the same speaker occurs in multiple parts of a vast data set is also difficult. Labelling expressivity and so on remains hard for reasons set out in Section~\ref{section:generation:research_trends_and_predictions:expressivity}. It is unlikely that speech datasets in \textit{current} use are polluted with synthetic speech but this is clearly a risk in the future (Section~\ref{section:risks:pollution}); text-based \gls{NLP} will encounter this problem long before speech, because the amount of \gls{LLM}-generated text being published online far exceeds the amount of synthetic speech. Fortunately, methods for detection of synthetic speech are already being developed, for use as \gls{deepfake} detectors: Section~\ref{section:sota:deepfake_detection}.

When used to train a generative model, errors in the data will lead to errors in generation. More generally, unlabelled variation in the data will lead to corresponding variation in generated output: see the case study in Section~\ref{section:SotA:TTS:natural_language_control}, for example. So a clear trend for the near future will be to automatically label very large datasets with a wider variety of properties (beyond the measures listed above for speech quality etc.), so that generative models can make use of that information. 

\paragraph{The use of multiple data sets for different phases of model training} Whilst a large amount of data might be required for training, it is not needed for post-training or finetuning an existing model (Section~\ref{section:generation:research_trends_and_predictions:acoustic_model}). Here, it is far more important to use high-quality data that are closely matched to the use case: the speech is similar to that which the model needs to generate, and/or is labelled with properties that need to be controlled during inference. A typical illustration of the former is Google's SoundStorm \citep{SoundStorm} which can generate impressive-sounding dialogues, because it is trained on \SI{100}{\kilo\hour} of precisely that kind of data. See also Section~\ref{section:generation:research_trends_and_predictions:acoustic_model}.

\subsubsection{Text processing and pronunciation}
\label{section:remaining:text}

Classical methods largely solved the \textit{basics} here, or at least provided mechanisms to do so, from normalising Non-Standard Words to specifying exceptions in a pronunciation dictionary. By "basics" we mean isolated sentences to be synthesised in a read-text speaking style. Some modern systems that do not explicitly normalise text, or do not predict pronunciation, continue to make unforced errors in these areas (Section~\ref{section:tts_errors}).

Recent advances in text-based \gls{NLP} have provided far more powerful tools for language understanding than were available when those conventional methods were devised, but there is very little work in integrating the two: for example, using modern representations of meaning to improve text normalisation. A resurgence of interest in improved evaluations, and especially those that target specific linguistic phenomena like homographs, might trigger new work in this area, if and when such evaluations reveal significant weaknesses.  A different way to take advantage of text-based models is to use one as the foundation for a speech generation model, as described in Section~\ref{section:SotA:acoustic_model}.

One aspect of pronunciation that has never received much attention is the (e.g., regional, social, or ethnic) accent of the speaker. This is understandable because hearing accent differences generally requires (near-)native expertise, accent is a niche subject within linguistics, and there are very few datasets labelled with accent. Nevertheless, datasets acquired using a typical data pipeline (Section~\ref{section:provenance_of_data}) will contain considerable variation in accent; annotating only the language is not sufficient. For example, labelling data as "Spanish" would fail to capture the /th/ vs. /s/ distinction between Castilian and Latin American varieties, not to mention the many other differences between Latin American varieties or regional accents of Spain.

\subsubsection{Controllable expressivity, emotion, and prosody}
\label{section:generation:research_trends_and_predictions:expressivity}

The disconnect between modern and classical approaches is particularly apparent in the area of expressive speech generation. Demonstrations of AI-generated spoken dialogues (e.g., SoundStorm \citep{SoundStorm}, Sesame on page~\pageref{page:sesame}, or Dia\footnote{\href{https://github.com/nari-labs/dia}{github.com/nari-labs/dia}}) are initially impressive, but are in essence just a generative model sampling from a large dataset of spoken dialogues. Very little control is possible, beyond the text, possibly speaker identities, and simple markup for fillers or laughter.

Control over expressivity is hard, but not because the models are limited or because the data do not exist. It is hard because it requires a specification of what should be controlled. At its most basic: if we do not know what labels to place on the data, we cannot expect to control a generative model. Linguistics has provided some theories and frameworks for \textit{describing} \gls{prosody} but falls short of anything resembling a framework for modelling the full range of expressivity possible in spoken language.

In the absence of a linguistic framework, an obvious response is to learn from data. It is important to make the distinction between \gls{prosody} and its \textit{acoustic correlates}, which include \gls{F0}, amplitude, and duration. Acoustic correlates can be easily estimated directly from the speech signal. Many models can learn from data how to control the acoustic correlates, but this is in itself of limited use. Other models sidestep the problem altogether and simply copy properties from a provided speech \gls{prompt}, as in so-called "prosody-transfer" and "zero-shot" \gls{TTS} (Section~\ref{section:SotA:TTS:zero-shot}).

So, learning prosody from data, without prosodic labels, implies using unsupervised methods. There are isolated pieces of work on this to be found in the literature. "Global Style Tokens" \citep{GST} is the most well-known example. \citet{hodariUsingGenerativeModelling2019} produced varied intonation by sampling from a Variational Autoencoder (VAE), and used unsupervised methods to discover "intonation codes" \citep{hodariPerceptionProsodicVariation2020}. \citet{shechtmanSupervisedUnsupervisedApproaches2021} learned low-dimensional parameters that provide control over word-level emphasis, in a Tacotron-like architecture. ProsodyFM \citep{ProsodyFM} claims to learn how to predict prosodic phrase breaks in an unsupervised fashion but in fact uses an external, pre-trained phrase break detector.  Overall, unsupervised discovery of prosody remains an under-explored area, and all existing methods suffer from the same problem, that any representation learned in an unsupervised manner is -- by its very nature -- hard to interpret and so not very useful as a control for \gls{TTS}.

\subsubsection{Acoustic modelling}
\label{section:generation:research_trends_and_predictions:acoustic_model}

The most obvious trend here is the increasing size of \glspl{acoustic_model}, especially of the \gls{SLM} variety. These are expensive to train from scratch, so we will see a paradigm emerging of pre-training a large model on a large dataset, followed by post-training and/or \gls{finetuning}. Public releases of such large models will greatly facilitate work in this area. Post-training is more extensive in scope than \gls{finetuning} and could involve the model learning a new task, such as post-training a \gls{TTS} model to perform speech editing. \Gls{finetuning} might involve, for example, creating a speaker-specific model from a multi-speaker one, or updating an existing model to drive a different \gls{NAC}.

Beyond the pre-training + post-training + \gls{finetuning} paradigm, we can also expect to see more extensive use of both speech and text \textit{foundation} models. These are general-purpose or multi-task models which can subsequently be specialised for, say speech generation. Llasa, described in Section~\ref{section:SotA:acoustic_model}, and Spark-TTS, described in Section~\ref{section:SotA:TTS:controllable}, are two examples of this.

A key unsolved problem in this new era of \gls{generative} models is how to deal with their intrinsically non-deterministic \gls{inference} behaviour: they can generate multiple renditions for any given, fixed input (see, for example, Section~\ref{section:SotA:TTS:natural_language_control}). The current way of dealing with this is to pass multiple renditions through external filters, such as \gls{ASR} to confirm that the text was spoken correctly, or \gls{ASV} to confirm that the desired speaker identity was achieved. Going a step further, one might "cherry pick" the best rendition \citep{adigweCanWeCherryPick2025}. Some of the \textit{unwanted} variation in model output is doubtless due to limitations in the training data (Section~\ref{section:provenance_of_data}).

Whilst we should expect to see an almost continuous flow of new \glspl{acoustic_model} being proposed, this may not be the best place to look for important advances. Rather, the availability of generic models (notably, but not exclusively, Transformer-based \glspl{SLM}) actually liberates researchers to return to core problems including data (Section~\ref{section:provenance_of_data}), text processing (Section~\ref{section:remaining:text}), controllable expressivity  (Section~\ref{section:generation:research_trends_and_predictions:expressivity}), and so on.

\subsubsection{Waveform generation}
\label{section:generation:research_trends_and_predictions:waveform_generation}

There are already some excellent solutions to this component of speech generation, falling into two broad classes: \glspl{neural_vocoder} that take a \gls{mel_spectrogram} as input, now decreasing in popularity; \glsreset{NAC}\glspl{NAC} that define their own \gls{acoustic_specification}, generally using discrete codes, and currently the dominant approach in the latest models.

What we are most likely to see next are gradual, incremental advances in waveform generators, making them (slightly) better suited for speech generation. We already saw the first steps in this direction in Section~\ref{section:SotA:waveform_generation} with developments including: dropping the residual nature of mainstream \gls{NAC} codebooks (which imposes an inconvenient requirement to generate them in a specific order); reducing the frame rate in order to reduce the sequence length, making modelling with a \gls{SLM} easier.

There is a link between \gls{SSL} representations from models like wav2vec2.0 \citep{wav2vec2.0} and the representations learned by a \gls{NAC}. Currently, \gls{SSL} representations are better for capturing \gls{linguistic_content} and are not optimised for waveform reconstruction. We may see a convergence of the two, an early hint of which can be found in StreamVC, described in Section~\ref{section:SotA:VC:streaming}.

As already observed from ASVspoof and similar challenges listed in Section~\ref{section:challenges}, and as will become clearer in Section~\ref{section:sota:deepfake_detection}, the \gls{SotA} in deepfake detection is essentially to detect low-level waveform artefacts. It is likely, therefore, that specialised waveform generators (and/or processors to be applied to generated waveforms) will be devised for evading deepfake detectors. These are, at least initially, likely to take an adversarial approach: it will be easier to \textit{mislead} the detector than to generate a waveform so natural that it is undetectable. See Section~\ref{section:current_approaches_deepfake:requirements:attacks}.

\mysummarybox{{of remaining problems, trends, and predictions for speech generation}}{
\begin{itemize}
\item data acquisition and curation pipelines can produce large-scale data, but:
\begin{itemize}
\item there has been very little work on understanding what is (and is not) in such datasets, and what biases they contain;
\item any model trained on such a dataset might be hard to evaluate.
\end{itemize}
\item There is still a gap between classical methods for text processing and pronunciation vs. what current models do.
\item There is a trend towards models that are pre-trained (e.g., on data from a pipeline), post-trained (to learn a new task), then finetuned (on high-quality data for the final use case).
\item Large, generic acoustic models with publicly-available pre-trained checkpoints will greatly facilitate research and dramatically lower the entry barrier for less well-resourced researchers. (This is already the case for neural vocoders.)
\item The generative nature of current models remains a two-edged sword, and more work is needed to fully understand where these models' capabilities and limitations come from (e.g., is it only about the data?).
\item The main barrier to creating controllable speech generation models is not technical, but a fundamental theoretical question which is much harder to answer.
\item Waveform generation already has excellent solutions, and these will continue to incrementally improve over time.
\end{itemize}
}

\newpage
\section{The current state-of-the-art in speech deepfake detection}
\glsresetall
\begingroup
\parindent=0em
\localtableofcontents
\endgroup
\label{section:SotA:detection}


\label{section:sota:deepfake_detection}

\myintrobox
{in this section}
{
\begin{itemize}
\item Speech deepfake detection builds on earlier work in \gls{anti_spoofing}.
\item \Gls{anti_spoofing} was narrowly-defined as protecting an \glsreset{ASV}\gls{ASV} system against attack, typically by introducing a \gls{countermeasure}.
\item \Gls{deepfake} detection has a broader and more fluid definition, reflecting:
\begin{itemize}
\item an expanding range of methods for creating those deepfakes, including generation and manipulation;
\item that a \gls{deepfake} may be targeted at an automatic system and/or human listeners.
\end{itemize}
\item This is a smaller and less mature field than AI speech generation, so this section is shorter than Section~\ref{section:SotA:generation}.
\end{itemize}
}

The brief history in Section~\ref{section:history:spoofing} and the timeline of challenges, notably the ASVspoof series, in Section~\ref{section:history:challenges} have already provided the context for this section. Some important lessons were learned from \gls{anti_spoofing}, which are worth summarising here:
\begin{itemize}
\item novel \textbf{features} can prove highly-effective, notably \gls{CQCC} devised shortly after ASVspoof1;
\item generalisation to \textbf{unseen methods} of speech generation is the hardest problem, and this includes generation methods that do not even produce high-quality speech;
\item \textbf{flaws} in dataset construction can and \textit{will} be exploited by machine learning, resulting to overly-optimistic evaluation results.
\end{itemize}
From this point onwards, we will focus only on the very recently-emerging \gls{SotA} in speech \gls{deepfake} detection. Note that this section has a simpler structure than Section~\ref{section:SotA:generation}.

Because of its origins in \gls{anti_spoofing}, the terminology used in the field of speech \gls{deepfake} detection is a little fluid. To protect an \gls{ASV} system against \gls{spoofing}, a \gls{countermeasure} is employed. This determines whether a particular speech sample presented at the input to the system is \gls{bona_fide} (spoken by a real person) or \gls{spoofed} (synthetic). For speech \gls{deepfake} detection, the same determination is made, but we might find the terms "real"/"genuine" and "fake" used in place of "\gls{bona_fide}" and "\gls{spoofed}".


\subsection{Current approaches to deepfake detection} 
\label{section:current_approaches_deepfake}

\myintrobox
{in this section}
{
\begin{itemize}
\item features and classifiers, which are the building blocks of all current deepfake detectors
\item related topics: tracing the method by which a deepfake was created, attacking deepfake detectors, explainability, and avoiding the need for deepfake detection.
\end{itemize}
}

This is a classification task, so the obvious architecture for any system is a feature extractor and a classifier: all current systems follow that design pattern. There are, of course, systems that use the \gls{waveform} directly as their input, but these either have an implicit feature-extraction sub-component within their architecture, or use an existing pre-trained \gls{SSL} model as a starting point. To squeeze out extra performance (i.e., higher classification accuracy), it is possible to combine multiple features and perhaps also multiple classifiers. This \textit{ensemble} approach is currently the most popular one, for example in the most recent \gls{deepfake} detection challenge, SVDD (page~\pageref{page:SVDD}).

\subsubsection{Classical features}

These are features that are extracted from the speech signal using expert-designed, hand-crafted methods. The most important features in the history of \gls{anti_spoofing} are \glsreset{CQCC}\gls{CQCC}, devised a decade ago and still useful today. They are a direct descendent of the \glsreset{MFCC}\gls{MFCC} features that dominated \gls{ASR} for several decades. The success of \gls{CQCC} has inspired the rediscovery of many classical features, for use in ensemble methods: Section~\ref{section:current_approaches_deepfake:ensembles}.

The reason that classical features remain useful is probably \textit{because} they are hand-crafted. They are not over-fitted to any specific property of \gls{deepfake} speech, but were originally devised for \gls{bona_fide} speech. This is in contrast to some more recent methods, coming up below, where the features themselves are made more specific to \gls{deepfake} speech (e.g., by \gls{finetuning} a SSL model on such speech); this may increase accuracy for seen types of synthetic speech, but harm generalisation to unseen types.

This goes to the heart of the problem in \gls{deepfake} detection, as it is currently approached: using examples of known fake speech, whether for selecting features, \gls{finetuning} feature extractors, or training classifiers, makes the detector too specific to those speech generation methods seen in the training set. One route to escape this would be to rely as little as possible on examples of fake speech, and instead to aim for accurate classification of \gls{bona_fide} speech. So-called "one-class" methods are one version of this: Section~\ref{section:current_approaches_deepfake:one_class}. Spoofprint \citep{Spoofprint} borrows ideas from \gls{ASV}. It uses only \gls{bona_fide} speech samples to train an embedding model, then uses \gls{spoofed} samples to learn the relatively small number of parameters of a simple transform to maximise discrimination between \gls{bona_fide} and \gls{spoofed} classes.

\subsubsection{Deepfake-specific features}

\citet{shiotaVoiceLivenessDetection2015} detected the "phenomenon of pop noise...when human breath reaches a microphone" in a liveness detector, but any detector relying too much on that acoustic cue could be attacked by adding pop noises to \gls{deepfake} speech. \citet{dengDetectionSyntheticSpeech2022}, in their entry to ADD2022, used rules to locate spectrogram "defects" including time regions with no energy (digital silence) or frequency bands with no energy (evidence of upsampling). Such specific features may work in narrow use cases, but would be easily defeated using some of the simpler attacks described in Section~\ref{section:current_approaches_deepfake:requirements:attacks}. They may, however, have value for explainability and attribution if they are able to pinpoint the properties of \gls{deepfake} speech signals that detectors are using.

\subsubsection{Direct waveform input}
\label{section:current_approaches_deepfake:waveform}

The AASIST \citep{AASIST} architecture comprises a feature extraction component based on RawNet2 from earlier work \citep{takEndtoEndAntispoofingRawNet22021} and a classifier network which is inspired by ensemble approaches. Conceptually at least, it falls into the common design pattern of deepfake detectors. At the time of its publication, AASIST was \gls{SotA}. More recently, it has appeared as a building block of other systems, such as the algorithm recognition system in Section~\ref{section:current_approaches_deepfake:requirements:xieGeneralizedSourceTracing2024a}.

\subsubsection{Features derived from Self-Supervised Learning (SSL)}
\label{section:current_approaches_deepfake:SSL_features}

Many pre-trained \glsreset{SSL}\gls{SSL} models of speech are available, and all are trained only on \gls{bona_fide} speech (assuming that current speech databases are not yet polluted with synthetic speech: Section~\ref{section:provenance_of_data}). They are strong models of speech that can provide features (\gls{latent} representations from within the model) that are useful for many downstream tasks. The entry by \citet{liuDeepSpectrotemporalArtifacts2022} into the partially-fake detection task of the ADD challenge (page~\pageref{page:ADD_challenge}) combines the latents from all layers of an \gls{SSL} model as input to a classifier, and performs \gls{finetuning} on the \gls{SSL} model jointly with the classifier.

An alternative to \gls{finetuning} all the parameters of an \gls{SSL} model, which can be prone to overfitting because this model has very high capacity, is to use an adaptation method that has far fewer free parameters \citep[e.g.,][]{wuAdapterLearningPretrained2024}.

\subsubsection{Meta-features}
\label{section:deepfake_detection:meta_features}

By "meta-feature", we mean the \textit{performance} (e.g., accuracy, output distribution, pattern of errors, etc.) of a model or algorithm, which indicates something about the data. For example, if an \gls{ASR} system has low accuracy for a particular speech signal, that tells us something about that signal: it might be noisy, reverberant, from a speaker type not well-represented in the training data, etc.

\citet{xueAudioDeepfakeDetection2022} \label{section:current_approaches_deepfake:technical:xueAudioDeepfakeDetection2022} uses \gls{F0} as a feature, based on observed differences in this feature's \textit{distribution} in \gls{bona_fide} vs. \gls{spoofed} speech.  It is worth inspecting Figure~1 in \citet{xueAudioDeepfakeDetection2022} where, although apparently unnoticed by the authors, there are clear signs of pitch tracking errors: the concentrations of very low \gls{F0} values around \qtyrange{25}{50}{\hertz}. The error pattern differs between the two types of speech. It is plausible that synthetic (i.e., \gls{spoofed}) speech is easier to pitch track than natural speech: the latter is likely to contain a higher proportion of irregular voicing, vocal fry, etc. So \gls{F0} \textit{as estimated by a pitch tracker} is a meta-feature.

There has not yet been enough exploration of meta-features for deepfake detection. There is a vast array of pre-trained models and classifiers for which there might be systematic differences in performance between \gls{bona_fide} and fake speech. Some illustrative ideas might be:

\begin{itemize}
\item codebook utilisation statistics, for a \gls{NAC}, or indeed any other encoder: perhaps fake speech occupies a different distribution of codes to \gls{bona_fide} speech;
\item statistics of code sequences for \gls{SSL} models such as HuBERT \citep{HuBERT}: perhaps fake speech is more or less predictable than \gls{bona_fide} speech;
\item objective measures (Section~\ref{section:SotA:evaluation}): perhaps fake speech obtains more or less consistent scores than \gls{bona_fide} speech;
\item speaker embeddings from an \gls{ASV} model: these might exhibit drift over time (or inconsistency utterance-to-utterance) for speech from a generative model, but not for \gls{bona_fide} speech;
\item audio codec reconstruction fidelity: perhaps repeatedly encoding-then-decoding fake speech leads to a different type of degradation than for natural speech, due to "multiplying" coding artefacts.
\end{itemize}
Of course, as and when each of these proves effective for \gls{deepfake} detection, then an obvious line of attack is opened up in which precisely these properties are manipulated in fake speech: see also Section~\ref{section:current_approaches_deepfake:requirements:attacks}.

Another variant on the theme of meta-features would be the agreement or disagreement between models. For example, a suite of objective speech quality metrics (Section~\ref{section:history:challenges:evaluation}) might all agree that \gls{bona_fide} speech is high quality, but may exhibit disagreement about the quality of a synthetic speech sample. Or, in an audiovisual setting, using the video should improve accuracy compared to audio-only \glsreset{ASR}\gls{ASR}; if it does not, then we infer there is an inconsistency between audio and video, indicating a \gls{deepfake}. There are likely to be many other ideas to be tried here.

\subsubsection{Ensembles of classifiers}
\label{section:current_approaches_deepfake:ensembles}
 
Many systems that perform well on \gls{deepfake} detection use ensemble approaches. The use of ensembles of features and/or classifiers started at least as early as the ASVspoof2 challenge with the system from \citet{ASVspoof2017SystemS02} which used 3 classical features (\gls{MFCC}, Perceptual Linear Prediction, \gls{CQCC}), derived further features from \gls{CQCC} using statistical transforms (normalisation, Principal Component Analysis), passed them to a collection of standard classifiers (\gls{GMM}, \gls{SVM}, etc), then combined their scores. The use of ensembles is well-motivated by the diversity of methods used to create the \gls{deepfake} speech used in typical evaluations.

\subsubsection{One-class classifiers}
\label{section:current_approaches_deepfake:one_class}

A rather different way to think about \gls{deepfake} detection is to frame the problem as \textit{anomaly detection}, with respect to natural \gls{bona_fide} speech. \citet{alegreOneclassClassificationApproach2013} is an early example, in which a one-class \gls{SVM} is trained only on \gls{bona_fide} speech, with expert intuition employed to devise the features. Although no examples of \gls{spoofed} speech are used to train the model, they are still required to tune it. This idea was rediscovered recently \citep{zhangOneClassLearningSynthetic2021} and the work by \citet{yeOneclassNetworkLeveraging2025} represents the current \gls{SotA} on the ASVspoof tasks from 2019 and 2021 (in the timeline on page~\pageref{page:ASVspoof2019}). Their system \textit{does} use \gls{spoofed} speech during training, and the "one-class" aspect is that the training objective aims to draw a tight decision boundary around \gls{bona_fide} samples, whilst pushing \gls{spoofed} samples away from this boundary but \textit{without} attempting to place a tight decision boundary around them. Interestingly, the system uses classical \gls{mel_spectrogram} features.

\mysummarybox{{of features and classifiers for deepfake detection}}{
\begin{itemize}
\item Classical features, extracted without machine learning, remain useful.
\item Using machine learning to extract features, whether by \glsreset{SSL}\gls{SSL} or in a \gls{waveform}-input model like AASIST, is also useful; these features are not guaranteed to outperform classical features so can be used alongside them in an ensemble.
\item Ensembles of multiple features and/or multiple classifiers have performed best in recent challenges.
\item Once the improvements from feature and classifier ensembles have been exhausted, the field may turn towards novel ways of optimising systems, with so-called "one-class" classifiers being the best current example of this way of thinking.
\end{itemize}
}

\label{section:current_approaches_deepfake:requirements}

\subsubsection{Testing the weaknesses of existing deepfake detectors}
\label{section:current_approaches_deepfake:requirements:weaknesses}

The majority of work in the field is of course directly on deepfake detection. But sometimes more can be learned by testing the weaknesses of existing detectors. This was the case in some of the ASVspoof challenges, where digital silence or telephony artefacts could be exploited by \glspl{countermeasure} to obtain very high accuracy, but which will of course not generalise well \citep{ASVspoof2017Analysis,ASVspoof2017ArtefactsJournal}. \citet{liComparativeStudyPhysical2022} also found flaws in the ADD challenge dataset: for example, very simple features such as intensity differ considerably between \gls{bona_fide} and fake speech \citep[Figure 1(b)]{liComparativeStudyPhysical2022}. Detectors naively trained on such data are vulnerable to some very straightforward attacks:   Section~\ref{section:current_approaches_deepfake:requirements:attacks}. \citet{mullerHumanPerceptionAudio2022} found that humans and machines not only have comparable accuracy in detecting \gls{deepfake} speech, but both struggle to identify certain types of attack.

\subsubsection{Source speaker tracing}
\label{section:current_approaches_deepfake:requirements:source_speaker_tracing}

\citet{SSTC2024_top_system} describe the best-performing system from the first Source Speaker Tracing Challenge (SSTC), denoted "Rachel\_W" in \citep[][Table~6]{SSTC2024Summary}. The \glsreset{SSV}\gls{SSV} task is to determine whether two \gls{VC} samples were created using the same source speaker. In contrast to work on \gls{deepfake} \textit{detection} where ensembles of features and classifiers are widely used, this system uses simple \gls{mel_spectrogram} features. A key feature of the system is the way it is trained progressively on a sequence of datasets: pre-training on a large set of natural source speech then a combined set of natural and \gls{VC} speech, followed by \gls{finetuning} to the \gls{SSV} task using a task-specific speaker-contrastive loss function.

\subsubsection{Deepfake algorithm recognition}

\citet{xieGeneralizedSourceTracing2024a}, \label{section:current_approaches_deepfake:requirements:xieGeneralizedSourceTracing2024a} a system entered into the ADD2023 Challenge (Section~\ref{section:challenges}), represents the \gls{SotA} in deepfake algorithm recognition\footnote{Iin the paper, the authors call it "source tracing", but this is \textit{not} source \textit{speaker} tracing.}. It uses the recently-rediscovered "one-class classifier" idea in which the training loss function is designed to push natural (\gls{bona_fide}) speech samples together into a tight distribution, and to disperse the fake (synthetic) samples widely in the same space. After \gls{bona_fide}/fake classification, a second classifier divides the fake samples into known classes (i.e., that were encountered during training) plus one class for novel (i.e., not encountered during training) fakes. The system is built from existing components used in various previous \gls{ASV} systems: wav2vec2.0 \citep{wav2vec2.0} and AASIST \citep{AASIST}. 

\citet{mullerAttackerAttributionAudio2022} compared embeddings derived from classical acoustic features (jitter, shimmer, etc) with learned embeddings, to create "attack signatures", finding that the learned embeddings led to much higher attack identification accuracy.

\subsubsection{Explainability and interpretability}

We define "explainability" as being able to understand the reasons for a system (e.g., performing source speaker tracing) reaching a particular decision, without necessarily "looking inside" the system. Interpretability is subtly different, and involves "looking inside" the system to understand what processes were used to reach that decision. Explainability is the most useful property for a user of a \gls{deepfake} detector. Interpretability is more useful to developers of such systems, trying to understand the strengths and weaknesses of their models.

The two previous tasks -- tracing the source speaker and identifying the algorithm -- are initial steps in the direction of an explainable \gls{deepfake} detector. For example, a detector might now output "This speech is classified as fake, because it came from \gls{VC} algorithm X and the source speaker was Y." These attribution tasks are currently framed as classification, with performance measured in terms of accuracy. The systems themselves do not need to be interpretable. Attribution is becoming a mainstream task; for example, \gls{TTS} system attribution was the topic of a recent IEEE student competition \citep{salviSyntheticSpeechAttribution2023}.

\citet{geExplainingDeepLearning2022} examined what evidence \glspl{countermeasure} use, and found that they do indeed rely on artefacts in the non-speech parts of the signal (as \citet{wangCanLargeScaleVocoded2024} also found) but also that some learn to focus on specific time regions or frequency bands, which are presumed to be signatures of particular \gls{spoofing} methods, providing some explainability.

\citet{moussaUnmaskingNeuralCodecs2024} used signal artefacts to determine if a speech \gls{waveform} has been generated by a \glsreset{NAC}\gls{NAC}, and if so which one. This has applications in speech \textit{forensics}, but is not directly useful as a \gls{deepfake} detector because \gls{bona_fide} speech might also legitimately be encoded-decoded by a \gls{NAC}, e.g., for internet-based voice calls.\label{page:moussaUnmaskingNeuralCodecs2024} 

\subsubsection{Attacks on deepfake detectors}
\label{section:current_approaches_deepfake:requirements:attacks}

By "attack" we mean creating a fake speech signal that will not be detected as such. One form of attack is conceptually very simple: to make modifications to already-generated \gls{deepfake} speech in order to hide the artefacts that the detector will rely on. We have already seen that detectors can over-rely on very specific cues, such as digital silence. \citet{huaImprovingSpoofingCapability2022} employed two trivial but highly-effective methods to attack such a detector: add noise to the entire signal; replace regions of digital silence with actual recorded silence (sometimes called "room tone"). In the ASVspoof 2019 and 2021 datasets, "Bonafide instances tend to have significantly longer leading and trailing silences than spoofed instances." \citep[Abstract]{mullerSpeechSilverSilence2021} which leaves detectors open to a silence-duration attack. So \citet{wangGeneralizableSpeechSpoofing2024} made their detector robust by trimming silence from the training data, resulting in a \gls{SotA} result for the ASVspoof3 (2019) dataset. 

There is an obvious future line of work in which selected cues taken from natural speech, including recording conditions (room tone, background noise, channel effects) and a range of speech production phenomena (e.g., lip smacks, in-breaths), are \textit{overlaid} on \gls{deepfake} speech (rather than produced by the AI speech generator). This would obviously then be followed by the development of detectors to identify signals constructed from elements that did not naturally co-occur.

Malafide \citep{Malafide} \label{page:malafide} employs convolutive noise (i.e., channel distortion) as a simple and elegant attack, using a learnable filter to remove the tell-tale signatures of a given speech \gls{deepfake} generator, so they are no longer available to a detector (which over-relies on those specific features). Malacopula \citep{Malacopula}, is another method for post-processing \gls{spoofed} speech, this time to make it have a more similar speaker embedding to \gls{bona_fide} speech, with the aim of increasing the false acceptance rate of an \gls{ASV} system.

\subsubsection{Alternatives to detecting deepfakes}

\Gls{deepfake} detection will never be perfect, so alternative, or complementary methods, are helpful. A "good actor" who uses AI speech for, e.g., video dubbing, could mark that speech as synthetic either overtly (by declaring this fact to the listener) or covertly (by injecting a watermark in the audio). Someone who does not wish their natural speech to be used for creating \glspl{deepfake} might  protect their published audio by rendering it unusable for that purpose.

\paragraph{Making AI speech detection easier}

WavMark \citep{WavMark} represents the latest method for applying a robust (i.e., hard to erase) watermark to speech \glspl{waveform}. It uses a neural model that is explicitly trained to be robust to known methods for watermark removal, including lossy compression, adding noise/reverberation, filtering, etc. Watermarking is a mature technology, although it is hard to know how widely deployed it is in AI speech generation products and services.

\paragraph{Rendering data unusable for deepfake creation}

RW-VoiceShield "applies adversarial perturbation to the spectrogram" \citep[][Section 3.1]{RW-VoiceShield} which substantially reduces the effectiveness of a \gls{VC} system that uses this protected speech data as the \gls{target_speaker}. This is a novel task, and it is as yet unclear whether such an approach could offer universal protection.

\mysummarybox{{for deepfake detection}}{
\begin{itemize}
\item Most research to date has been directly on the \gls{deepfake} detection task, exploring different features, classifiers, and ensembles of both.
\item Interest is growing in going beyond mere detection and in providing some degree of explainability.
\item So far, this has meant attributing a \gls{deepfake} to a particular source speaker or algorithm.
\item Associated areas of research are appearing, and we can expect new ones to emerge over the next few years:
\begin{itemize}
\item synthetic speech generation or manipulation specifically designed to attack a detector, primarily by evading detection but potentially also being attributed to an incorrect source;
\item sidestepping the need for \gls{deepfake} detection, for example by watermarking synthetic speech, or protecting \gls{bona_fide} speech from misuse.
\end{itemize}
\end{itemize}
}

\subsection{Review and survey articles on deepfake detection}
\label{section:deepfake:reviews_and_surveys}

This is a much shorter list than for AI speech generation. This is not because there are few reviews to choose from: on the contrary, this is a fashionable area and there are far too many review articles. Regrettably, most lack authority and have little to recommend them.

\begin{itemize}
\item If looking for a single authority on \gls{deepfake} detection, then read the work of Jianhua Tao, starting with \citet{yiAudioDeepfakeDetection2023}.
\item \citet{liSurveySpeechDeepfake2025} is a traditional literature review on the subject, cataloguing all the necessary elements and
\citet{mubarakSurveyDetectionImpacts2023} is another traditional literature review, with comprehensive tables of methods and an extensive bibliography covering both visual and audio \gls{deepfake} detection, plus a list of existing review articles by other authors.
\item \citet{zhangAudioDeepfakeDetection2025} contains another useful table listing other reviews, in addition to the authors' own.
\end{itemize}


\subsection{Datasets for deepfake detection}

A comprehensive inventory of datasets is beyond the scope of this report. Instead, a few selected datasets are described, to: illustrate current thinking; understand that creating and distributing a dataset looks simple but is not; reveal limitations of current approaches to \gls{deepfake} detection.

Historically, before the advent of the term \gls{deepfake}, the ASVspoof challenge series (Section~\ref{section:history:challenges}) was the primary source of data for training an \gls{anti_spoofing} algorithm. The data were either collated from other challenges on \gls{VC} and \gls{TTS}, or purposely-created in the case of replay detection. An important lesson learned at that time was that flaws in a dataset could be exploited by the \gls{anti_spoofing} algorithm, giving an over-optimistic impression of performance. The creators of future datasets would be well-advised to carefully understand how these flaws occurred. 

More recent datasets still take the same basic approach: collate examples of the type of speech to be detected, plus some \gls{bona_fide} speech. These are then divided into disjoint subsets for \gls{training} and testing. Where a public challenge is anticipated, two disjoint test sets are needed: one for participating teams to use during development, and an entirely unseen test set reserved for final system evaluations. 

Whether to distribute such a dataset publicly is not a simple decision. The advantages of Open Source (e.g., reducing entry barriers, a level playing-field, enabling fair cross-system comparisons)  need to be weighed against some quite severe disadvantages. As we saw in the two "current state-of-the-art" Sections \ref{section:SotA:generation} and \ref{section:SotA:detection}, many current approaches start from large pre-trained models, trained once (e.g., an \gls{SSL} model released by well-resourced organisation such as Meta) and used many times (e.g., by a \glsreset{SME}\gls{SME} creating a \gls{deepfake} detection product). Any easily-accessible data is vulnerable to being included in automatically-collected data (e.g., using a pipeline such as Emilia described  on page~\pageref{page:emilia}) and hence contaminating these large models (see also Section~\ref{section:risks:pollution}). This makes it difficult to obtain genuinely unseen test examples, and increases the risk that researchers are inadvertently testing their systems on examples that (even if only indirectly, via a pre-trained model) was included in the \gls{training} data. Some challenges mitigate this by only allowing systems to be trained on specified data, but this is unsatisfactory because it disallows the use of pre-trained models.

\noindent Some illustrative datasets are:

\begin{itemize}

\item The lab-based RedDots~Replayed dataset \citep{RedDotsReplayed} was found to have flaws including digital silence, something that subsequent datasets need to avoid.

\item MLAAD (Multi-Language Audio Anti-Spoofing Dataset) \citep{MLAAD} covers 23  languages, to complement the single-language datasets that otherwise dominate this field.

\item \citet{mullerDoesAudioDeepfake2022} took a harder but more ecologically-valid approach for their InTheWild dataset: they collected actual \glspl{deepfake} instead of simulating them in the lab; importantly, if unsurprisingly, they found that these 'in the wild' deepfakes are much harder to detect than those from typical challenges.

\item CodecFake \citep{CodecFakeEnhancing} passes recorded natural speech through several \glspl{NAC} to create data for training a \gls{deepfake} detector, which performs well on unseen examples from a \gls{TTS} model that uses a \gls{NAC}. This confirms that, like most \gls{SotA} detectors, it relies on low-level artefacts from the \gls{NAC} and not imperfections in the \gls{TTS}.

\item WaveFake \citep{WaveFakeDataSet} claims to be for training \gls{deepfake} detectors, but only contains speech from a single speaker (LJ~Speech) passed through a handful of \glspl{NAC}. This is an example of a dataset being created by well-intentioned researchers from an adjacent field (in this case, in "IT Security").

\item FakeOrReal (FoR) \citep{FakeOrRealDataSet} is another example of a dataset created by researchers outside the mainstream of \gls{deepfake} detection. It only contains samples for a few commercial \gls{TTS} services from major vendors which prove very easy to detect when seen during training.

\end{itemize}

\noindent Although it is little more than a call-to-action, \citet{akhtarVideoAudioDeepfake2024} does contain useful lists of image, video, and speech datasets. \citet{liSurveySpeechDeepfake2025} also contains a useful list.

\paragraph{Data augmentation} In many  classification tasks, such as \gls{ASR}, modest but consistent improvements in accuracy are often gained by augmenting the \gls{training} data: this involves manufacturing further data from a fixed, existing dataset. RawBoost \citep{RawBoost} applies expert-designed perturbations that simulate channel effects including telephone, encoding, recording, noise, etc. to make a \gls{deepfake} detector more robust. \citet{geSpoofingAttackAugmentation2024} \label{page:geSpoofingAttackAugmentation2024} show that some classes of \gls{deepfake} detector are not fully robust even to \textit{previously-seen} forms of attack (i.e., synthetic speech) if the attack model is perturbed (training data, number of channels in the \gls{mel_spectrogram}, even the random seed for \gls{inference}), which is further evidence that detectors over-rely on very specific artefacts; they propose augmenting the training data with such perturbations to improve robustness. \citet{kwakLowqualityFakeAudio2022} use masking and other augmentation techniques with the same motivation: preventing the detector relying on anything too specific in the synthetic speech. These are all useful, if obvious, techniques to incrementally improve detector robustness.

\mysummarybox{{of datasets for deepfake detection}}{
\begin{itemize}
\item Evaluating a \gls{deepfake} detector obviously requires a dataset comprising examples of known \gls{deepfake} speech, and \gls{bona_fide} speech.
\item Constructing such a dataset, and using it to conduct a fair and meaningful evaluation, are harder than it first appears.
\item Developing a detector also appears to require a similar dataset for \gls{training} the model, but this brings a fundamental risk of overfitting to the forms of \gls{deepfake} represented in this data.
\item Overfitting is a widespread problem across machine learning, and there are some standard ways to mitigate this, including large and diverse \gls{training} data, data augmentation, and model \gls{training} paradigms.
\item But \gls{deepfake} detection is different to classifying naturally-occurring samples (e.g., \gls{ASR} for speech, or image classification for photographs) because entirely new forms of \gls{deepfake} will continue to be invented.
\end{itemize}
}

\newpage

\section{Concluding remarks}
\glsresetall

\subsection{Access to AI speech generation technology capable of creating deepfakes}
\label{section:access}

Practically all corporations and many \Glspl{SME} offer what is called "custom voice building", meaning the creation of a fully-functional \gls{TTS} system from customer-supplied recordings with little or no skill required from the customer. Such a system could of course be used to make low-effort \glspl{deepfake}, but these would be readily detectable since they are from a known model.

To obtain a \gls{deepfake} that has the potential to evade a detector requires creating it with a \textit{novel} method, or hiding its artefacts using post-processing. Just how novel that method needs to be, remains an open question. On page~\pageref{page:geSpoofingAttackAugmentation2024} we saw that \citet{geSpoofingAttackAugmentation2024} found some detectors to fail even for synthetic speech from systems that were included in the detector's \gls{training} data. This occurred when the speech generation model had undergone almost trivial changes, well within the capability of a typical undergraduate computer science student working with an Open Source model and freely-available cloud computing resources.

It is likely that detectors can be made robust to such medium-effort \glspl{deepfake}. Creating a high-effort  \gls{deepfake} at the present time requires enough expertise to make fairly significant modifications to an existing model, or to create a new model (e.g., re-using existing components from multiple existing models), and perhaps to train that model from scratch (this currently requires computing resources costing a few \$1000s).

Many model \glspl{checkpoint}, and the necessary code to run them, are not made freely available, to prevent their misuse. For example, Meta say "because of the risks of misuse, we are not making the Voicebox model or code publicly available at this time"\footnote{\href{https://voicebox.metademolab.com}{voicebox.metademolab.com}} even though they also built a detection model which was claimed to be highly-effective. Releasing \textit{code} would allow substantial changes to be made, with the potential to evade that detection model. But there are already plenty of other Open Source codebases available, from which a bad actor might create a system.

\subsection{Benefits of AI speech generation}
\label{section:benefits}

\paragraph{Mass-market products and services} include digital assistants, games, social media, and content localisation. As we saw in the \gls{TTS} timeline in Section~\ref{section:TTS_timeline}, the first major commercial breakthrough came with using \gls{unit_selection} for \gls{IVR} in the late 1990s, followed by \gls{hybrid_unit_selection} in digital assistants from around 2010. The size of the market and the range of use-cases will continue to widen as the capabilities of AI speech generation improve.

\paragraph{Niche applications} may have lower commercial returns but give immense benefit to certain individuals, including reading text out loud to people with a visual impairment, or using voice cloning to provide personalised synthetic voices to people who are going to lose their own voice (e.g., the company SpeakUnique).\label{page:speakunique} 

\paragraph{Voice privacy} was historically achieved by distorting the signal, but can now be done using a specialised variant of \gls{VC}, in real-time if required. The putative application of voice privacy is hiding  identity when interacting with automated systems (e.g., \gls{ASR}), but there are other applications such as hiding accent and therefore ethnicity for call-centre workers (e.g., the company Sanas).

\pagebreak

\subsection{Threats from AI speech generation}
\label{section:threats}

The high quality of synthetic speech that is now possible brings a number of threats. Some threats come about simply through the use of the technology for ordinary commercial applications, but others come from its misuse.

\paragraph{The expanding use of AI speech generation} brings obvious threats that are simply the inevitable commercial consequences of progress in technology, such as displacing \glspl{voice_actor} with \gls{TTS}. These are part of the wider story of AI advancement and will play out in the market over the coming years.

\paragraph{Modern \glsreset{NAC}\glspl{NAC}} may supersede classical compression methods (e.g., mp3) and previous-generation machine learning \glspl{audio_codec} (e.g., Opus) for audio compression in online videos and social media, and for internet voice calling. This means that merely detecting \gls{NAC} artefacts (mentioned on page~\pageref{page:moussaUnmaskingNeuralCodecs2024}) will not be useful.

\paragraph{The increasing quantity of AI-generated speech online} poses a more subtle threat, which was alluded to in the description of data acquisition and curation pipelines in Section~\ref{section:provenance_of_data}. Future web-scraped data will become polluted with synthetic samples which will in turn contaminate models trained on that data -- probably a greater concern for \gls{deepfake} detectors than speech generation models. A solution used in text-based \gls{NLP} is to restrict data to sources that existed prior to the recent AI explosion. This is not a viable long-term solution: because language usage evolves, we need to regularly update the data used to train models. \label{section:risks:pollution}

\paragraph{Genuine (bona fide) speech may be incorrectly labelled as fake.} Deepfake detection is already a necessary form of protection for consumers of media in general, and social media in particular.
But all current detectors are imperfect: when deploying them, an operating point must be selected which will \textit{inevitably} result in incorrectly labelling some \gls{bona_fide} speech as fake.

\paragraph{The interplay between speech generation and deepfake detection} will continue, as a natural continuation of the interplay between \gls{spoofing} and \gls{anti_spoofing} for \glsreset{ASV}\gls{ASV}. We should expect increasing sophistication in the attack methods, starting with approaches that hide low-level signal artefacts from the detector, which should stimulate the development of detectors that move beyond artefact detection.

\subsection{Wrapping up}

This report devoted a considerable amount of space to describing the history of AI speech generation, especially the timelines for \gls{TTS}, \gls{VC}, and the challenges that play an important role in setting the research agenda. This was time well-spent, because it allowed us to understand that the current \gls{SotA} is not really a technological revolution, but just another step on a long journey. 

But there is a reason that the most recent developments could still be called revolutionary. This is not because of the technology (which will always keep evolving), but because the naturalness that is now possible has passed an important threshold: it can sometimes deceive listeners into believing they are hearing natural speech. Two other important thresholds were passed earlier in history: \gls{hybrid_unit_selection} reaching the naturalness required for consumers to elect to use digital assistants; \gls{SPSS} deceiving the first generation of \glsreset{ASV}\gls{ASV} systems.

Synthetic speech is not yet undetectable by automatic \gls{deepfake} detection systems. Could this be the next threshold that will be passed? 

\newpage
\label{page:acronyms}
\printglossary[type=\acronymtype,title={Acronyms}, toctitle={List of acronyms}]

\printglossary[title={Technical terms}, toctitle={Glossary of technical terms}]

\pagebreak
\printbibliography[heading=bibintoc]

\fancyfoot[R]{Page~\thepage~(last page)}

\end{document}